%% file: main.tex
\documentclass[10pt,journal,compsoc]{IEEEtran}
\usepackage{color}
\usepackage{graphicx}
\usepackage{multirow}
\usepackage{amsmath}
\usepackage{amssymb}
\usepackage{amsthm}
\usepackage{syntax}
\usepackage{xcolor,listings}
\usepackage{fancyvrb}
\usepackage{enumitem}
\usepackage{alltt}
\usepackage{booktabs}
\usepackage[skip=0pt]{caption}
\usepackage[aboveskip=-2pt,belowskip=0pt]{subcaption}
\usepackage[linesnumbered,ruled,noend]{algorithm2e}

\newtheorem{definition}{Definition}

\ifCLASSOPTIONcompsoc
  \usepackage[nocompress]{cite}
\else
  \usepackage{cite}
\fi
\ifCLASSINFOpdf
\else
\fi
\begin{document}
%
\title{An Optimized Tri-store System for Multi-model Data Analytics (Technical Report)}
%
%
%
%

\author{Xiuwen Zheng, Subhasis Dasgupta, Arun Kumar, Amarnath Gupta
\IEEEcompsocitemizethanks{\IEEEcompsocthanksitem This work was partly funded by NSF Award \# 1909875.}
}

%
%

\markboth{IEEE Transactions on Knowledge and Data Engineering}%
{Shell \MakeLowercase{\textit{et al.}}: Bare Demo of IEEEtran.cls for Computer Society Journals}
%



\IEEEtitleabstractindextext{%
\input{sections/0-abstract.tex}
}

\maketitle

\IEEEdisplaynontitleabstractindextext

%
\IEEEpeerreviewmaketitle

\input{sections/1-introduction}
\input{sections/2-relatedwork}

\input{sections/3-language}
\input{sections/3.2-workload.tex}
\input{sections/4-system-overview}
\input{sections/5-validation}

\input{sections/6.2-physical-planning}
\input{sections/6.1-logical-query-planning}
\input{sections/7-costmodel}

\input{sections/8-experiment}

\input{sections/2-relatedwork2}

\input{sections/9-conclusion-and-future-work}
\bibliographystyle{IEEEtran}
\bibliography{IEEEabrv,ref}
\vskip -2\baselineskip plus -1fil
\begin{IEEEbiographynophoto}{Xiuwen Zheng}
    Xiuwen Zheng is  a Ph.D. student of CSE at UC San Diego, advised by Amarnath Gupta and  Arun Kumar. Her research interests  are in  polystore database and  machine learning systems.
\end{IEEEbiographynophoto}
\vskip -2\baselineskip plus -1fil
\begin{IEEEbiographynophoto}{Subhasis Dasgupta}Subhasis Dasgupta is a researcher of SDSC at UC San Diego. His research interests are in information security, grid computing and query processing.
    
\end{IEEEbiographynophoto}
\vskip -2\baselineskip plus -1fil
\begin{IEEEbiographynophoto}{Arun Kumar}
Arun Kumar is an Associate Professor of CSE and HDSI at UC San Diego. His research interests are in data management and systems for ML/AI workloads.
\end{IEEEbiographynophoto}
\vskip -2\baselineskip plus -1fil
\begin{IEEEbiographynophoto}{Amarnath Gupta}
Amarnath Gupta is a Research Scientist of SDSC at UC San Diego. His research interests are in scientific data modeling, information integration and  multimedia databases.
\end{IEEEbiographynophoto}


%

\newpage
\appendices
\input{appendix}
\end{document}

%% file: sections/0-abstract.tex
\begin{abstract}
Data science applications increasingly rely on heterogeneous data sources and analytics. 
This has led to growing interest in polystore systems, especially analytical polystores. 
In this work, we focus on a class of emerging multi-data model analytics workloads  that fluidly straddle relational, graph, and text analytics. 
Instead of a generic polystore, we build a ``tri-store'' system that is more aware of the underlying data models to better optimize execution to improve scalability and runtime efficiency.
We name our system AWESOME (Analytics WorkbEnch for SOcial MEdia). It features a powerful domain-specific language named ADIL. ADIL builds on top of underlying query engines (e.g., SQL and Cypher) and features native data types for succinctly specifying cross-engine queries and NLP operations, as well as automatic in-memory and query optimizations. 
Using real-world tri-model analytical workloads and datasets, we empirically demonstrate the functionalities of AWESOME for scalable data science applications and evaluate its efficiency.
\end{abstract}


%% file: sections/1-introduction.tex
\IEEEraisesectionheading{\section{Introduction}\label{sec:intro}}

Multiple modalities of data are increasingly common in data science workloads across myriad domains, including social sciences, \cite{wolff2021information,moustaka2018systematic}, enterprises, healthcare, and cybersecurity~\cite{mehta2018concurrence,mageto2021big,lu2018internet}. For instance, in our ongoing multi-year collaboration with political scientists at UCSD, we find that analyzing large volumes of tweets, microblogs, and news corpora enable them to get more insights into sociopolitical phenomena such as disinformation on social media during elections or debates over criminal justice reform~\cite{wu2019constructing,zheng2019social}. 

Naturally, there is growing demand for systems for ``multi-model'' analytics in both sciences and industry~\cite{state-of-the-art-analytics,lu2019multi}. Such systems must support cross-data model queries and analytics~\cite{zhang-benchmark-2021}. 
\textit{Polystore systems} have emerged recently to meet this need~\cite{duggan2015bigdawg2,khan2019one,lu2019multi, shrestha2020survey, guo2020multi}.
They are typically a \textit{middleware} to access multiple underlying data stores but which give users the illusion of a single engine. But as the complexity of analytical workloads grows, it is critical to not just support cross-store queries for retrieval or simple analysis but also \textit{complex analytical operations}.
In this work, we focus on a large emerging class of multi-model workloads that require complex analytics across the three canonical data modalities of graphs, relations, and text. 



\subsection{Motivating Workloads}
We start with two illustrative workloads that motivated us to design an analytical tri-store system. 

\begin{figure}[t]
    \begin{subfigure}{.5\textwidth}
      \centering
      \includegraphics[width=0.85\linewidth]{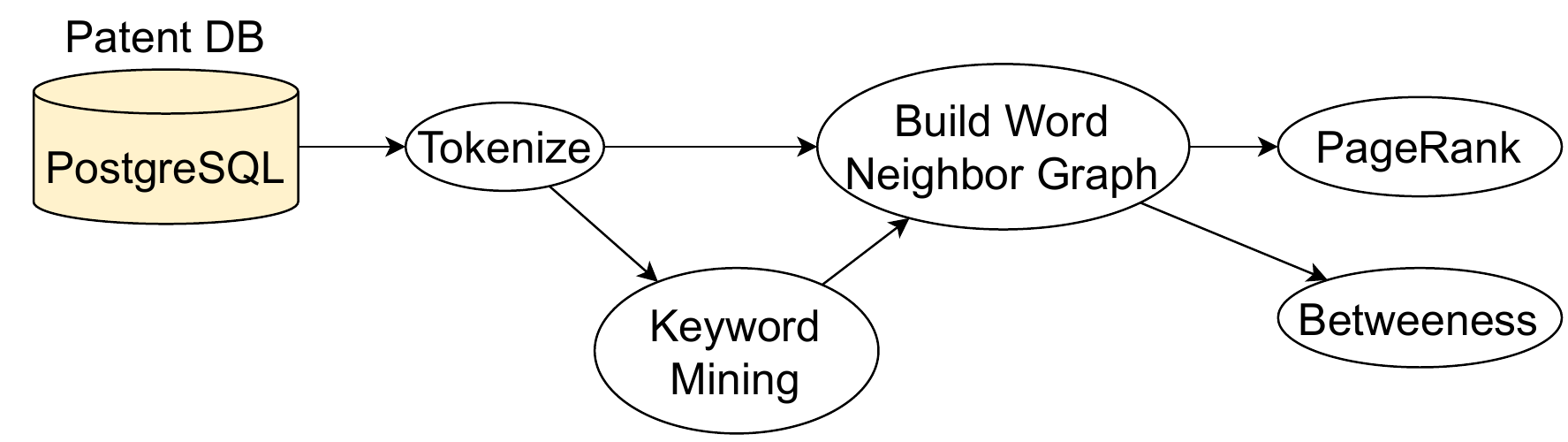}
       \caption{In-memory execution Plan.}\label{fig:pa1}
    \end{subfigure}%
    
    \begin{subfigure}{.5\textwidth}
      \centering
      \includegraphics[width=0.85\linewidth]{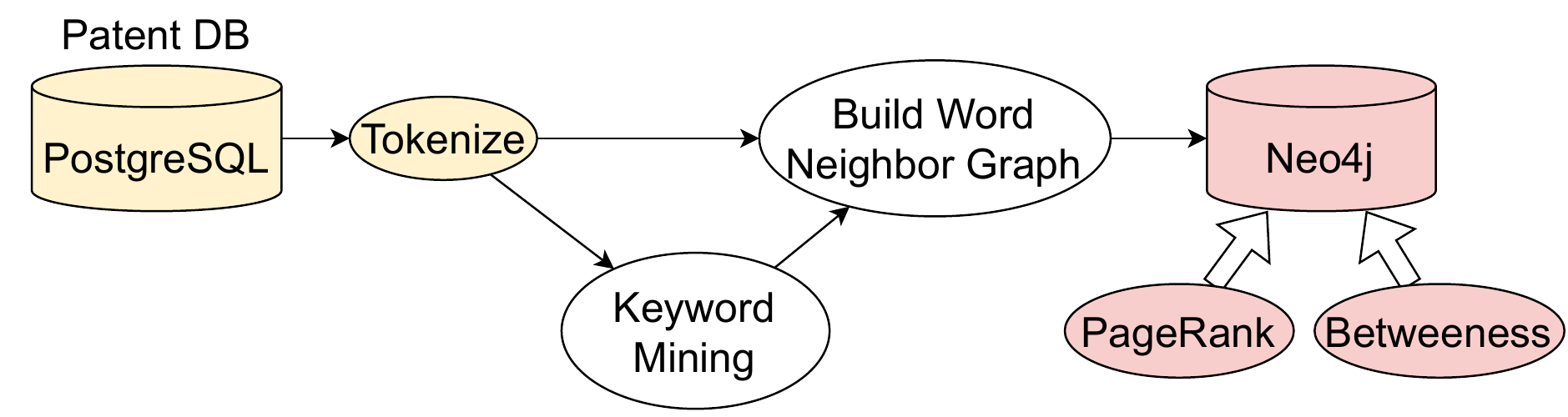}
      \caption{In-DBMS execution Plan.}
    \end{subfigure}%
    \caption{Illustration of \textit{PatentAnalysis} workload.} \label{fig:pa}
\end{figure}

Fig.~\ref{fig:pa} presents the \textbf{PatentAnalysis} workload, first introduced and used in~\cite{ampornphan2020exploring}. It mines influential terms/phrases from patent corpus. The analysis pre-processes patent text (e.g., tokenization), extracts keywords, and creates a word neighbor graph to capture word co-occurrence in abstracts. Finally, graph algorithms (betweeness and PageRank) obtain representative words/phrases.

\begin{figure}[t]
\centering
     \includegraphics[width=0.48\textwidth]{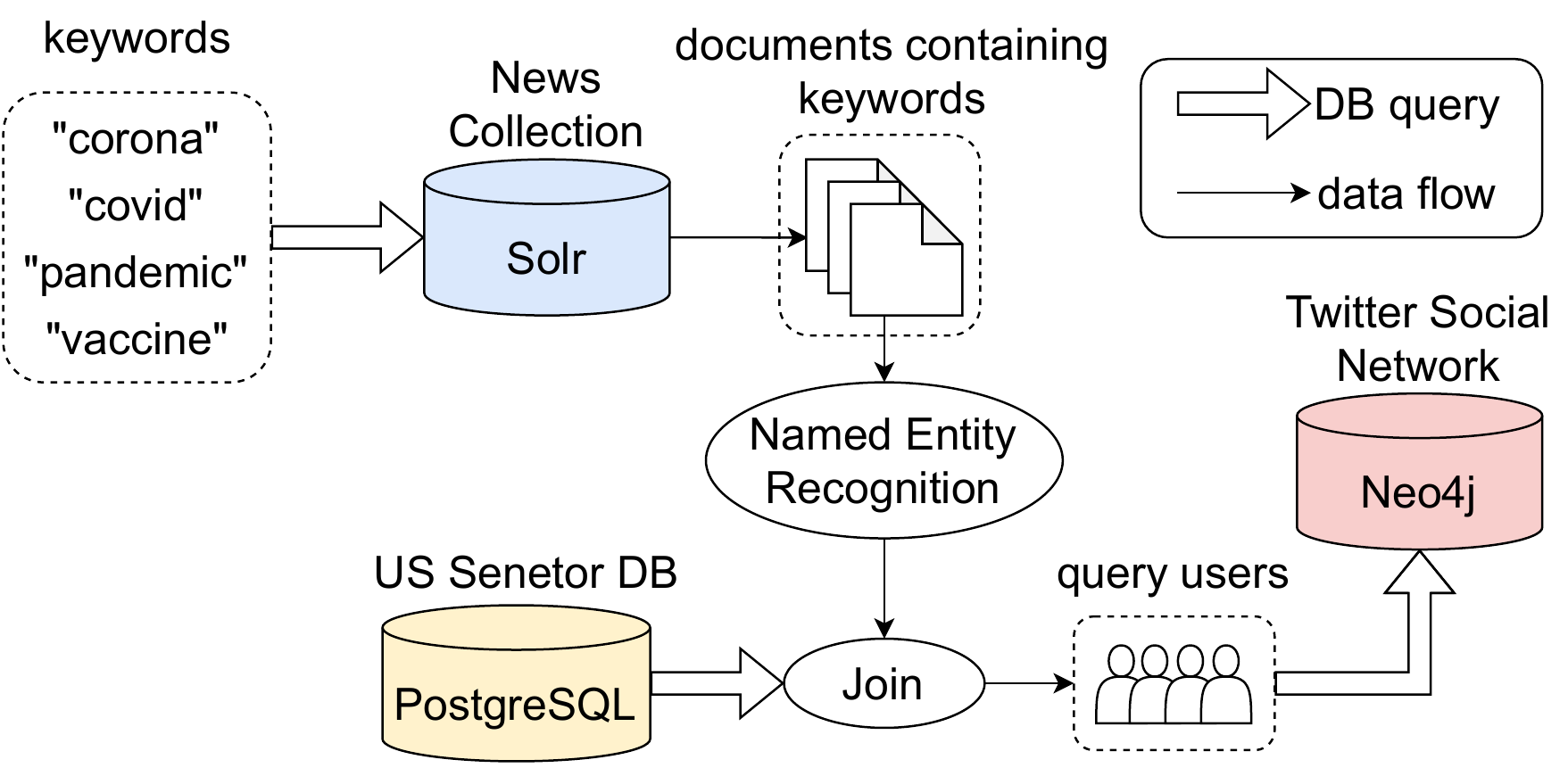}
     \vspace{2mm}
    \caption{Illustration of \textit{PoliSci} workload.}
    \vspace{2mm}
     \label{fig:poliscidiagram}
\end{figure}

Fig.~\ref{fig:poliscidiagram} presents the \textbf{PoliSci} workload, based on our collaboration at UCSD. It jointly analyzes news articles and tweets. It retrieves news reports related to COVID-19 via text queries on Solr. Then it uses named entity recognition (NER) to identify entities (e.g., "President Trump") in the corpus. That entity list is then joined with a table of Twitter handles of US Senators, stored in PostgreSQL. Finally, the Twitter social graph, stored in Neo4j, is queried to obtain users who mentioned any of those Twitter users and all tweets that contain any of those Senators' names.

Taken together, these workloads exemplify an emerging class of workloads, especially on social media data or other network-associated data, with two defining characteristics:

\noindent\textbf{C1:} They straddle three modalities: graph (e.g., Neo4j), RDBMS (e.g., PostgreSQL), and text search (e.g., Solr).

\noindent\textbf{C2:} They involve a wide range of \textit{analytical operations}, such as key phrase mining, named entity recognition.

Given the growing importance of such workloads, in this work we propose an optimized ``tri-store'' system straddling relational, graph, and text analytics. We name our system AWESOME: Analytics Workbench for Social Media Data. Table~\ref{tab:relatedwork} compares AWESOME's capabilities against prior art (more in Section 8). We now explain the system desiderata.

\begin{figure*}[ht]
    \centering
    \includegraphics[width=0.85\textwidth]{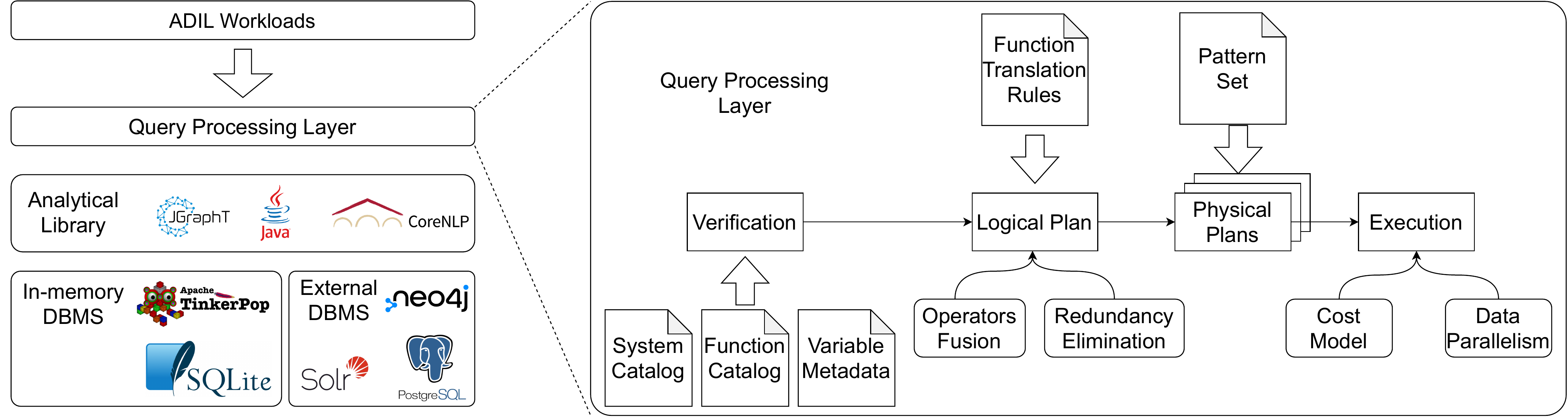}
    \vspace{2mm}
    \caption{Illustration of the AWESOME system architecture.}
    \vspace{2mm}
    \label{fig:sysoverview}
\end{figure*}

\subsection{Prior Art vs.~Our Desiderata}

We find that most prior polystore systems, due to thier high generality needs, focus mainly on \textit{logical level} of optimizations and tend to loosely couple the ``unistores'' of the data models. Many also do not account for (expensive) complex analytical operations in data science, especially NLP functions. One can aim to reimplement such analytics operations in a DBMS as user-defined functions (UDFs), but that is tedious, error-prone, and typically inefficient compared to the native analytics libraries. Based on these observations, we have our first desideratum:

\vspace{1mm}
\noindent\textbf{D1:} The system must better optimize across DBMSs for graphs, relations, and text IR--a tri-store rather than a polystore--without modifying their system code, while also exploiting relevant NLP analytics libraries.

\vspace{1mm}
A recent polystore, RHEEM~\cite{agrawal2018rheem}), does unify multiple DBMSs and and also analytics platforms such as Spark. 
But users are required to learn and write analyses its own \textit{platform-agnostic} programming language (RHEEMLatin). RHEEM also has its own low level operations that limit how much the power of the underlying unistores can be harnessed, resulting in a loss of semantic completeness and potentially more coding burden on users. Based on these observations, we have two more desiderata:

\vspace{1mm}
\noindent\textbf{D2:}  The system must preserve semantic completeness and power of underlying unistores and analytical tools.

\vspace{1mm}
\noindent\textbf{D3:} The system must minimize additional coding learning curve on top of underlying unistores as much as possible.




\subsection{Design Decisions in AWESOME}
\label{sec:design-decisions}

\noindent \textbf{Tri-Model Dataflow Language.} To enable users to express tri-store analytics succinctly, we craft a unified high-level language to support full unistore queries, e.g., SQL over RDBMS (fulfills \textbf{D2}), high-level analytics functions from libraries (fulfills part of \textbf{D1}), and basic Python-inspired data types to handle query arguments and intermediate data, as well as some control flow such as iteration (fulfills \textbf{D3}).
We do rigorous semantics checks at compile time.

\vspace{1mm}
\noindent \textbf{Complex Analytical Functions.} For analytics on heterogeneous data (\textbf{C2} and part of \textbf{D1}), we integrate popular analytical libraries for NLP and graph mining. Extensibility via UDFs is also supported. For example, in the \textit{PoliSci} workload, the Solr query result is sent to an \texttt{NER} operator from the CoreNLP~\footnote{https://stanfordnlp.github.io/CoreNLP/} library.

\vspace{1mm}
\noindent \textbf{Tri-Store Middleware.} AWESOME works transparently with 3 underlying uni-stores: RDBMS, graph DBMS, and text IR systems (as per \textbf{C1}). We use PostregSQL, Neo4j, and Solr for the prototype but our system design is general. Intermediates are handled automatically without user exports/imports. For example, in the \textit{PoliSci} workload, the named entities are joined with a table in PostgreSQL and that results is sent to a Neo4j Cypher query.

\vspace{1mm}
\noindent \textbf{Holistic Optimization.} To fulfill \textbf{D1}, AWESOME performs \textit{physical level} optimizations too, such as in-memory caching across, data parallelism for operators, and unistore/library selection when there is a choice. Such optimizations are planned \textit{holistically}, instead of individually for operators, to reduce data movement and maximize resource utilization.

\subsection{Summary of Technical Contributions and Novelty}

\begin{itemize}[leftmargin=*]
    \item We present the formal description of ADIL, a dataflow language straddling relations, graphs, and text IR with rigorous semantics. This paper significantly extends the initial version of ADIL in~\cite{gupta:awesome:2016}, which lacked the expressive power for complex analytical workloads. 

    \item We implement the prototype of AWESOME, the first scalable and optimized tri-store system for data science workloads spanning relations, graphs, and text IR. 
    Figure~\ref{fig:sysoverview} illustrates the system architecture of AWESOME. 

    \item We formalize how AWESOME creates query plans and executres them. We devise a suite of transparent query optimizations, including rewrites, to reduce runtimes by reducing data movement, carefully apportioning resources, pattern-based physical plan generation, cost-based plan selection, and placing computation in memory as needed.

    \item Two specific novel technical aspects in AWESOME are its capability-based rewriting over in-memory vs.~persistent unistores and pattern-based physical plan optimization for tri-model workloads. 
    
    On the former, AWESOME supports in-memory data handling as needed and careful orchestration of intermediates across backend unistores to minimize total cross-DBMS data movement and optimize overall runtimes. Likewise, it can automatically select between in-memory analytics libraries and in-DBMS functions depending on data sizes, e.g., between JGRaphT and Neo4j for betweenness centrality. As we will show later in Section~\ref{sec:exp}, such cross-engine rewrites can save up to 88\% run times.

    On the latter, we implement in-depth physical-level execution optimizations for cross-model analysis plans. To this end, we adopt a \texttt{pattern-based} optimization policy where a pattern is a sub-plan, i.e., a subtree of (retrieval and/or analytical) operations that can be \textit{optimized as a unit}. As we will show later in Section~\ref{sec:exp}, such pattern-based planning can be significantly faster than operator-level only planning. AWESOME also implements data parallelism for most operators to raise resource utilization and reduce runtimes.

    \item We present an extensive empirical evaluation using real-world datasets and workloads to demonstrate AWESOME's support for tri-model analytics, higher scalability, and lower runtimes compared to prior art and strong baselines.
    For example, AWESOME is \textbf{4.6$\times$} faster than the best baseline for the \textit{PoliSci} workload.
\end{itemize}

%% file: sections/2-relatedwork.tex
\vspace{-1.5ex}
\begin{footnotesize}
\begin{table*}[ht!]
\centering
\caption{Major technical features of existing polystore systems and AWESOME.}\label{tab:relatedwork}
\begin{tabular}{|c|ccccc|ccccc|}
\hline
\multicolumn{1}{|l|}{\multirow{2}{*}{}} & \multicolumn{5}{c|}{Language}& \multicolumn{5}{c|}{System Design} \\ \cline{2-11} 
\multicolumn{1}{|l|}{}& \multicolumn{1}{c|}{\begin{tabular}[c]{@{}c@{}}DBMS \\ Query\end{tabular}} & \multicolumn{1}{c|}{\begin{tabular}[c]{@{}c@{}}UDF \end{tabular}} & \multicolumn{1}{c|}{\begin{tabular}[c]{@{}c@{}}Graph  \\ Analytics\end{tabular}} & \multicolumn{1}{c|}{\begin{tabular}[c]{@{}c@{}}Text  \\ Analytics\end{tabular}}  & \multicolumn{1}{c|}{\begin{tabular}[c]{@{}c@{}}Control \\ Flow\end{tabular}} & \multicolumn{1}{c|}{\begin{tabular}[c]{@{}c@{}} Native Tri-\\Data Model\end{tabular}} & \multicolumn{1}{c|}{\begin{tabular}[c]{@{}c@{}}RDBMS \\ Support\end{tabular}} & \multicolumn{1}{c|}{\begin{tabular}[c]{@{}c@{}} Graph DBMS \\ Support\end{tabular}} & \multicolumn{1}{c|}{\begin{tabular}[c]{@{}c@{}}Text DBMS \\ Support\end{tabular}} & \begin{tabular}[c]{@{}c@{}}In-memory\\  DBMS Support\end{tabular} \\ \hline
BigDAWG~\cite{duggan2015bigdawg2,she2016bigdawg}  & \multicolumn{1}{c|}{\checkmark} & \multicolumn{1}{c|}{}&\multicolumn{1}{c|}{}&
\multicolumn{1}{c|}{}& & \multicolumn{1}{c|}{} & \multicolumn{1}{c|}{\checkmark} & \multicolumn{1}{c|}{} & \multicolumn{1}{c|}{\checkmark}& \\ \hline
Rheem~\cite{lucas2018rheemstudio,agrawal2018rheem,kruse2020rheemix} & \multicolumn{1}{c|}{}& \multicolumn{1}{c|}{\checkmark}& \multicolumn{1}{c|}{\checkmark}& \multicolumn{1}{c|}{\checkmark}& \checkmark& \multicolumn{1}{c|}{}& \multicolumn{1}{c|}{\checkmark}& \multicolumn{1}{c|}{}  &\multicolumn{1}{c|}{}& \\ \hline
Estocada~\cite{alotaibi2019towards,alotaibi2020estocada} & \multicolumn{1}{c|}{\checkmark}& \multicolumn{1}{c|}{}&\multicolumn{1}{c|}{}&\multicolumn{1}{c|}{}& & \multicolumn{1}{c|}{}& \multicolumn{1}{c|}{\checkmark}& \multicolumn{1}{c|}{\checkmark}&  \multicolumn{1}{c|}{\checkmark} &  \\ \hline
Tatooine~\cite{bonaque2016mixed} & \multicolumn{1}{c|}{\checkmark}& \multicolumn{1}{c|}{}&\multicolumn{1}{c|}{}&\multicolumn{1}{c|}{}& & \multicolumn{1}{c|}{}& \multicolumn{1}{c|}{\checkmark}& \multicolumn{1}{c|}{\checkmark}& \multicolumn{1}{c|}{\checkmark} & \\ \hline
Myria~\cite{wang2017myria}& \multicolumn{1}{c|}{\checkmark}& \multicolumn{1}{c|}{\checkmark}&\multicolumn{1}{c|}{\checkmark}&\multicolumn{1}{c|}{}& \multicolumn{1}{c|}{\checkmark} & \multicolumn{1}{c|}{}& \multicolumn{1}{c|}{\checkmark}& \multicolumn{1}{c|}{}& \multicolumn{1}{c|}{}& \\ \hline
Hybrid~\cite{simmons2017hybrid,podkorytov2019hybrid}& 
\multicolumn{1}{c|}{\checkmark}& \multicolumn{1}{c|}{\checkmark}& \multicolumn{1}{c|}{} & \multicolumn{1}{c|}{}& \multicolumn{1}{c|}{}& \multicolumn{1}{c|}{}& \multicolumn{1}{c|}{\checkmark}&\multicolumn{1}{c|}{}& \multicolumn{1}{c|}{}& \multicolumn{1}{c|}{\checkmark}\\ \hline
\textbf{AWESOME}& \multicolumn{1}{c|}{\checkmark}&\multicolumn{1}{c|}{\checkmark}&\multicolumn{1}{c|}{\checkmark}&\multicolumn{1}{c|}{\checkmark}&\multicolumn{1}{c|}{\checkmark}& \multicolumn{1}{c|}{\checkmark}& \multicolumn{1}{c|}{\checkmark} & \multicolumn{1}{c|}{\checkmark}& \multicolumn{1}{c|}{\checkmark}& \multicolumn{1}{c|}{\checkmark}\\ \hline
\end{tabular}
\end{table*}
\end{footnotesize}

%% file: sections/3-language.tex
\section{ADIL: A Dataflow Language}
\label{sec: language}
\begin{figure}[t]
     \centering
     \includegraphics[width=0.46\textwidth]{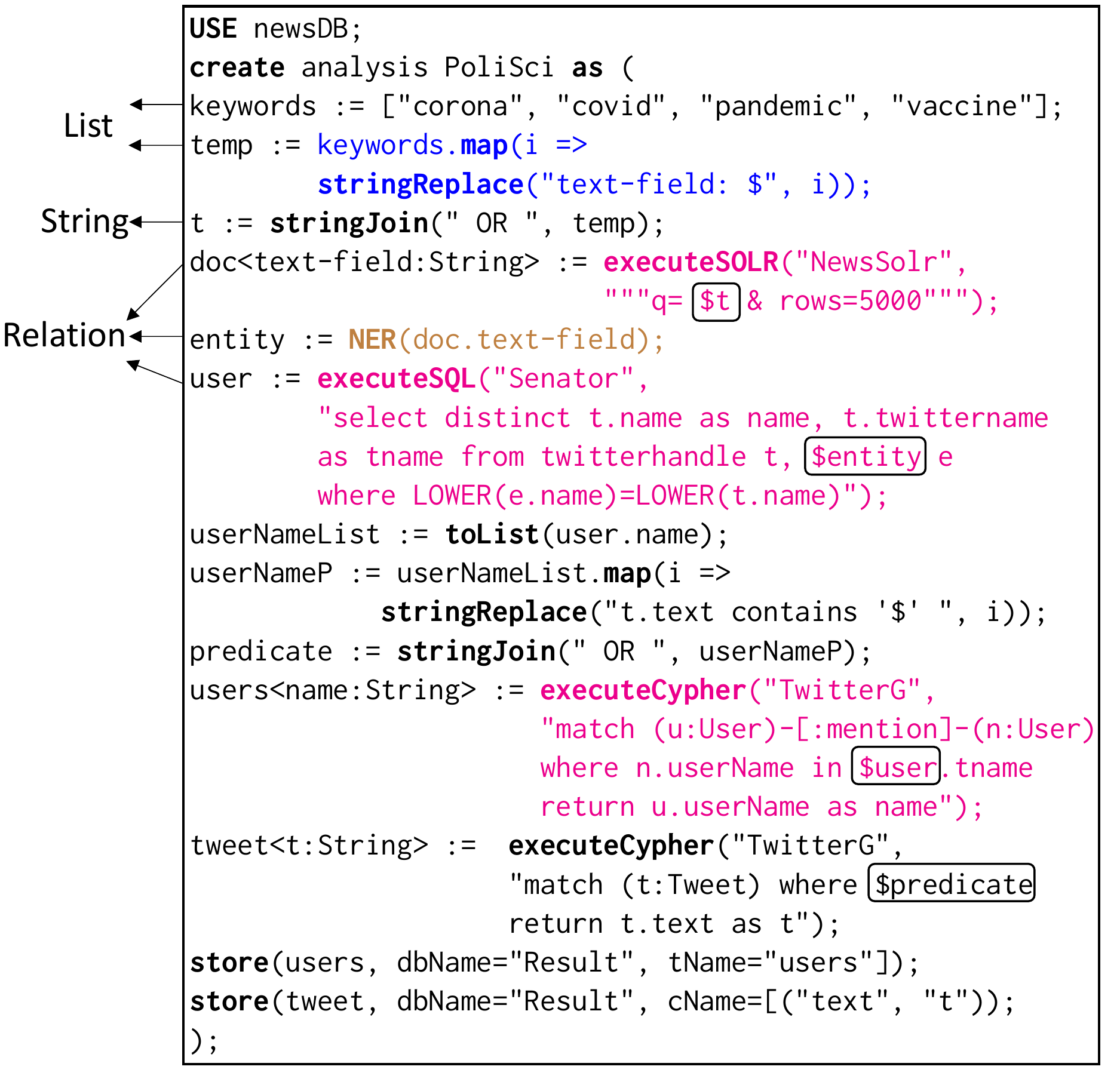}
     \caption{\textit{PoliSci} represented in ADIL.}
     \label{fig:polisciscript}
\end{figure}

ADIL, the surface language for AWESOME, is designed as a dataflow language. The user expresses an analysis workload in ADIL as a sequence of assignment statements where the LHS of the assignment is a variable or multiple variables and the RHS  is an expression.  Figure~\ref{fig:polisciscript} presents the ADIL script for the \textit{PoliSci} workload. 

\subsection{Data Types}
\label{sec:datatype}
ADIL  supports the following data types in native. We annotate the data types for some variables in Figure~\ref{fig:polisciscript}.
\begin{itemize}[leftmargin=*]
    \item \textbf{Primitive types}: \textsf{Integer}, \textsf{Double}, \textsf{String}, and \textsf{Boolean}. 
    \item \textbf{Relation and Record}: A \textsf{Relation}  variable represents a relational table and a \textsf{Record}  variable  is a single tuple of a relation. 
    \item \textbf{Property Graph and Graph Element}:  Users can  construct,  query against, or apply analytical functions (e.g.,  PageRank) on property graphs. A \textsf{GraphElement} variable can be either a node or an edge with labels and properties. 
    \item \textbf{Corpus and Document}: A \textsf{Corpus} is a collection of documents, and each \textsf{document} consists of document content (\textsf{String}), a document identifier (\textsf{Integer}) and tokens (\textsf{List<String>}).
    \item \textbf{Matrix}: We support \textsf{Matrix} data type and commonly-used matrix operators such as dot products on matrix-valued variables. In addition, an AWESOME matrix has  optional \textsf{row map} and \textsf{column map} properties which are semantic mappings from matrix row (resp. column) indices  to values in another  data type. For example, for a document term matrix, the row map is a mapping from row indices to the document ids and the column map is a mapping from column indices to terms (i.e., tokens).
    \item \textbf{Collection}: A \textsf{List} is a collection of indexed elements with homogeneous type; a \textsf{Tuple} is a finite ordered sequence of elements with any type.    \textsf{List} data type is strictly homogeneous: each element should have the same type. However,  there can be  heterogeneous objects in a  \textsf{Tuple} variable. For example, the following tuple $T$ contains a relation, a graph, a list and constant values. 

\begin{Verbatim}[commandchars=\&\~\!]
R := &textbf~executeSQL!(..., ...); //produces relation R
G := &textbf~BuildGraphFromRelation!(...); //produces graph G
T := {R, G, [1, 2, 3], "string", 2 }; 
\end{Verbatim}
\end{itemize}
In this paper, \textsf{Relation}, \textsf{PropertyGraph} and \textsf{Corpus} types are collectively referred to as the ``constituent data models'' because they correspond to the data models of underlying stores. 


\subsection{ADIL Workload Structure} 
\label{subsec:adil-syntax}
 An ADIL script starts by declaring a polystore instance registered in  AWESOME system  catalog:
\begin{Verbatim}[commandchars=+\[\]]
+textbf[USE] newsDB;
+textbf[create] analysis NewsAnalysis +textbf[as] {/*main code block*/}
\end{Verbatim}
AWESOME system  catalog is a file that maintains the metadata for each user-defined polystore instance including  the alias, connection detail, and schema of data stores in this instance. For underlying data store which admits a schema (e.g., PostgreSQL, Solr), a copy of the schema is maintained in the catalog. For stores that do not admit a schema (e.g., Neo4j), a set of schema-like information (e.g., node/edge labels/properties) is maintained. In the above example,  the  metadata of polystore instance \textit{newsDB} will be retrieved from the system catalog 
which contains the information of all DBMSs used in the workload named \textit{NewsAnalysis}. 

The main code block contains a sequence of assignment statements (Section \ref{subsubsec:assignment}) and store statements (Section \ref{sec:store}). 
\subsection{Assignment Statement}\label{subsubsec:assignment}
An ADIL assignment statement evaluates an RHS expression and assigns the result to one or more LHS variables. The grammar for assignment statement is shown as follows.

\begin{grammar}
<assignment-statement> ::= <var1> `, ' <var2> `, ' $\cdots$  `:=' <assign>

\vspace{-8pt}
<assign> ::= <basic-expr> | <ho-expr>
\end{grammar}
\vspace{-3pt}
The RHS expression (<assign>)  can be ``basic'' or ``higher-order'' explained  by the following grammar fragments,  
\begin{grammar} 
\label{grammar:expr}
<basic-expr> ::= <const>|<query>|<func> 

\vspace{-8pt}
<ho-expr> ::= <assign> `>' | `==' | `<' <assign>
\alt <var>`.map(' <lVar> `->' <assign> `)'
\alt <var>`.reduce((' <lVar1> `,' <lVar2> `) ->' <assign> `)'
\alt <var> ` where ' <assign> 
\end{grammar}


\subsubsection{Basic Expression}
<basic-expr> includes three types:

\noindent \textbf{Constant Expression (<const>):} A constant expression evaluates to a constant of any allowed data type. The expression can itself be a constant, e.g., \texttt{['x', 'y', 'z']},  or a prior constant variable, or an element of a prior collection  variable, e.g., \texttt{a[1]}.


\noindent \textbf{Query Expression (<query>):} A query expression executes a query against a data store  or against an AWESOME variable with a constituent data model. It uses standard query languages: SQL-93 for relational queries, OpenCypher~\cite{francis2018cypher} for property graph queries, and Lucene~\cite{mccandless2010lucene} for retrieval from text indices. In Figure~\ref{fig:polisciscript}, three query expressions are marked in pink and they use \texttt{executeSOLR}, \texttt{executeSQL} and \texttt{executeCypher} keywords respectively. The first argument of a query expression is the alias of target DBMS registered in the polystore instance.
If the query is  against a variable  created in prior statements, the first argument is left empty. 
The second argument is a standard Lucene/SQL/Cypher query with the exception of the \texttt{\$} followed by a variable name (highlighted by the rounded rectangles in the figure). ADIL uses \texttt{\$} as a prefix of the variable  passed as a parameter to a  query. 

\noindent \textbf{Function Expression (<func>):} 
AWESOME supports a rich native library for common data analytical tasks. The   expression   includes  function name with required positional parameters followed by optional  and  named parameters. A parameter can be a constant  or a variable. The expression can return a single or multiple variables. The \texttt{NER} function expression marked as brown in Figure~\ref{fig:polisciscript} takes a relation variable as parameter and returns a relation variable. 


\subsubsection{Higher-Order Expression.}\label{sec:highorder}
A higher-order  expression is recursively defined where another  expression  serves as its sub-expression.
The following  snippet from \textit{NewsAnalysis} workload shows an example statement where the RHS is a nested higher-order  expression:
\begin{Verbatim} [commandchars=+\[\]]
wtmPerTopic := topicID.+textbf[map](i =>
        WTM +textbf[where] getValue(_:Row, i) +textbf[>] 0.00);
\end{Verbatim}
\texttt{topicID} is a list of Integers and \texttt{WTM} is word-topic matrix where each row presents a word's weights on all topics. For each topic, it produces a word-topic matrix consisting of words with weights higher than 0 on this topic.
This snippet contains map, filter and binary comparison which are explained as follows. 



\noindent \textbf{Map Expression:} 
A map expression operates on a collection variable, evaluates a sub-expression for each element in the collection, and returns a new collection object. 
The sub-expression can be a constant, a query, a function or another higher-order expression. In this snippet, it takes a list of integers (\textit{topicID}) as input, and for each, applies another higher-order expression (a filter expression) on the \texttt{WTM} matrix to generate a matrix. Thus the returned variable (\textit{wtmPerTopic}) is a list of matrices. 

\noindent \textbf{Filter Expression:} The filter expression is indicated by the \texttt{where} clause -- its sub-expression is a predicate; it returns a new collection with values which satisfy the given predicate.  Since a  matrix can be  iterated by rows or  by columns, users need to specify the iteration mode: the underscore sign ($\_$) is used to represent  every  single element in the matrix, and the colon ($:$) followed by the type specify the element type. In the example snippet, it applies a binary comparison predicate on each row of the matrix and returns a new matrix consists of the rows satisfying the predicate.  

\noindent \textbf{Binary Comparison and Logical Operations:} A binary comparison accepts two expressions and  compares their values to return a Boolean value. 
In the example above, \begin{verbatim}
getValue(_:Row, i) > 0.00
\end{verbatim}
checks whether the $i$-th element of a row vector is positive.  More generally, ADIL supports any binary logical operators such as AND, OR and NOT over predicates. 

\noindent \textbf{Reduce Expression:} A \texttt{reduce} operation  aggregates results from a collection by passing a commutative and associative binary operator as its sub-expression. 
For example, the following snippet
\begin{Verbatim} [commandchars=+\[\]]
R := relations.+textbf[reduce]((r1,r2) => join(r1,r2,on="id"))
\end{Verbatim}
takes a list of relations as input and then joins each two tables and returns a new table at the end.




\subsection{Store Statement}
\label{sec:store}
A store statement specifies the variables to be stored to a persistent storage, which can be an underlying DBMS registered in the system catalog or the AWESOME file system; it also includes the instructions for how to store the variable. In Figure~\ref{fig:polisciscript}, the last two lines store \texttt{users} and \texttt{tweet} variables to   relational DBMS,   and specifies the DBMS alias (\texttt{dbName} parameter), table name (\texttt{tName} optional parameter) and mapping between  the targeted  column names to the relational variables' column names (\texttt{cName} optional parameter). 



\subsection{Some Properties of ADIL} 
\label{sec:expressiveness}
A full discussion of the formal properties of  ADIL  is beyond the scope of this paper. Here we provide a few  properties that will be useful in validating and developing logical plans from ADIL scripts.
\begin{enumerate}[leftmargin=*]
    \item ADIL does not have a \textsf{for loop} or a \textsf{while} operation. Instead, it uses the \textsf{map} operation to iterate over a collection and apply function over each element, the \textsf{filter} operation to select out elements from a collection that satisfies predicates, the \textsf{reduce} operation to compute an aggregate function on  a collection. In ADIL, the collection must be \textit{completely constructed} before the \textsf{map} (resp. filter or reduce) operation can be performed. Therefore, these operations are guaranteed to terminate.
    \item ADIL is strongly typed.
    \item In an assignment where the RHS expression is a query in a schemaless language like OpenCypher, the user must specify a schema for the LHS variable in the current system.
    \item The data type and some metadata information of any LHS variable can be uniquely and correctly determined by analyzing the RHS expression (see Section \ref{sec: check}).
\end{enumerate}


%% file: sections/3.2-workload.tex
\section{AWESOME Workloads}
\label{sec: workload}
We introduce five different types of queries supported by AWESOME. An ADIL workload can be a combination of  these different types of queries. We introduce some optimization opportunities for each query type and  present   three example workloads to be evaluated.
\subsection{Query Types}
ADIL supports a wide variety of analytical workloads that utilize and analyze heterogeneous data, such as relation, graph, and text data, through the following five query types:

\begin{itemize}[leftmargin=*]
    \item Single-DBMS query. A query written  in a standard query  language for a data model, e.g., SQL and Cypher,  with variables passed to it.  A variable can be a String, Integer or  List data type. For a SQL query, the variable passed  can be a Relation or Column type. In the following snippet from \textit{PoliSci}, variable \texttt{namedentity} is a Relation variable:   
\begin{Verbatim}
user := executeSQL("News", 
    "select distinct name, twittername from 
    twitterhandle t, $namedentity e 
    where LOWER(e.name)=LOWER(t.name)");
\end{Verbatim}
    \item Cross-DBMSs query. The query involves a series of  native queries delegated to at least two  underlying stores with different data models.  The result of one query  will be passed to another query on a different store. In the following snippet from \textit{PoliSci}, the result from a SQL query \texttt{user} is passed to a Cypher query to join with the graph:
\begin{Verbatim}[commandchars=\&\~\!]
user:=executeSQL("Senator", 
    "select distinct t.name as name, 
    t.twittername as tname 
    from twitterhandle t, $entity e 
    where LOWER(e.name)=LOWER(t.name)");}
users<name:String>:=executeCypher("TwitterG", 
    "match (u:User)-[:mention]-(n:User) 
    where n.userName in $user.tname
    return u.userName as name");
    \end{Verbatim}
    \item Transformation query. ADIL supports conversion  for some data types, e.g., conversion between a Relation and a Property Graph. The snippet from \textit{PatentAnalysis} shows a transformation function to create a property graph from a relation:
\begin{Verbatim}
graph := ConstructGraphFromRelation(wordsPair, 
(:Word {value: wordsPair.word1})
-[:Cooccur{count:wordsPair.count}]
->(:Word{value:wordsPair.word2}));
\end{Verbatim}
    \item Analytical query. ADIL supports a set of analytical functions and developers can extend AWESOME to support more black box functions with AWESOME native data types as input and output. In \textit{PatentAnalysis} workload, \texttt{pageRank} and \texttt{betweenness} functions are applied on graph to get pagerank and betweenness centrality:
\begin{Verbatim}
betweenness := betweenness(graph);
pagerank := pageRank(graph);
\end{Verbatim}
    \item Iteration query for collections. This query operates on a collection, e.g., a List. It manipulates each element in the collection, or filters elements based on predicates, or aggregates the collection to return a single value. The following snippet from \textit{NewsAnalysis} applies pagerank algorithm to each graph in a collection: 
\begin{Verbatim}
scores := graphPerTopic.map(g => 
    pageRank(g, topk=true, num=20));
\end{Verbatim}
\end{itemize}

\subsection{Optimization Opportunities}
Different optimization opportunities exist for  different types of queries. We describe the optimization ideas for each.

\noindent \textbf{Single-DBMS query:} The location of execution should be transparent to users. If the datasets queried are all in one store and the variables passed to the query are constants, or if the user explicitly specifies a store to use, then the query will be delegated to that store. However, if the datasets queried are AWESOME variables or the datasets include both data stored in database and AWESOME variables, then there will be multiple ways to execute it depending on where the query execution will happen. AWESOME supports in-memory query engines for the supported data models, such as Tinkerpop for graph query  and SQLite for relational query. The in-memory ones save data movement cost for AWESOME variables, and their performance can vary significantly from the on-disk ones on some types of queries. To select the best engine to store data and execute the query, we calibrate the performance of the in-memory and on-disk DBMSs on some complex queries. A cost model built on the calibration results will be used for optimization.

\noindent \textbf{Cross-DBMSs query:} The query optimization problem will be much more complex than the problem in a single model setting, as it involves cross-model operations such as joins between heterogeneous data from different stores. A uniform set of rewriting rules and cost estimation models need to be developed for optimization. We leave it as future work.

\noindent \textbf{Transformation queries and other types of queries that create variables with a constituent data model:} Where to store the variables needs to be determined. They can be stored in the backend on-disk database, in-memory database, or in the native AWESOME data structure. The decision will not be made independently; instead, all the subsequent  queries or analytical functions on the  created variables will be considered together to make a holistic decision. 

\noindent \textbf{Analytical queries:} An analytical function can be supported by different platforms, for example, pagerank algorithm is supported by some libraries including JGraphT and Tinkerpop, and also some graph DBMSs including Neo4j. Using which platform to execute the function needs to be determined. Each analytical query will be decomposed into a series of logical operators and each logical operator can have multiple mappings to executors. The cost model will be used to decide a mapping.

\noindent \textbf{Iterative queries:}  They can distribute collection elements to multiple cores, taking advantage of data parallelism.

 \subsection{Three Workloads and Optimization Opportunities}
We introduce three workloads and their optimization opportunities, which will be further explored in later sections.

\subsubsection{\textit{PatentAnalysis} workload.} As shown in Figure~\ref{fig:pa}, this workload consists of a single-DBMS query, a transformation query, and a series of analytical queries. The analytical function \textit{Tokenize} has two mappings: one maps to the in-DBMS implementation, and the other maps to the AWESOME native implementation. The best mapping will be decided by the cost model. The transformation query creates a graph for keywords from documents, and then  analytical functions are applied on it.  The graph can be stored and analyzed in a  graph analytics platform such as JGraphT, or  in  a database such as  Neo4j. The decision  will be based on the total cost consisting of  the cost of data movement and  analytical functions.

\subsubsection{\textit{PoliSci} workload.} 
\begin{figure}
    \centering
    \includegraphics[width=0.45\textwidth]{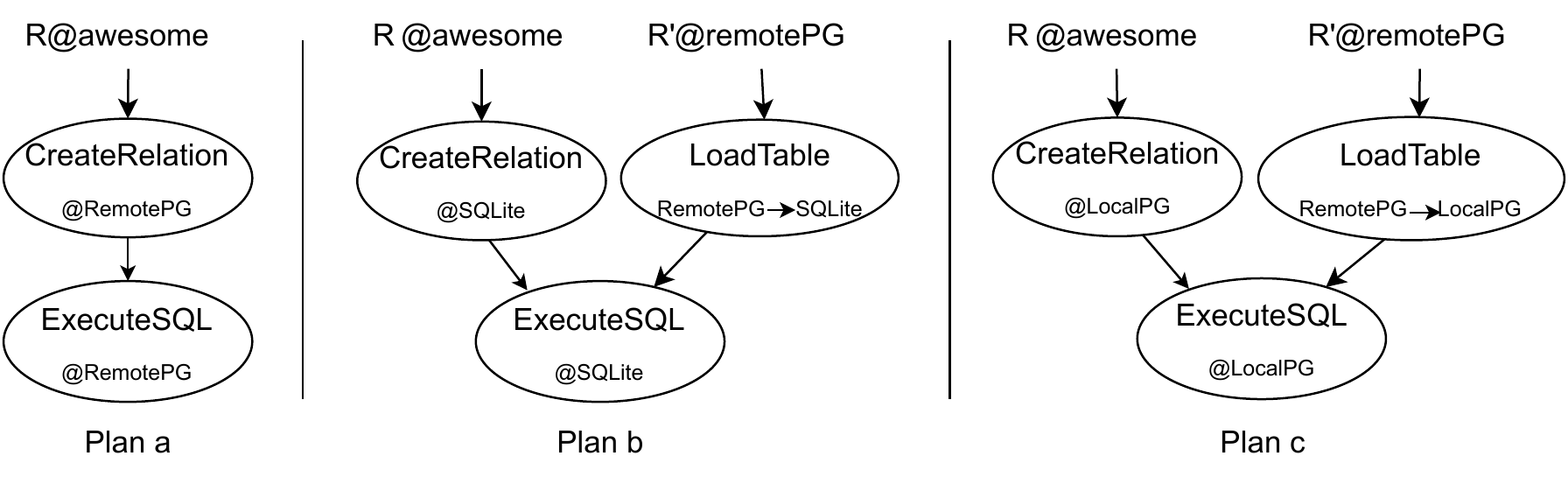}
    \caption{Execution sub-plans for cross-engine \textit{ExecuteSQL}.}\label{fig:executeSQL}
\end{figure}

As shown in Figure~\ref{fig:poliscidiagram}, this workload mainly consists of single-DBMS queries. One SQL query is to do a join between an AWESOME relation $R$ returned by the NER function and a relation $R'$ in a remote PostgreSQL database. There are three different plans  shown in Figure~\ref{fig:executeSQL} for this kind of cross-engine queries: (a) AWESOME tables are stored to the remote PostgreSQL server; (b) AWESOME tables and needed columns  from PostgreSQL tables are stored in the  in-memory SQLite server; (c) all these tables, remote tables or AWESOME tables, are moved in a local PostgreSQL server. Then the query will be executed in the server hosting the data.   AWESOME calibrates  this type of SQL queries which join two tables on an attribute in both  Postgres and in-memory SQLite, and the cost model considers the estimated query  execution time along with the cost of  data movement operators to decide the best plan holistically.

\subsubsection{\textit{NewsAnalysis} workload.} 
\begin{figure}
    \centering
    \includegraphics[width=0.45\textwidth]{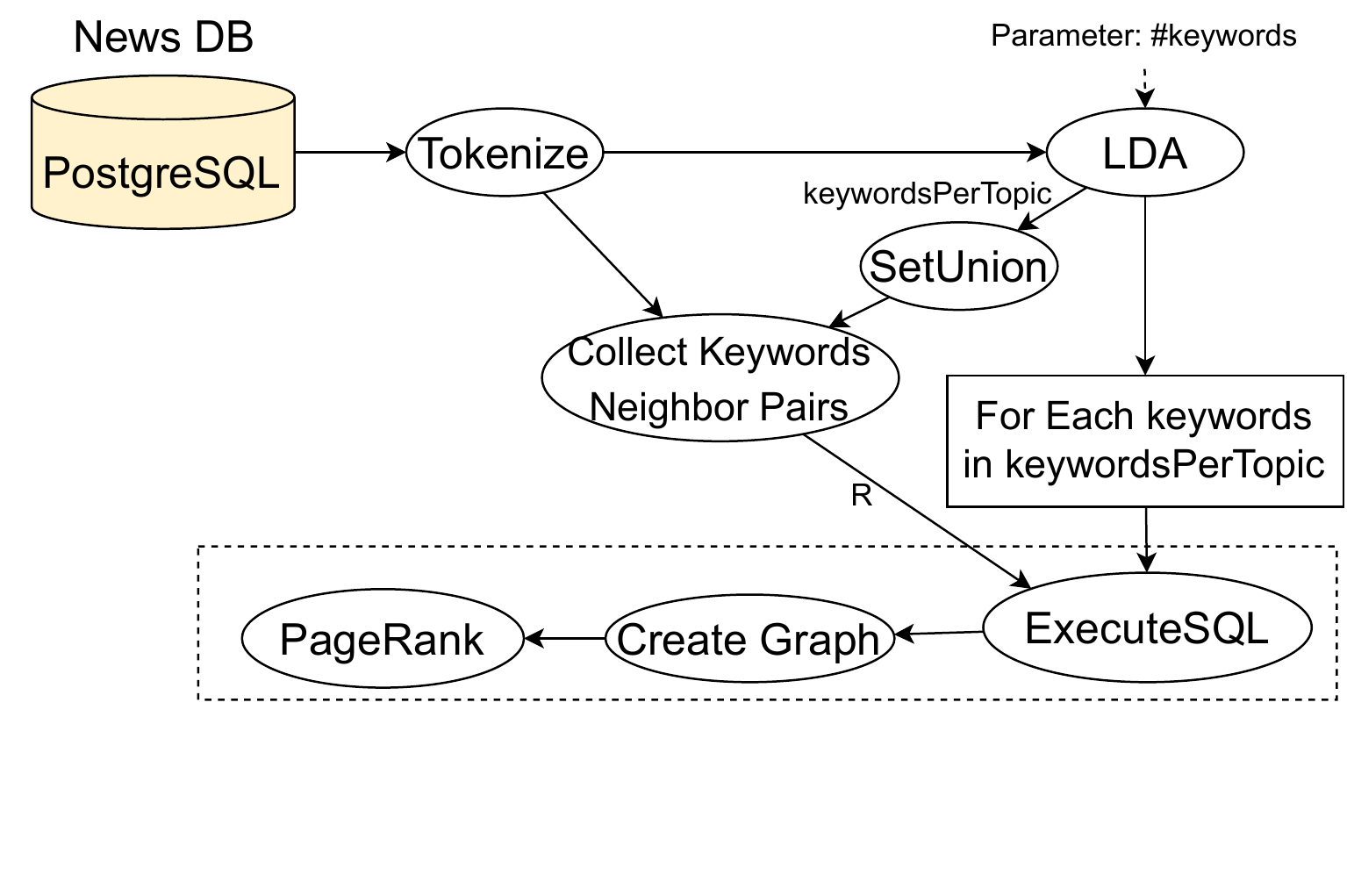}
    \caption{Illustration of \textit{NewsAnalysis} workload.}\label{fig:newsanalysis}
\end{figure}

As demonstrated in Figure~\ref{fig:newsanalysis}, this workload consists of single-DBMS queries, analytical queries, transformation queries, and iterative queries. It implements the method proposed in \cite{DBLP:conf/ictir/GollapalliL18} which uses PageRank for characterizing topic quality in LDA. Newspapers are extracted from a PostgreSQL database and then preprocessed.  LDA is applied to generate keywords for each topic. As the dashed rectangle in the figure shows, to evaluate the quality of each topic, it collects the word neighbor pairs from newspapers for its keywords, creates a word neighbor graph from the pairs, and computes pagerank value for nodes where are words. These steps are conducted for each topic. To simplify computation, we only iterate through the newspaper once to collect word neighbor pairs for the union of all topics' keywords, which is stored as a relation. Then, for each topic, a SQL query is initiated to get its keywords pairs.
The transformation query that creates a keyword neighbor relation $R$ from documents will be optimized together with the \textit{ExecuteSQL} queries to decide where to store $R$ and execute queries. The transformation query that creates a graph from each relation will be optimized together with the following \textit{PageRank} analytical function. The iterative queries on a collection will take advantage of the modern multi-core system and be executed in a data-parallel fashion.

%% file: sections/4-system-overview.tex
\vspace{-1ex}
\section{System Architecture}\label{sec:architecture}
The system architecture of AWESOME is shown in Figure \ref{fig:sysoverview}. The primary architectural components are:

\noindent \textbf{(a) Data Stores.} AWESOME supports on-disk DBMSs, including Neo4j, Postgres, and Solr, as well as in-memory DBMSs, such as Tinkerpop and SQLite.

\noindent \textbf{(b) Analytical Capability.} AWESOME incorporates existing analytical libraries for NLP and graph algorithms, such as CoreNLP and JGraphT, and native functions written in Java.

\noindent \textbf{(c) Query Processing.} In AWESOME, a ``query'' is essentially a multi-statement analysis plan that consists of data retrieval, transformation, function execution, and management and storage of intermediate results. The query processor verifies an ADIL script, creates the optimal logical plan (Section~\ref{sec:logicalplan}), generates a set of physical plans (Section~\ref{sec:physical}), and applies cost model (Section~\ref{sec:costmodel}) and data parallelism mechanisms to create an execution plan.

%% file: sections/5-validation.tex
\vspace{-1.5ex}
\section{Validating ADIL Scripts}\label{sec: check}
An ADIL script is  complex with many expensive operations. 
To reduce the  avoidable run-time errors, AWESOME implements a strict compile-time semantics check mechanism which consists of two parts:  1) \textit{Validation}  refers to  determining the semantic correctness of each expression, 2) \textit{Inference} refers to inferring the data type and metadata of the variables generated from each expression.
 

\vspace{-1ex}
\subsection{Validation}
For different RHS expressions, the validation process is different.


\noindent\textbf{System catalog based validation.}
To validate a  query expression (<query>) against an external DBMS, the system catalog is used to get the schema information. For example, for a SQL query,  it checks if the relations and columns  in the query exist in the database.

\noindent\textbf{Function catalog based validation.} For a function expression, AWESOME checks if  the data types of the input variables/constant values are consistent with the parameters information registered in the function catalog. 

\noindent\textbf{Validation with Variable Metadata.}
Variable metadata map stores the key properties of  variables and is built through inference process. It is looked up for every expression containing a variable.
For a query expression, if it queries on  relation-valued variables, their schema is found from the variable metadata map instead of the system catalog. For a function expression, if an input parameter is a variable, its data type will  be  found in the map. 

\noindent\textbf{Validation Example.} Usually, more than one types of validation need to be used. 
We use the example snippet from Sec.~\ref{sec:highorder}  to show how to validate a nested higher-order expression. To validate the Map expression,  it gets the data type and element type of  \texttt{topicID} from the variable metadata map, then it checks 
if the variable has a collection  type and the element type will be used to validate the sub-expression which is a Filter expression; to validate the Filter expression, similar to the Map expression, the data type of   \texttt{WTM} is checked and the element type is used to validate the sub-expression which is a binary comparison expression, besides, it also checks if the return type of the sub-expression is a  Boolean; to validate the binary comparison expression, it validates if the two operands have the same data type and the data type  is comparable: in this example, the type of the left operand can be inferred based on the function catalog; At the end, it checks the \texttt{getValue} function using the element   type information of \texttt{WTM} and \texttt{topicID}.


\vspace{-1ex}
\subsection{Inference}\label{sec:infer}
Inference refers to  building variable metadata map. Table~2  in the technical report shows the variable types,  and their corresponding metadata properties. 
For each statement in an  analysis plan, the RHS expression is validated and then the type and metadata of the LHS variables are inferred as much as possible and be stored in the  map.

\begin{table}[t]
\caption{Metadata for different data types.}
\label{tab:meta}
\footnotesize
    \begin{tabular}{ll}
    \toprule
    \textbf{Data Type}      & \textbf{Metadata}  \\ \midrule
    \texttt{Relation}       & Schema $S = \{\textit{ColName}:\textit{Type}\}$ \\ \hline
    \texttt{Property Graph} & \begin{tabular}[l]{@{}l@{}}Node labels set $NL$\\ Node properties map $NP =\{\textit{PropName}:\textit{Type}\}$
    \\ Edge labels set $EL$ \\ Edge properties map $EP=\{\textit{PropName}:\textit{Type}\}$ \end{tabular} \\ \hline
    \texttt{List}           & Element type, Element metadata, Size \\ \hline
    \texttt{Tuple}          & Each element's type and metadata, Size \\ \hline
    \texttt{Map} & \begin{tabular}[l]{@{}l@{}}
    Key type, Key metadata, Value type,\\ Value metadata, Size
    \end{tabular} \\ \hline
    \texttt{Matrix}         & Row (and column) count, Element type \\ \bottomrule
    \end{tabular}
\end{table}
The  inference mechanisms are different for different   expressions. For a SQL query expression, the schema of the returned relation is inferred by parsing the SELECT clause and looking up the system catalog/variable metadata map to get column types. 
For function expressions, the return types reside in the function catalog. 
For example, the following expression invokes function \texttt{lda}. 
By querying the function catalog, we know that it outputs two matrix variables.
Thus the data types of \texttt{DTM} and \texttt{WTM} will be set as Matrix. 
\begin{Verbatim}[commandchars=+\[\]]
DTM, WTM := +textbf[lda](processedNews,
              docid=true, topic=numTopic);
\end{Verbatim}
For nested expressions, the inference is handled from the innermost expression to the  outermost expression. Taking  the  snippet shown in Sec.~\ref{sec:highorder} as an example,   the LHS variable's type and  metadata is inferred by the following steps: 1) the Filter expression returns a matrix since \texttt{WTM} is a matrix; and 2) Map expression will return a list of matrices since its sub-expression returns a matrix.

%% file: sections/6.2-physical-planning.tex
\vspace{-1.5ex}
\section{Physical Plan}\label{sec:physical}
Section~\ref{sec:logicalplan} will introduce the logical plan generation and rewriting rules.  
Based on the optimized logical plan DAG, we introduce the physical planning details of AWESOME. 
As shown in Algorithm~\ref{algo:physicalplan}, there are mainly two steps to generate the candidate physical plans. AWESOME also provides buffering option for intermediate results and the detail can  be found in Section 6 of technical report.     

\begin{algorithm}[t]
\small
\caption{\emph{Physical Plan Generation}} \label{algo:physicalplan}
\SetKwFunction{CandidatePhsicalPlanGen}{CandidatePhsicalPlanGen}
\SetKwFunction{AddDataParallelism}{AddDataParallelism}
\SetKwFunction{AddBuffering}{AddBuffering}
\SetKwProg{Fn}{Function}{:}{}
\KwIn{A logical plan $G = (V, E)$, a boolean flag $buffer$.}
\KwOut{Candidate physical plans: $candPlans$}
$candPlans \gets$ \CandidatePhsicalPlanGen($G$)\; 
$candPlans \gets$ \AddDataParallelism($candPlans$)\;
\lIf{$buffer$}{$candPlans \gets$ \AddBuffering($candPlans$)}
\end{algorithm}

\subsection{Definitions}
To begin with, we provide some definitions as follows. 
\begin{definition} [Logical Plan DAG]
A logical plan directed acyclic graph (DAG) is a graph in which each node represents a logical operator. There are two types of edges: a data flow edge connects the predecessor node generating data to the successor node consuming data, and a sub-operator consumption edge connects a node which is a sub-operator in a higher-order expression to another node which represents the higher-order operator, such as a Map operation.  
\end{definition}

\begin{definition} [Physical Plan]
    A physical plan is similar to a logical plan DAG, with the only difference being that the logical operators are replaced by physical operators. Each physical operator is mapped to a piece of  code that  executes it.  
\end{definition}

A logical operator is platform-agnostic, while a physical operator  is platform specific  and maps to the actual execution code. For instance, \textit{CreateGraph} is a logical operator which represents creating a graph, while \textit{CreateNeo4jGraph}, \textit{CreateTinkerpopGraph}, \textit{CreateJGraphTGraph}, etc., are  physical operators that create a graph in Neo4j, Tinkerpop and JGraphT, respectively. A comprehensive list of logical operators and physical operators can be found in Appendix~\ref{appendix:ope}. A logical plan DAG will be translated to candidate physical plans defined as the follows which contain all possible physical plans: 

\begin{definition}[Candidate Physical Plans]
Candidate physical plans  consist of a DAG $PG = \{OP^p, E\}$ consists of physical operators and virtual inflated nodes, and a map $PM: I \to \{OP^p, E\}$ where each  key is a  virtual node ID and each  value is a set of different physical sub-plans.
\end{definition}

AWESOME generates the candidate physical plans by matching patterns in the logical DAG rather than translating each individual logical operator to its corresponding physical operators. To achieve this, AWESOME maintains a pattern set, defined as follows:
\begin{definition}[Pattern Set]
A pattern set $Pat: \{\{OP^l, E^l\} \to \{ \{OP^p, E\} \}\}$ is a  mapping where a key is a logical sub-plan and a value is a set of physical sub-plans. The pattern set is ordered by the sizes of keys i.e., the numbers of nodes in the logical sub-plans, to ensure that larger patterns in a logical plan are matched earlier in the translation procedures. 
\end{definition}

    

\subsection{Candidate Physical Plans Generation}
We propose the pattern-based transform algorithm, Algorithm~\ref{algo:candplan}, for generating candidate physical plans from a logical plan DAG.

\begin{algorithm}[t]
\footnotesize
\caption{\emph{Candidate Physical Plan Generation}} \label{algo:candplan}
\KwIn{An ordered  pattern set $Pat$; An optimal logical plan DAG $G = (V, E)$. 
}
\KwOut{Candidate physical plans: $PG$ and $PM$.}
\SetKwFunction{FindMatchPattern}{FindMatchPattern}
\SetKwFunction{SingleOpeTrans}{SingleOpeTrans}
\SetKwFunction{PatternTrans}{PatternTrans}
\SetKwProg{Fn}{Function}{:}{}
$PG \gets G, PM \gets \{\}$\;
\tcc{match patterns from the largest to the  smallest.}
\For{$pat \in Pat$} {
    $pSubs \gets Pat[pat]$\;
    $lSubs \gets$ \FindMatchPattern($PG, pat$) \;
    \For {$sub \in lSubs$} {
        \tcc{If the pattern has only one physical sub-plan, replace the pattern with the DAG.}
        \If {$pSubs.size == 1$}{
            $PG \gets$\SingleOpeTrans($PG, sub, pSubs$)\;
        }
        \tcc{If the pattern has several candidate physical sub-plans, transform $sub$ to a virtual node and add the node ID and physical sub-plans to map $PM$.}
        \Else{
            $PG, PM \gets$ \PatternTrans($PG, PM, sub, pSubs$)\;
        }
    } 
}
\end{algorithm}

\begin{figure}[ht]
   \centering
   \includegraphics[width=0.5\textwidth]{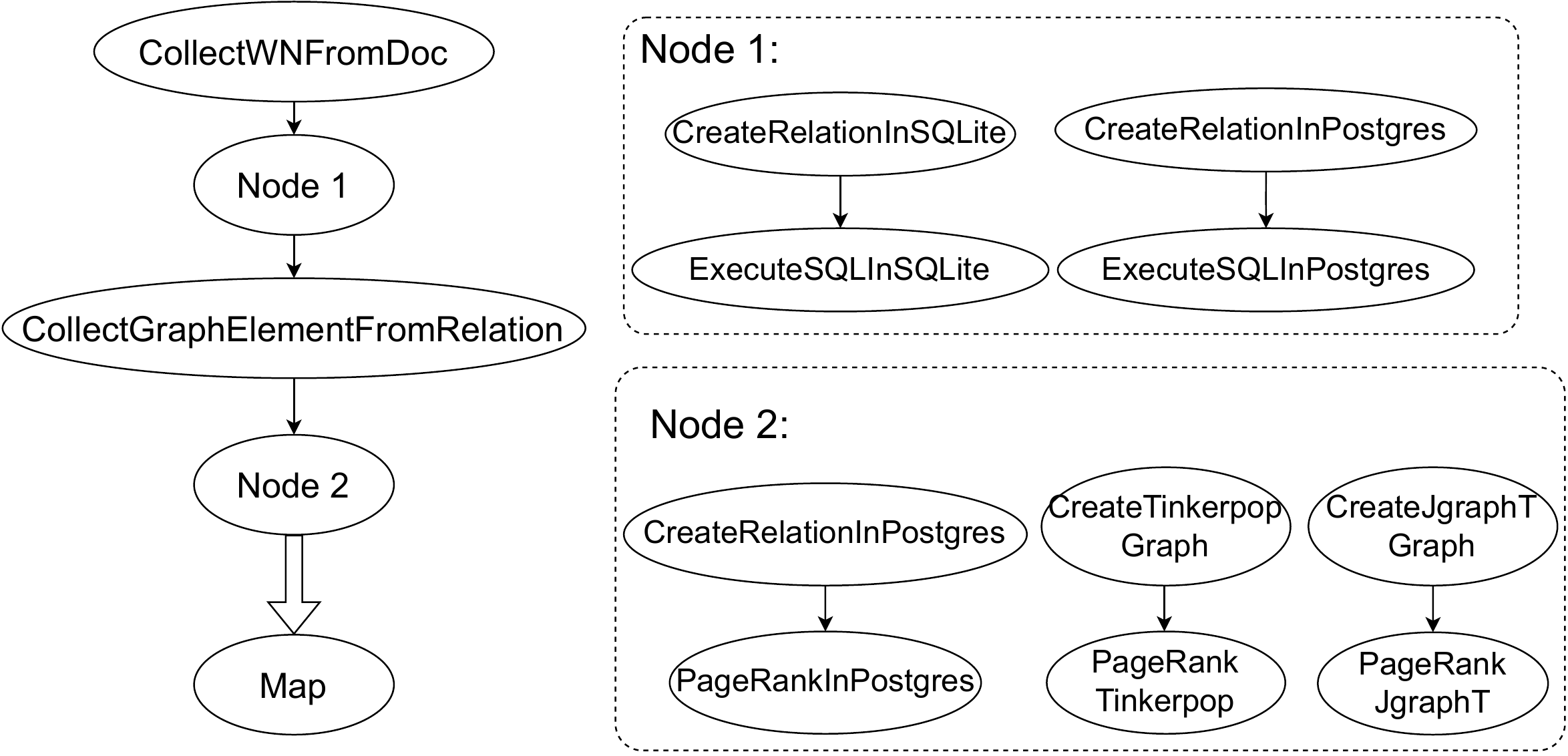}
    \caption{Candidate physical plans illustration.}  \label{fig:physicalplan}
\end{figure}

The algorithm takes as input an ordered pattern set, $Pat$, and an optimal logical plan DAG, $G = (V, E)$, and produces candidate physical plans, $PG$, and a map, $PM$. The algorithm matches patterns in the logical plan DAG, starting with the largest pattern, and checks if any sub-DAG in the logical plan matches the pattern. If a pattern corresponds to only one physical sub-plan, the matched sub-DAG is replaced with the physical sub-plan (lines 6-7). If a pattern has multiple candidate physical sub-plans, the matched sub-DAG is transformed into a virtual node, and the node ID and physical sub-plans are stored in the map $PM$ (lines 8-9). Figure~\ref{fig:physicalplan} shows an example of candidate physical plans for a logical plan sub-DAG shown in Figure~\ref{fig:mapfusion}. The dashed rectangle marked by Node 1 shows the candidate physical plans for the logical sub-DAG \textit{CreateRelation} $\to$ \textit{ExecuteSQL}. 


\vspace{-1.5ex}

\vspace{-1.5ex}
\subsection{Partitioned Data Parallelism}
AWESOME takes advantage of data parallelism to exploit modern multi-core systems. To achieve this, it adds partition and merge operators to the physical plan. Appendix E  presents some physical operators with their data parallel capabilities, where \textbf{ST} means single-threaded operators that cannot be executed in a data parallel fashion, \textbf{PR} means data parallelizable operators, and \textbf{EX} means operators provided by external libraries. The execution of \textbf{EX} operators is fully supported by external libraries and can utilize multi-core features in their native implementation. As a result, they are excluded from the subsequent AWESOME optimizations that are based on data parallelism.

For a \textbf{PR} operator with multiple inputs, it is associated with a $capOn$ attribute that specifies the input on which it has data parallelism capability. For example, the \textit{FilterStopWords} operator takes a corpus and a list of stop-words as input, and it can be executed in parallel by partitioning the corpus input. In this case, the $capOn$ attribute is set to the ID of the corpus variable. Every \textbf{PR} operator is executed in parallel by partitioning the $capOn$ input data. 

\begin{figure}[t]
    \centering
    \includegraphics[width=0.4\textwidth]{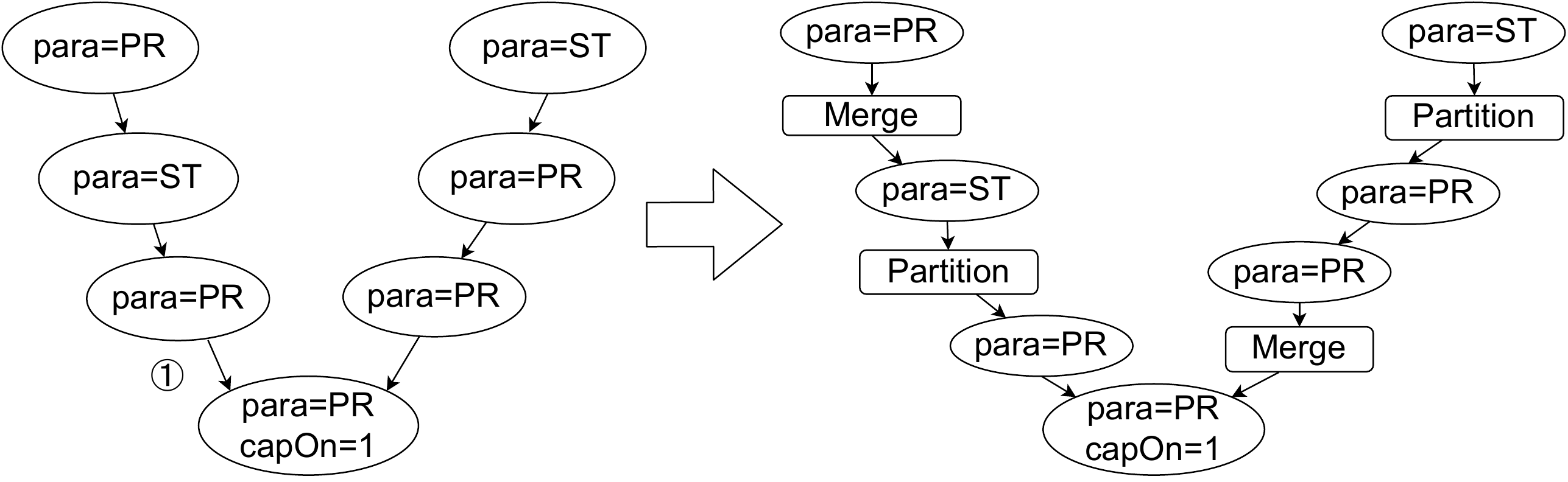}
    \caption{Illustration of data parallel execution.}
    \label{fig:parallelexe}
\end{figure}

Figure~\ref{fig:parallelexe} illustrates the concept. The left sub-figure shows the original physical plan DAG, and the right one shows the plan after considering data parallelism. When an operator with \textbf{PR} capability gets its input, if its $capOn$ input was not partitioned, a Partition step is added to generate partitioned results. If a non-$capOn$ input was partitioned, a Merge step is added to collect the data into a single collection. When an operator with \textbf{ST} capability gets data from an operator with \textbf{PR} capability, a Merge step is added.

\subsection{Buffering Mechanism}
AWESOME employs a buffering mechanism to avoid storing unnecessary intermediate results in memory. Different from pipeline, buffering mechanism does not utilize multiple cores to execute different operators simultaneously.
Some operators can process input in a batch-by-batch manner, and some can generate output in a batch-by-batch manner. 
We refer data with this manner as stream hereafter.  There are four types of buffering capabilities: 
\begin{enumerate}[leftmargin=*]
    \item $SI$ (Stream-Input): the input can be passed as stream to the operator, but it produces a whole inseparable result at once;
    \item $SO$ (Stream-Output): the operator takes an inseparable input but can produce result progressively as stream;
    \item $B$ (Blocking): both the input and output need to be a whole;
    \item $SS$ (Stream-Stream): both the input and output can be a stream.
\end{enumerate}

Each physical operator is associated  with its buffering capability. Table \ref{tab:operator} presents it for some physical operators.   Similar to data parallelism capability, there is another $capOn$ attribute associated if the operator has more than one input. The physical DAG will be partitioned to a collection of chains. Inside each chain, the intermediate result is  not stored in memory; the upstream operator produces stream output to be consumed by the downstream operator. 
The data across chains has to be stored in memory.

\begin{figure}
    \centering
    \includegraphics[width=0.25\textwidth]{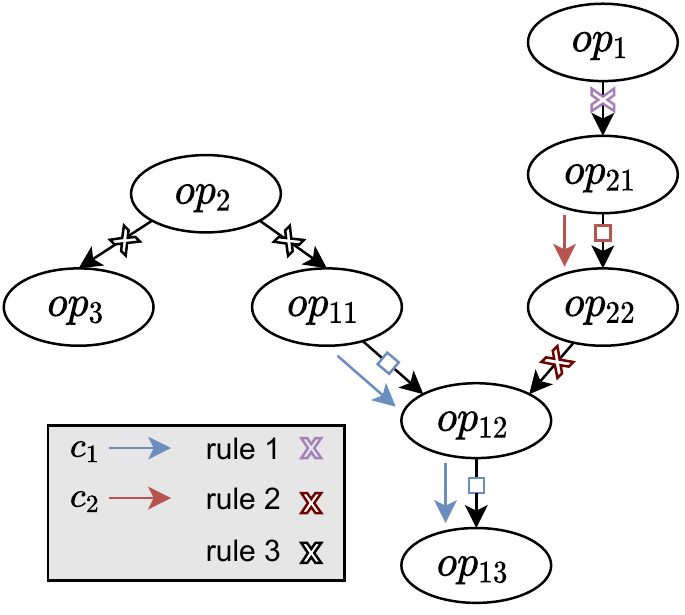}
    \caption{Illustration of buffering rules.}
    \label{fig:cutgraph}
\end{figure}

The  collection of chains is collected from the physical DAG by partitioning it based on the  partition rules which are shown below and also illustrated in Fig.~\ref{fig:cutgraph}:
\begin{itemize}[leftmargin=*]
    \item For an edge $e = (op_{e1}, op_{e2})$, if $op_{e1}$ can't generate stream result or $op_{e2}$ can't take stream input, $e$ will be cut. For example, in Fig.~\ref{fig:cutgraph}, the edge between $op_1$ and $op_{21}$  is cut. 
    \item For an edge $e = (op_{e1}, op_{e2})$, if data from $op_{e1}$ to  $op_{e2}$  is not the $capOn$ input of $op_{e2}$, $e$ will be cut. In Fig.~\ref{fig:cutgraph}, the edge between $op_{22}$ and $op_{12}$ is cut. 
    \item For an operator $op$, if it has more than one outgoing edges, then all outgoing edges will be cut. In Fig.~\ref{fig:cutgraph}, the outgoing edges from $op_2$ are all cut. 
\end{itemize}

AWESOME users have the option to turn on buffering mechanism.  The buffering mechanism would be very helpful if : 1) the analysis plan contains many  data flow edges which buffer can be added to; or 2) AWESOME is running on a memory-limited machine; or 3) users care about the responsiveness of the system and expect to get some initial results back soon  without waiting for the complete results. 


\vspace{-1.5ex}
\subsection{Failed Attempt: Pipeline + Data Parallelism}
We built a framework that hybridizes pipeline (i.e., task parallelism) and data parallelism, however, the experimental results reveal that such framework is not suitable for AWESOME. We briefly introduce this framework and explain why this technique did not boost performance to provide some insights for future researches. 

Similar to the buffering mechanism, an AWESOME physical DAG is partitioned into a list of  chains based on the partition rules. Then each chain will form a pipeline where operators can be executed simultaneously using multi-cores. Once the upstream operator produces a batch of results, the downstream operator will be executed on that batch immediately and simultaneously. Both pipeline and data parallelism utilizes multi-cores to increase resources utilization, thus we define a scheduling problem to allocate a specific amount of cores (the number of cores in an OS) to operators in each pipeline chain. A simple solution is to allocate cores to match the produce and consume rates of data.  

However, from the experimental results, this framework is not more efficient than  data parallelism framework even under the best allocation strategy due to two properties of AWESOME operators. We theoretically explain the reason why this framework does not outperfrom data parallelism framework.  For a simple pipeline chain with two operators: $op_1 \to op_2$, suppose that there are a total of $n$ cores and it costs $t_1$ for $op_1$ to produce a batch of data and $t_2$ for $op_2$ to consume the batch, then there will be $t_1n/(t_2+t_1)$ cores assigned to $op_1$ and the rest of cores assigned to $op_2$.

Suppose that $op_1$ will produce $m$ batches in total, then the execution time of applying data parallelism solely $T_1$ and of applying pipeline + data parallelism $T_2$ can be computed as, 
\begin{equation}
    \begin{split}
    T_1 & = \frac{(t_1 + t_2) m}{n} + agg*n\\
    T_2 & = \max\{\frac{t_1 m}{n_1}, \frac{t_2 m}{n - n_1}\} + agg * n_1, 
    \end{split}
\end{equation}
where $n_1$ is the number of cores assigned to $op_1$, and $agg * \#core$ is the sequential aggregation cost of data parallelism. Since for AWESOME aggregation operators such as \textit{SUM}, the aggregation cost is usually very small and can be  negligible comparing to other time-consuming analytical functions, we can  prove that $T_1 	\approx \frac{(t_1 + t_2) m}{n} \leq \max\{\frac{t_1 m}{n_1}, \frac{t_2 m}{n - n_1}\} \approx T_2$ always holds where the equality is achieved when the above optimal allocation solution is applied. Thus, the pipeline and data parallelism framework can't outperform data parallelism if all operators in a chain are data parallel-able.


Appendix D  presents the data parallel capability and buffering capabilities for most AWESOME operators. In the future, when there are more operators with different properties are added to AWESOME, this framework may have chance to outperform the solely data parallelism framework.

%% file: sections/6.1-logical-query-planning.tex
\vspace{-1.5ex}
\section{Logical Plan}
\label{sec:logicalplan}
This section introduces the optimized logical plan generation. A complete list of  logical operators is  in Appendix E.

\vspace{-1.5ex}
\subsection{Logical Plan Creation}
A raw logical plan is directly translated from the parsing results. 
In some  cases, each expression corresponds to a single  logical operator. 
For example, an \texttt{ExecuteSQL}  expression will be mapped to an \textit{ExecuteSQL} logical operator. For some function expressions, each function will be decomposed to a series of more fine-granular operators. For  higher-order expressions, each will be translated to two operators connected by sub-operator consumption edge. 

\noindent\textbf{Function Translation.}  For analytical functions, the corresponding logical operators can vary based on different function inputs. 
For example, the function \texttt{LDA} can take either a \texttt{Matrix} variable or a \texttt{Corpus} variable as input, which will be translated to  logical operators \textit{LDAOnTextMatrix} and \textit{LDAOnCorpus} respectively. 

\noindent\textbf{Higher-order Expressions to Sub-plans.}
For higher-order expressions (e.g., map expressions), a single expression will be translated to a sub-plan consisting of several operators.  Figure~\ref{fig:mapfusion} (a) presents a logical plan for a series of higher-order expressions from  workload \textit{NewsAnalysis}.   
There are two types of edges denoting data flow (solid edges) and sub-operator consumption (arrows), respectively. 

\vspace{-1.5ex}
\subsection{Logical rewriting}
AWESOME applies a set of rewriting rules to the raw logical plan to generate an optimized logical plan.

\noindent\textbf{Rule 1: Keywords decomposition.}
Some ADIL keywords  can be decomposed into several low-level logical operators. For example, the \texttt{NER} function, which recognizes named entities in a corpus, is translated into a series of \textit{NLPAnnotator} operators with different annotation sub-operators. The \texttt{buildWordNeighborGraph} function will be decomposed to \textit{CollectWNFromDocs} and \textit{CreateGraph}.  

\noindent\textbf{Rule 2: Redundancy elimination.}
The same operators with the same input are executed only once. Some keywords  may share common operators that will be  merged.

\begin{figure}[ht]
   \centering
   \includegraphics[width=0.45\textwidth]{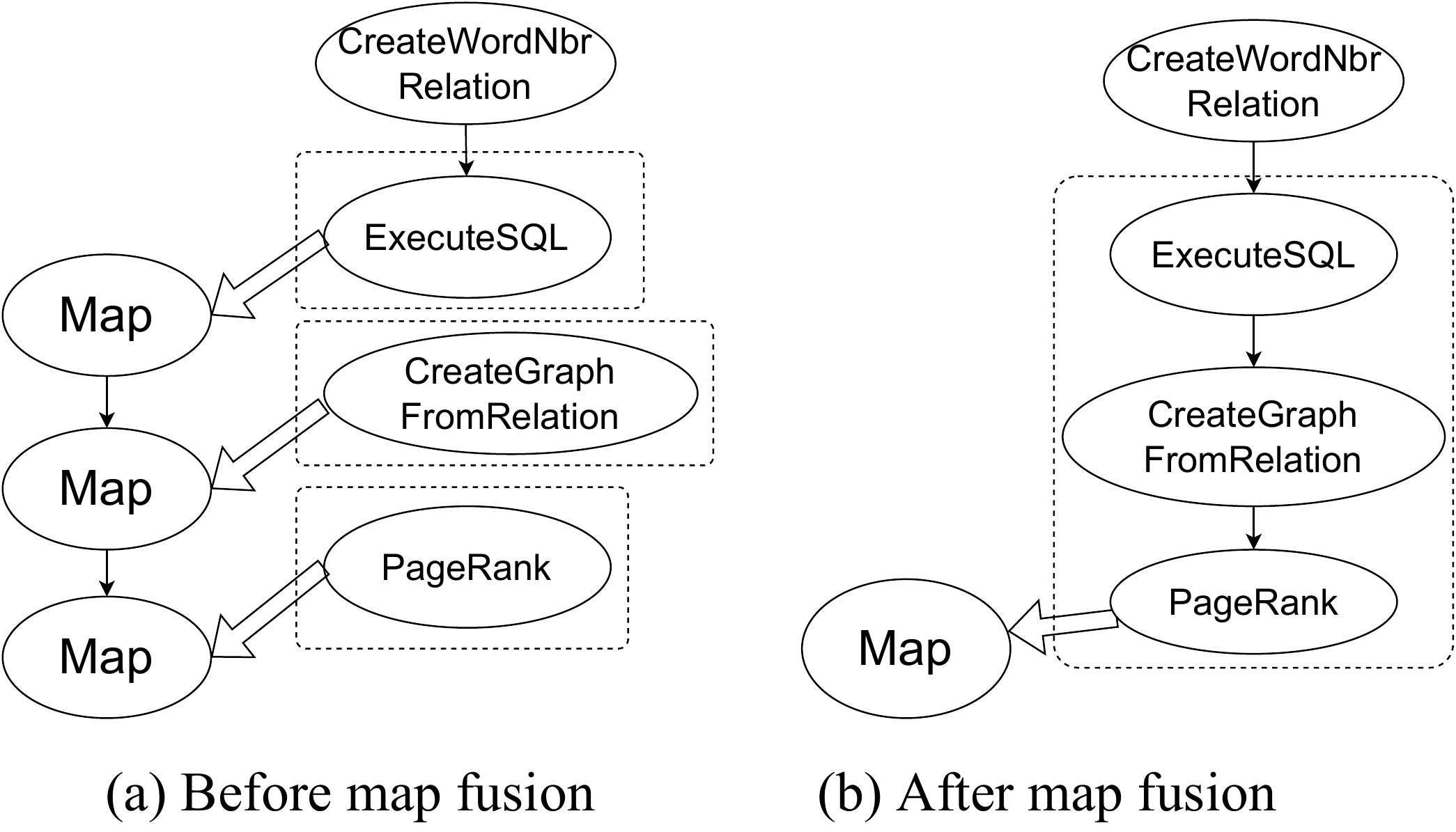}
   \caption{Illustration of map fusion.}
   \label{fig:mapfusion}
\end{figure}

\noindent\textbf{Rule 3: Operator fusion.}
There are two special operators that apply a sub-operator to each element of a collection variable: \textit{Map} and \textit{NLPAnnotator}. A series of \textit{Map} or \textit{NLPAnnotator} operators are fused, and their sub-operators are connected. This is termed as  Map Fusion and NLP Annotation Pipeline. Figure~\ref{fig:mapfusion} (b) shows the plan after map fusion. This rewriting has two advantages: 1) intermediate results are not materialized, which saves memory, and 2) it benefits physical plan generation by connecting all the sub-operators of a series of Maps so that a larger sub-DAG can be matched with a pattern in the pattern set, resulting in more efficient candidate plans. Figure~\ref{fig:physicalplan} shows the candidate physical plans for the plan after map fusion.

\vspace{-1.5ex}

%% file: sections/7-costmodel.tex
\vspace{-1.5ex}
\section{Learned Cost model} \label{sec:costmodel}
The query planning stage generates candidate physical plans, and in the execution stage, the optimal plan is chosen at run-time based on a learned cost model.

For each virtual node, the cost model is applied to each sub-plan to estimate the execution cost, and the sub-plan with the lowest cost is chosen. We use a learned cost model instead of a rule-based model. Cost should be decided at the sub-plan level rather than the operator level, making rule-based optimization difficult to design. For a single logical operator, it is easy to design rules to decide which physical implementation should be chosen under specific circumstances. However, a logical sub-plan may consist of several logical operators, and each of them may be transferred to multiple different physical operators, leading to a large space of rules.
\vspace{-1.5ex}
\subsection{Cost Model}
Suppose that a physical sub-plan $S$ consists of multiple operators $\{op_1, \cdots, op_n\}$, 
the overall cost estimation is given as the sum of the estimated cost of all operators since AWESOME does not apply task parallelism, i.e., $est_{S} = Cost(op_1) + \cdots + Cost(op_n)$ where $Cost(\cdot)$ is a trained linear regression model with the polynomial of raw features (of degree 2) as variables that predicts the execution cost of a physical operator, i.e.,
\begin{equation}
\begin{split}
Cost(op) = w_0 + w_1 f_1 + \cdots + w_n f_n + w'_1 f^2_1 +  \cdots + w'_n f^2_n \\
+ w_{(1,2)} f_1f_2 +  \cdots + w_{(n-1, n)} f_{n-1} f_n,
\end{split}
\end{equation}
where $f_{1}, \cdots, f_n$ are the raw  features for $op$. $Cost(op)$ is trained based on training data collected from  calibration for  $op$. The features vary for different physical operators. For example, for some relation-related operators, the raw features include the tables sizes; for graph-related operators, node count or  edge count is a raw feature and for some graph or relation queries  the predicate size is another  raw feature.


\vspace{-1.5ex}
\subsection{Calibration}
To train the individual cost model $Cost(\cdot)$, we designed a set of synthetic datasets that vary in some parameters, and ran each operator on them to collect a set of execution times. 

\noindent\textbf{Operators and features.} We mainly train the cost model for graph- or relation-related operators. For graph-related operators, we evaluated common operators such as creating graph in different platforms  and computing PageRank from different platforms. Graph size serves as one feature for the cost estimation. Additionally, we calibrated some typical types of Cypher queries, and two example types are:

\noindent\textbf{Type I}: Queries with a series of node or edge property predicates. For example, \textit{MATCH (n)-[]-(m) WHERE n.value IN L1 AND m.value IN L2}, where L1 or L2 is  a  list of strings. The number of predicates and the sizes of L1 and L2 are other raw features that determine the query cost.

\noindent\textbf{Type II}: Full-text search queries. In this kind of query, there is a node/edge property that contains long text, and the query finds nodes/edges whose text property contains specific strings. For example, \textit{MATCH (n)-[]-(m) WHERE n.value CONTAINS string1 OR n.value CONTAINS string2 OR ...}. The number of OR predicates is another raw feature of the cost function.

For relation-related operators, we evaluated common operators such as \textit{CreateRelationInPostgres} and \textit{CreateRelationInSQLite}. Relation size serves as one parameter. Besides, we also calibrated two types of SQL queries:

\noindent\textbf{Type I}: Queries with WHERE IN predicates. For example, \textit{SELECT name FROM R WHERE city IN L}, where L is a list of strings. The size of $L$ is another raw feature that determines the query cost.

\noindent\textbf{Type II}: Join between two tables. For example, \textit{SELECT R1.a, R2.b FROM R1, R2 WHERE R1.c = R2.c}, where the sizes of the two tables are raw features.


\noindent\textbf{Datasets.} \noindent\textbf{Datasets.} We design a set of graph datasets and relation datasets  which are used for  graph- and relation-related operators respectively. We present the statistics in Table~\ref{tab:synthetic}. 
\begin{table}[b]
    \centering
    \footnotesize
    \caption{Parameters of synthetic datasets for cost model.}~\label{tab:synthetic}
    \begin{tabular}{ccc}
    \toprule
    &\textbf{Parameter} & \textbf{Value} \\ \midrule
        \multirow{4}{*}{\textbf{graph dataset 1}} &edge size & 500, 1k, $\cdots$, 800k \\
        &avg. density & 2 \\
        &node property & value: String\\
        &keyword size & 50, 100, 500, 1k, 2k\\\hline
        \multirow{3}{*}{\textbf{graph dataset 2}} & node size & 5k, 10k, $\cdots$, 500k \\
        & node property & tweet: String \\
        &keyword size & 50, 100, 500, 1000\\\midrule
        \multirow{2}{*}{\textbf{relation dataset}} & PostgreSQL table row count & 100, 1k, 10k, 100k\\
        & Awesome table row count & 100, 1k, 10k, 100k\\ \bottomrule
    \end{tabular}
\end{table}

For graph datasets, there are two types of graphs: The first type of datasets is used to test operators including \textit{CreateGraph}, \textit{PageRank} and the Type I Cypher queries: We created several property graphs  with different edge sizes, and to simplify the model we kept the density of graphs as a constant value 2; each node (or edge) has a value property which is a unigram and we make sure each node's (or edge's) property is unique, then we created keywords lists with different sizes  from the values set as the predicates. The second dataset is designed for the Type II Cypher queries: We created graphs with different node sizes and each node has a \textit{tweet} property whose value  is a tweet text collected from Twitter; All the unigrams are collected from these tweets and after removing the most and the least frequent words, we randomly selected words to create different sizes of keywords lists which will be used to do text search.

\begin{figure}[t]
    \begin{subfigure}{.24\textwidth}
      \centering
      \includegraphics[width=\linewidth]{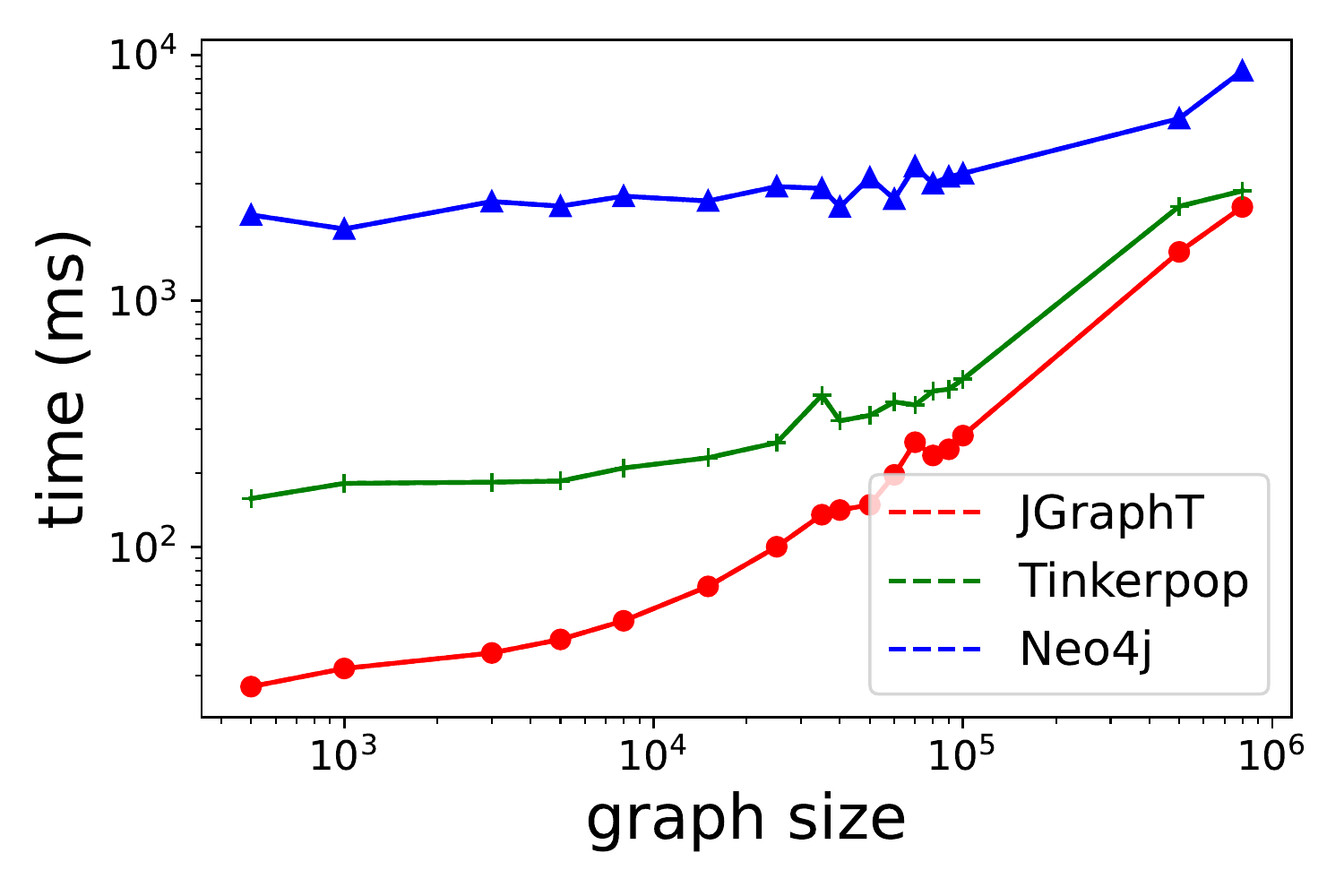}
      \caption{Graph Creation}
    \end{subfigure}%
    \begin{subfigure}{.24\textwidth}
      \centering
      \includegraphics[width=\linewidth]{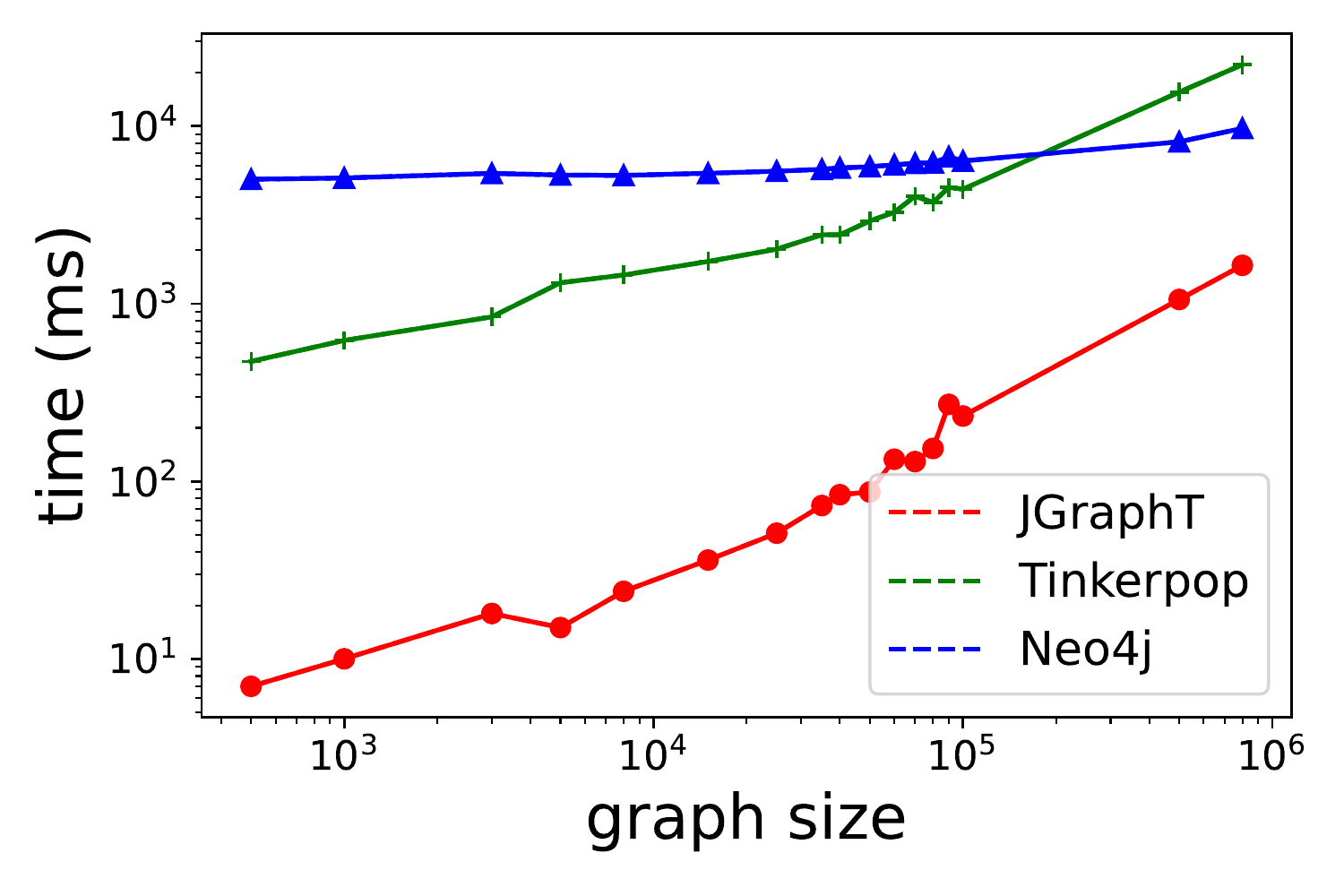}
      \caption{Page Rank}
    \end{subfigure}
        \begin{subfigure}{.24\textwidth}
      \centering
      \includegraphics[width=\linewidth]{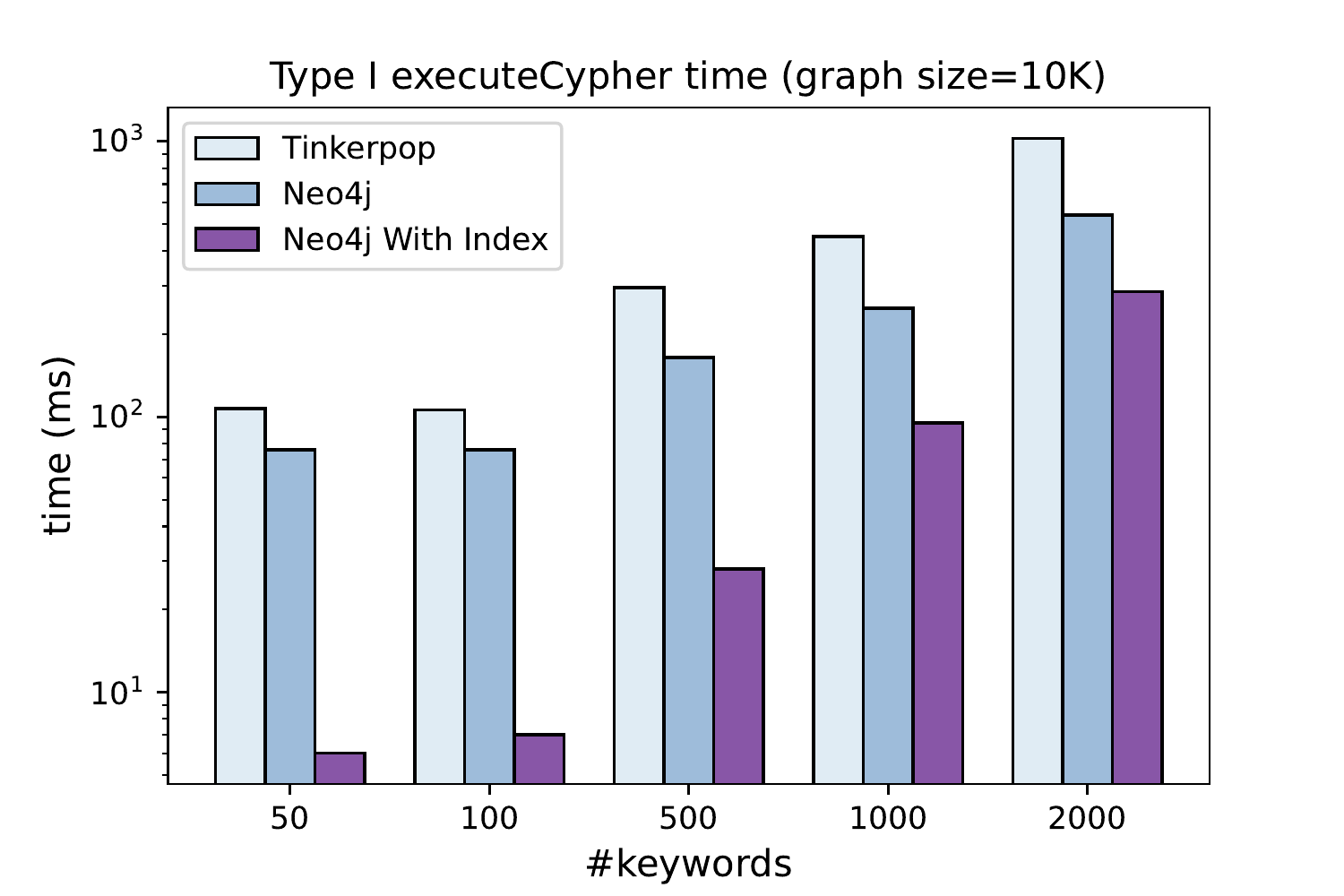}
      \caption{Type I Cypher Query}
    \end{subfigure}%
    \begin{subfigure}{.24\textwidth}
      \centering
      \includegraphics[width=\linewidth]{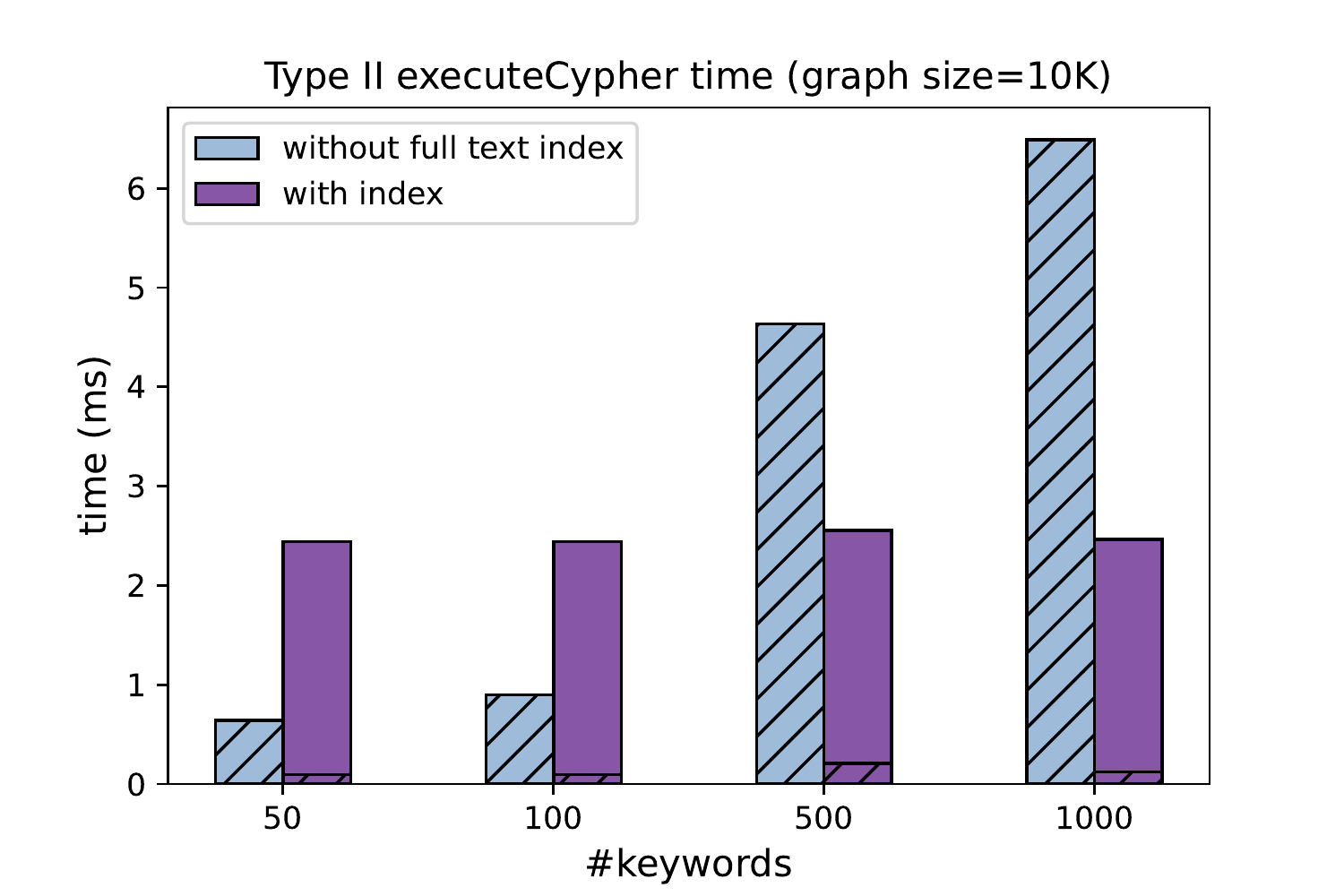}
      \caption{Type II Cypher Query}
    \end{subfigure}
\caption{Calibration results for graph operators.  The part with '//' in (d) denotes query execution time .}
\label{fig:cali}
\end{figure}

\noindent\textbf{Calibration Results.}  
Figure~\ref{fig:cali}  shows the calibration results for some graph-related  operators.  More results can be found in Appendix D.

\subsection{Training and Cost Estimation}
The individual cost model for each operator is trained based on the calibration results  to minimize the loss function, i.e., mean squared error. 
At run-time, based on the input of a virtual node, the features are collected and passed to the  cost model  to compute the cost for each candidate physical sub-plan. The best sub-plan with the lowest cost will be selected for that virtual node.

%% file: sections/8-experiment.tex
\section{Experiments}\label{sec:exp}
In this section, we first empirically validate whether AWESOME is able to improve the efficiency of analytical polystore workloads. 
Then, we drill into how the cost model of AWESOME contributes. 

\begin{figure*}[th]
\minipage{0.9\textwidth}
    \includegraphics[width=\linewidth]{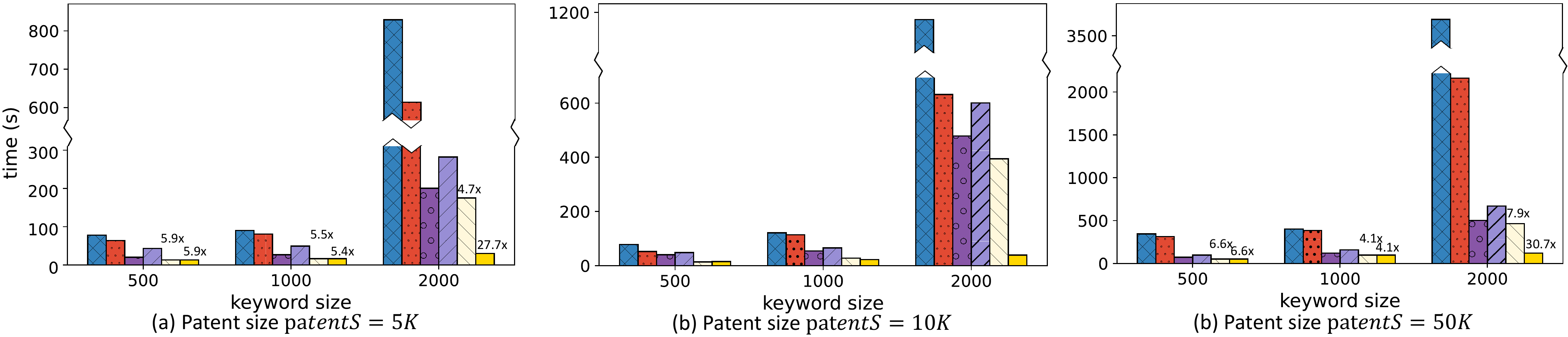}
      \caption{Time for  \textit{PatentAnalysis} w.r.t.~different $patentS$.}
    \label{fig:patent-exp}
\endminipage\hfill
\minipage{0.9\textwidth}
    \includegraphics[width=\linewidth]{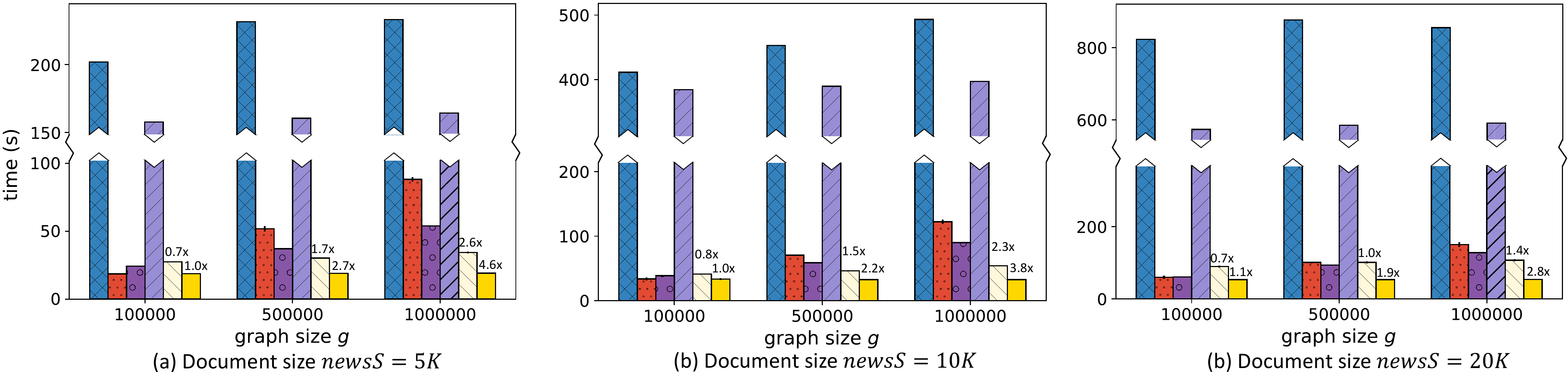}
      \caption{Time for \textit{PoliSci} w.r.t.~different $newsS$.}
    \label{fig:polisci-exp}
\endminipage\hfill
\minipage{0.9\textwidth}
    \includegraphics[width=\linewidth]{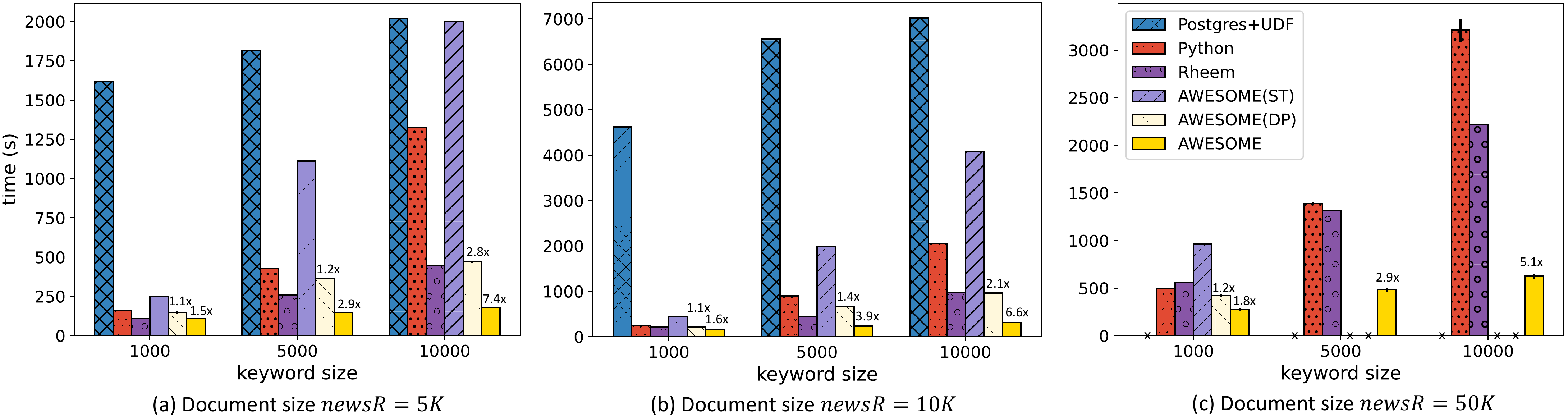}
    \caption{Time for  \textit{NewsAnalysis} w.r.t.~different $newsR$.}
    \label{fig:newsanalysis-exp}
\endminipage\hfill
\minipage{0.9\textwidth}%
    \includegraphics[width=\linewidth]{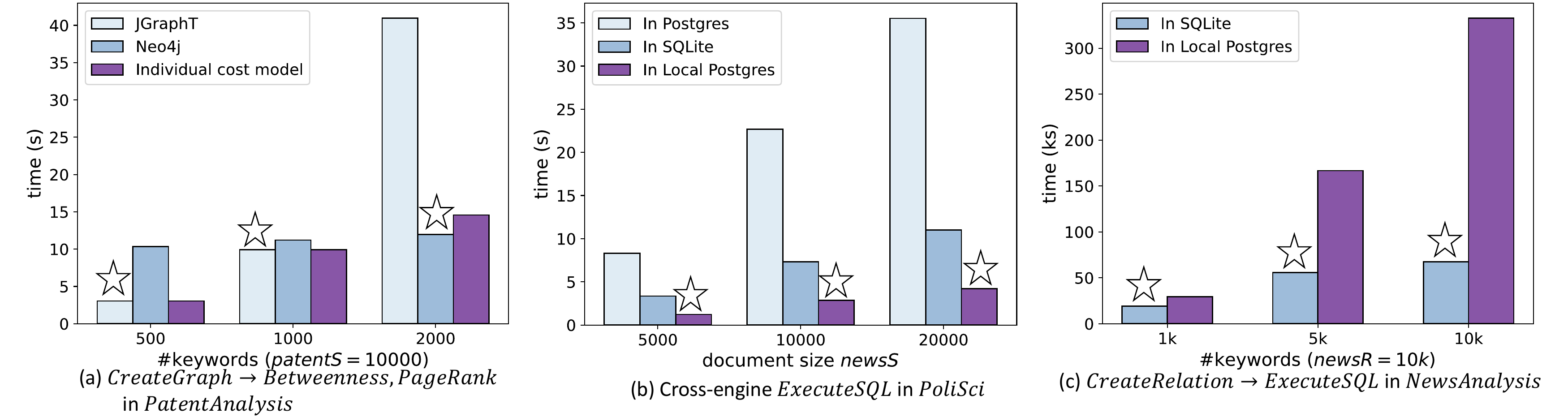}
    \caption{Execution time of different physical plans. }
    \label{fig:costmodel}
\endminipage
\end{figure*}

\subsection{Experimental Setup}
We focus on the single-machine multi-cores setting, and the distributed version of AWESOME will be our future work. The experiments were run on CloudLab \cite{duplyakin2019design}, a free and flexible cloud platform for research. 
The machine has 24-core AMD 7402P CPUs, 128 GB memory and 1.6 TB disk space. It runs Ubuntu 18.04. 

\noindent\textbf{Datasets.}
We  collect five real world datasets stored in different data stores for  the workloads. 
\begin{itemize}[leftmargin=*]
\item \textit{SbirAwardData}: a PostgreSQL relation of about 50K patents with information such as abstract. 
    \item \textit{Newspaper}: A PostgreSQL relation with over 1M news articles with an average length of 500 words collected from the Chicago Tribune newspaper. 
    \item \textit{SenatorHandler}: A PostgreSQL relation of about 90 United States senators with their names and twitter user names.
    \item \textit{NewsSolr}: A collection of news stored in the Solr engine with size around 20 GB.
    \item \textit{TwitterG}: Attribute graphs stored in Neo4j that represent the Twitter social networks of different sizes. 
    A node in \textit{TwitterG} is labeled as either ``User'' or ``Tweet''. 
    A User node has a property \textit{userName}, and a Tweet node has a property \textit{tweet} storing  tweet content. A User node  connects to another User node by a directed edge if the first one \textit{mentions} the second, and a User node  connects to a Tweet node by a directed edge if the user \textit{writes} it.
    \end{itemize}
\vspace{-0.5em}
\noindent\textbf{Workloads and parameter settings.}
We evaluate three analytical workloads and vary some parameters in the workloads.
\textbf{PatentAnalysis}, as shown in Figure~\ref{fig:pa}, is  an analytical workload focusing on text and graph analytics. It uses patent abstracts from \textit{SbirAwardData} dataset. We vary two parameters: 1) the number of patent  abstracts extracted from the dataset, denoted as $patentS$, by changing the SQL query script. 2) the number of selected keywords $k$ by changing the keyword mining function argument.  
\textbf{PoliSci}, shown in Figure~\ref{fig:poliscidiagram},  focuses on the polystore aspect of the system where input data is  stored in heterogeneous remote  data stores. It queries on the \textit{NewsSolr}, \textit{SenatorHandler} and \textit{TwitterG} datasets. We  change  two parameters: 1) $newsS$, the number  of news  selected from NewsSolr dataset.  2) the  size (number of nodes) of the  graph used from graph collection   $TwitterG$.
\textbf{NewsAnalysis}, as shown in Figure~\ref{fig:newsanalysis}, is  a complex text analytical task,  focuses on analytical functions including graph algorithm like PageRank and NLP functions like LDA. It uses  \textit{Newspaper} dataset. 
We vary the number of news selected, denoted as $newsR$ and the number of keywords in each topic, denoted as  $k$.


\noindent\textbf{Compared Methods.}
We implement some baselines: For single-DBMS baseline, we implement \textbf{Postgres+UDF} based on  PostgreSQL with MADLIB extension and Python extension for UDF. For programming languages, we choose Python. For prior polystore system, we choose RHEEM because of its  ease to extend for analytical functions. We also implement a few versions of AWESOME. We provide the SQL scripts for the \textbf{Postgres+UDF} method and the Python code in Appendix A and C, respectively, and the RHEEM implementation code is available on our GitHub repository. The \textbf{Postgres+UDF} implementation requires around 300-1000 tokens for each  script (we separate tokens by space and newline) and   \textbf{Python} implementation consists of approximately 380-600 tokens each, while ADIL scripts have only around 100 tokens each.

\begin{itemize}[leftmargin=*]
\item \textbf{Postgres+UDF}: It stores all datasets in a single store, Postgres, and uses pure SQL scripts with user-defined functions written in Python or implemented by the MADLIB toolkit \cite{hellerstein2012madlib}.
\item \textbf{Python}: The workloads were implemented in Python. To make fair comparison, we implemented multi-threads for some functions.
\item \textbf{RHEEM}: RHEEM supports Spark for parallel execution. To imitate a distributed environment for Spark, we chose one core as the master node and set up four worker nodes, with each worker node assigned 5-6 cores.
\item \textbf{AWESOME(ST)}: This version of AWESOME does not use any AWESOME features including  candaidate plan generation and data parallel execution. AWESOME native operators are executed using single core, while the external functions  executed by external platforms  can use multiple cores. 
\item \textbf{AWESOME(DP)}: This version of AWESOME applies only the data parallelism feature and uses a default execution plan without using AWESOME's planning strategy.
\item \textbf{AWESOME}: This is the full version of AWESOME, incorporating candidate plan generations, cost model, and data parallelism.
\end{itemize}

\subsection{End-To-End Efficiency}
For \textbf{AWESOME}, we measure the end-to-end execution time, which includes the total time from taking a workload as input, parsing, validating, logical planning, physical planning, and executing.
For \textbf{Postgres+UDF} baseline, we report only the execution time, which excludes the data movement cost of moving data from various stores to a single Postgres store, to ensure a fair comparison. For \textbf{RHEEM} baseline, we reformatted the TwitterG datasets as CSV files and rewrote the Cypher queries on the property graph using RHEEM's keywords and our extended RHEEM functions since RHEEM does not support graph databases.

Figures~\ref{fig:patent-exp} to ~\ref{fig:newsanalysis-exp} present the end-to-end execution costs of the six compared methods, with the numbers on top of bars denoting the speed-up ratio to the \textbf{Python} baseline. The black \textbf{X}s indicate that the program either pops out an error or cannot finish within 3 hours.

\noindent\textbf{Comparison with single DBMS and UDF} Our experience with implementing \textbf{Postgres+UDF} has shown that this single DBMS with UDF setting fails to qualify for polystore analytical tasks due to three reasons: 1) Data movement cost, requiring ad-hoc code to move data from various stores to a single store. 2) Programming difficulty, as it is not flexible to program with pure SQL. For workload \textit{NewsAnalysis}, even with the MADLIB toolkit, which implements LDA and PageRank UDFs, hundreds of lines of SQL need to be written (as shown in Appendix A). 3) Efficiency, since in-DBMS implementation of some analytical functions such as LDA and PageRank is much less efficient than using mature packages from Java or Python. AWESOME dramatically speeds up the execution time over \textbf{Postgres+UDF}, achieving up to 31X and 15X speedup for \textbf{PatentAnalysis} and \textbf{PoliSci} workloads, respectively. For \textit{NewsAnalysis}, when $newsR=50K$, \textbf{Postgres+UDF} pops out a server connection lost error message when running in-Postgres LDA function.

\noindent\textbf{AWESOME vs. Python} AWESOME implementations, \textbf{AWESOME(DP)} and \textbf{AWESOME}, show great scalability compared to Python, significantly improving efficiency when parameters get large. For \textbf{PatentAnalysis}, \textbf{AWESOME(DP)} achieves up to 8X speedup, and \textbf{AWESOME} achieves up to 31X. For workload \textit{PoliSci}, as the graph size $g$ increases, AWESOME achieves an increasingly large speedup over the Python implementation. When $g=100K$, \textbf{AWESOME(DP)} is slightly slower than \textbf{Python}, but with the cost model, \textbf{AWESOME} performs similarly to \textbf{Python}. When $g$ is increased to $1M$ and $newsS=5K$, \textbf{AWESOME(DP)} and \textbf{AWESOME} speed up the execution time by around 3x and 5x, respectively. For workload \textit{NewsAnalysis}, when $newsR=50K$ and the keyword size is $10K$, \textbf{AWESOME} can finish in around 10 minutes, which is about a 5x speedup over Python.

\noindent\textbf{AWESOME vs. RHEEM} In our \textbf{RHEEM} implementation, we use RHEEM's native keywords as much as possible, but for some functions or operators that cannot be supported natively by RHEEM, we extend RHEEM with black box functions. \textbf{RHEEM} has similar performance to \textbf{AWESOME(DP)} since its Spark engine executes many operators in a data parallel fashion, but there are no fine-grained optimization opportunities for black box functions, making it less competitive than \textbf{AWESOME}. For the \textbf{PoliSci} workload, \textbf{RHEEM} shows slightly better results than \textbf{AWESOME(DP)} when parameters are small, but it shows worse scalability than AWESOME. Since RHEEM does not have graph database support, we read graph files and wrote Cypher queries in RHEEM's keywords and functions, which hurt scalability. However, AWESOME directly delegated the Cypher queries to the Neo4j database. For the \textbf{NewsAnalysis} workload, we extended RHEEM with a PageRank function. Since this extension uses a more efficient package than the default execution plan of  \textbf{AWESOME(DP)}, it outperforms \textbf{AWESOME(DP)}. However, \textbf{RHEEM} has worse performance than \textbf{AWESOME} with the cost model feature under all parameter settings for every workload.

\subsection{Drill-Down Analysis}

We drill down  to  analyze  the effectiveness of AWESOME data parallelism and how AWESOME's  planning with  a cost model help improve efficiency and scalability.  

The effectiveness of data parallelism varies by workload, depending on the features of the operators in the workload and how much these data parallel-able operators contribute to the total execution time. \textbf{AWESOME(DP)} achieved up to an 11X and 5X speedup over \textbf{AWESOME(ST)} for \textit{PoliSci} and \textit{NewsAnalysis} workloads respectively, because there are computationally expensive data parallel-able operators  such as \textit{CollectWNFromDocs}. However, for  \textit{PatentAnalysis} workload, the bottleneck was the Betweenness function which is an external operator, so \textbf{AWESOME(DP)} only achieved up to a 2X speedup.

To analyze how AWESOME planning works, we present snippets from each workload and show the execution time of different candidate plans for these snippets in Figure \ref{fig:costmodel}. The bars with stars on top are the  execution plans selected by AWESOME's cost model.

 Figure~\ref{fig:costmodel} (a) shows the execution time for different sub-plans related to graph creation, Betweenness and PageRank computation of the graph in \textit{PatentAnalysis}. The results indicate that when working with small graphs, JGraphT outperforms Neo4j by reducing data movement costs. However, as the number of selected keywords increases and graph becomes larger and denser, the scalability and advanced graph algorithms support of Neo4j enable it to outperform JGraphT by a considerable margin. The cost model examines the performance of the entire sub-plan including both graph creation and algorithm computation to determine the optimal physical sub-plan. Even though the creation time of JGraphT is much less than that of Neo4j, the cost model correctly decided to use Neo4j considering the total time.

Figure~\ref{fig:costmodel} (b) presents the execution time for different sub-plans (shown in Figure~\ref{fig:executeSQL}) regarding the cross-engine \textit{ExecuteSQL} logical operator in \textit{PoliSci}. $SenatorHandler$ table from PostgreSQL server is joined with the AWESOME named entity table on an attribute. As the number of  documents increases, the size of the named entity table also increases, then the local Postgres execution plan is much more effective than the remote Postgres plan by avoiding moving large amounts of data to a remote server. While the in-SQLite execution plan saves time on data movement, it is not as efficient as the local Postgres implementation due to SQLite's slower performance for this type of SQL query (Type-II SQL query calibrated).

Figure~\ref{fig:costmodel} (c) displays the execution time for the ExecuteSQL logical operator in \textit{NewsAnalysis}, which executes a Type I SQL query calibrated. Node 1 of Figure~\ref{fig:physicalplan} depicts the possible physical sub-plans. The in-SQLite execution plan is particularly dominant when the number of keywords passed to the WHERE IN predicate in the query increases.

These results on small snippets demonstrate the effectiveness  of our  cost model.

%% file: sections/2-relatedwork2.tex
\vspace{-1ex}
\section{Related Work}
\label{sec:related}
In today’s big data era, data heterogeneity is becoming one of the biggest challenges
for modern data management systems. In order to support the management of heterogeneous data more efficiently, two approaches have been developed over the last decades.

The first approach  is to build a multi-model DBMS, which is a single DBMS that supports multiple data models natively. Many popular commercial databases, such as PostgreSQL and MongoDB, support multiple models by either extending their native storage or designing a completely new storage to support a new data model \cite{lu2019multi}. However, this approach still faces the issue of extensibility, as it can be difficult to extend an existing multi-model database to support arbitrary data models. Our focus is on the second approach: building polystore systems. Unlike multi-model databases, polystore systems are DBMSs built on top of multiple different underlying stores for different data models. AWESOME is an example of a polystore system.

We present  features comparison of selected  prior polystore  and AWESOME in Table~\ref{tab:relatedwork}. 
Besides systems shown here, there are  other existing work~\cite{khayyat2015bigdansing, gog2015musketeer, doka2016mix}. We conclude some important language and system design features for analytical polystores based on our experience working with domain scientists.   As the table suggests, none of the prior polystore systems has all of these features. We analyze them from language and system design perspectives. 

\vspace{-1ex}
\subsection{Polystore Languages}
In terms of polystore languages, several existing systems provide various levels of support for querying and analysis. BigDAWG, Estocada and Tatooine  all support querying backend DBMSs using their  native languages but do not provide support for analytical functions or control flow logic such as IF-ELSE or For loop. Myria   provides a hybrid imperative-declarative query language called MyriaL that supports SQL queries, comprehensions, and function calls, but SQL is the only supported query language. Hybrid.Poly  offers a hybrid relational analytical query language based on their extended relational model, which allows for the use of analytical functions in the SELECT clause. However, it does not provide built-in native support for text or graph analytical functions and  like MyriaL, SQL is the only supported query language.  RheemLatin, the language for Rheem  is an extension of PigLatin \cite{olston2008pig}. It provides a platform-agnostic way of specifying analytical plans using its own keywords each of   which can be translated to different execution platforms. However, this approach has some limitations. Firstly, it lacks support for native  query languages, while it shows that it is possible to express a SQL query using its syntax and keywords, it can be challenging to  keep the semantic completeness of query languages  such as Cypher or Lucene.  Secondly, the RheemLatin keywords are fine-grained, requiring users to write lines of code using these keywords to implement analytical functions such as SGD, which demands a significant amount of programming expertise. 

\vspace{-1.5ex}
\subsection{Polystore Systems}
Many polystore systems \cite{alotaibi2019towards, alotaibi2020estocada, wang2017myria, bonaque2016mixed} only support DBMSs as the backend platforms. BigDAWG  is one such system. It organizes storage engines into a number of islands, each consisting of a data model, a set of operations, and a set of candidate storage engines. It supports heterogeneous data models including relation and array, and supports cast functions to migrate data between two stores in the same island or from 
one island  to another island with a different model. While BigDAWG is highly extensible to support more DBMSs,  the focus on DBMSs as backend platforms restricts the range of analytical capabilities and application scenarios that BigDAWG can support.  Besides,   BigDAWG does not prioritize query optimization.  Estocada is another such polystore system   but takes a different perspective on query optimization. They take advantage of materialized views, even if they are in different data models than the queried datasets. 
Estocada primarily focuses on logical plan-level optimization, and their approach can be incorporated into other polystore systems. They have incorporated this rewriting approach into BigDAWG, significantly improving its efficiency. However, AWESOME primarily focuses on physical plan and execution-level optimization, and Estocada's rewriting can easily be incorporated into AWESOME as well.




The Rheemix system  is a more general-purpose polystore system that supports a wider range of applications by building over diverse engines, including not only DBMSs but also other analytical systems such as Spark, Flink, and Java Streams. Users write analytical tasks using  a workflow of Rheemix keywords  (e.g., \texttt{filter, map, groupBy}) which are directly mapped to operators of the underlying platforms through operator inflation. A cost model is designed to choose the best mapping. However, the use of primitive (fine-granular) operators can limit its expressiveness while making optimization easier.

Our general observation is that, while there is a clear growth in creating and using polystores for different applications, no existing polystore system supports all requisite  features needed by domain scientists to solve a variety of real-world analytical workloads, and it is hard to extend an existing system to support all these features.

%% file: sections/9-conclusion-and-future-work.tex
\vspace{-1ex}
\section{Conclusion and Future Work}\label{sec:conclusion}
In this work, we empower an emerging class of large-scale data science workloads that naturally straddle analytics over relations, graphs, and text.
In contrast to complementary work on polystores that aim for high generality, we build a more specific tri-store system AWESOME that offers a succinct domain-specific language on top of standard unistore engines, automatically handles intermediate data, and performs automatic query optimizations. We empirically demonstrate the functionality and efficiency of AWESOME.
As for future work, we plan to pursue new cross-model query optimization opportunities to make AWESOME even faster and also scale to distributed execution.

%% file: appendix.tex
\onecolumn
\section{SQL script for workloads: UDF + Pure SQL}
\subsection{\textit{NewsAnalysis}}
\counterwithin{lstlisting}{section}
\definecolor{dkgreen}{rgb}{0,0.6,0}
\definecolor{gray}{rgb}{0.5,0.5,0.5}
\definecolor{mauve}{rgb}{0.58,0,0.82}
\lstdefinelanguage{ASM}{
    morekeywords={Use, create analysis, range,  as, executeSQL, executeCypher, 
    executeSolr, preprocess, lda, map, union, buildWordNeighborGraph, 
    ConstructGraphFromRelation, pagerank, sum, store, rownames,  stringreplace, stringJoin, toList, tokenize, keyphraseMining, betweenness 
    },
    sensitive=false, 
    morecomment=[l]{//}, 
    morecomment=[s]{/*}{*/}, 
    morestring=[b]" 
}
\lstdefinestyle{myCustomSQLStyle}{
frame=single,
language=sql,
basicstyle=\ttfamily,
showstringspaces=false,
numbers=none,
commentstyle=\color{gray},
columns=flexible,
basicstyle={\small\ttfamily},
numbers=none,
numberstyle=\tiny\color{gray},
breaklines=true,
breakatwhitespace=true,
tabsize=3,
}
\lstdefinestyle{myCustomPythonSQLStyle}{
frame=single,
language=python,
basicstyle=\ttfamily,
showstringspaces=false,
numbers=none,
commentstyle=\color{gray}
showstringspaces=false,
columns=flexible,
basicstyle={\small\ttfamily},
numbers=none,
numberstyle=\tiny\color{gray},
keywordstyle=\color{blue},
commentstyle=\color{dkgreen},
stringstyle=\color{mauve},
breaklines=true,
breakatwhitespace=true,
tabsize=3,
}
\lstdefinestyle{myCustomAwesomeStyle}{
frame=single,
language=ASM,
basicstyle=\ttfamily,
showstringspaces=false,
numbers=none,
commentstyle=\color{gray}
showstringspaces=false,
columns=flexible,
basicstyle={\small\ttfamily},
numbers=none,
numberstyle=\tiny\color{gray},
keywordstyle=\color{blue},
commentstyle=\color{dkgreen},
stringstyle=\color{mauve},
breaklines=true,
breakatwhitespace=true,
tabsize=3,
}
\lstnewenvironment{sqlpyls}[1]{\lstset{basicstyle=\tiny, style=myCustomPythonSQLStyle}}{\captionof{lstlisting}{#1}}
\lstnewenvironment{sqlls}[1]{\lstset{basicstyle=\tiny, style=myCustomSQLStyle}}{\captionof{lstlisting}{#1}}
\lstnewenvironment{asmls}[1]{\lstset{basicstyle=\tiny, style=myCustomAwesomeStyle}}{\captionof{lstlisting}{#1}}
\begin{sqlpyls}{Python UDF}
create function buildgraphfromtext(text character varying[], distance integer) 
returns character varying[] language plpython2u
as
$$
count = {}
for i in range(len(text)-distance):
    for j in range(1, distance):
        temp = (text[i], text[i+j])
        if temp in count:
            count[temp] += 1
        else:
            count[temp] = 1
result = []
for key in count:
    result.append([key[0],key[1],count[key]])
return result
$$;
\end{sqlpyls}

\begin{sqlls}{SQL UDF}
create function unnest_2d_1d(anyarray) returns SETOF anyarray 
immutable strict parallel safe language sql as
$$
SELECT array_agg($1[d1][d2]) FROM 
generate_subscripts($1,1) d1, generate_subscripts($1,2) d2
GROUP BY d1 ORDER BY d1
$$;
\end{sqlls}

\begin{sqlls}{SQL Code}
drop table if exists tokenizednews, graph, topicgraph CASCADE;
drop MATERIALIZED VIEW if exists graphelement;
-- set execution begin time
INSERT INTO timenow(type, starttime, stoptime)
SELECT 'Newsanalysis_5k_1k_start', now(), clock_timestamp();
---- tokenize and build word neighbor graph
CREATE table tokenizednews as (
    select id as docid, news from newspaper 
    where  src = 'http://www.chicagotribune.com/' 
    order by id limit 5000);
ALTER TABLE tokenizednews ADD COLUMN words TEXT[];
UPDATE tokenizednews SET words = regexp_split_to_array(
    lower(regexp_replace(news, E'[,.;]','', 'g')), E'[\\s+]');
create MATERIALIZED VIEW graphelement as (
    with temp as (select unnest_2d_1d(buildgraphfromtext(words, 5)) as n from tokenizednews), temp2 as (select n[1] as word1, n[2] as word2, n[3]::INTEGER as cnt from temp)
    select word1, word2, sum(cnt) from temp2 group by word1, word2);   
---- LDA
SELECT madlib.term_frequency(
    'tokenizednews',         -- input table
    'docid',                 -- document id column
    'words',                 -- vector of words in document
    'news_tf',               -- output table with term frequency
    TRUE);                   -- TRUE to created vocabulary table
SELECT madlib.lda_train( 
    'news_tf',               -- test table in the form of tf
    'lda_model',             -- model table created by LDA training   
    'lda_output_data',       -- readable output data table
    200000,                  -- vocabulary size
    10,                      -- number of topics
    1000,                    -- number of iterations
    1,                       -- Dirichlet prior for the per-doc topic multinomial
    0.01                     -- Dirichlet prior for the per-topic word multinomial
    );
SELECT madlib.lda_get_topic_desc( 
    'lda_model',              -- LDA model generated in training
    'news_tf_vocabulary',     -- vocabulary table that maps wordid to word
    'helper_output_table',    -- output table for per-topic descriptions
    20000);
INSERT INTO timenow(type, starttime, stoptime)  SELECT 'Newsanalysis_5k_LDA', now(), clock_timestamp();
--- create a text network graph   
create table graph as (select word1, word2 from graphelement where word1!='' and word2!='' group by word1, word2);
---- build graph for each one
create table topicgraph as (
    with topicwords as (select word, wordid from helper_output_table 
        where prob > 0 and topicid = 0 order by prob desc limit 1000),
    temp as (select wordid, word2  from graph, topicwords where word1 = word)
    select temp.wordid as word1, topicwords.wordid as word2, 1 as topic from temp, topicwords where temp.word2=word);
insert into topicgraph(word1, word2, topic) (
    with topicwords as (select word, wordid from helper_output_table 
        where prob > 0 and topicid = 1 order by prob desc limit 1000),
    temp as (select wordid, word2  from graph, topicwords where word1 = word)
    select temp.wordid as word1, topicwords.wordid as word2, 2 as topic from temp, topicwords where temp.word2=word);
insert into topicgraph(word1, word2,  topic) (
    with topicwords as (select word, wordid from helper_output_table 
        where prob > 0 and topicid = 2 order by prob desc limit 1000),
    temp as (select wordid, word2  from graph, topicwords where word1 = word)
    select temp.wordid as word1, topicwords.wordid as word2, 3 as topic from temp, topicwords where temp.word2=word);
insert into topicgraph(word1, word2,  topic) (
    with topicwords as (select word, wordid from helper_output_table 
        where prob > 0 and topicid = 3 order by prob desc limit 1000),
    temp as (select wordid, word2  from graph, topicwords where word1 = word)
    select temp.wordid as word1, topicwords.wordid as word2, 4 as topic from temp, topicwords where temp.word2=word);
insert into topicgraph(word1, word2,  topic) (
    with topicwords as (select word, wordid from helper_output_table 
        where prob > 0 and topicid = 4 order by prob desc limit 1000),
    temp as (select wordid, word2  from graph, topicwords where word1 = word)
    select temp.wordid as word1, topicwords.wordid as word2, 5 as topic from temp, topicwords where temp.word2=word);
insert into topicgraph(word1, word2,  topic) (
    with topicwords as (select word, wordid from helper_output_table 
        where prob > 0 and topicid = 5 order by prob desc limit 1000),
    temp as (select wordid, word2  from graph, topicwords where word1 = word)
    select temp.wordid as word1, topicwords.wordid as word2, 6 as topic from temp, topicwords where temp.word2=word);
insert into topicgraph(word1, word2,  topic) (
    with topicwords as (select word, wordid from helper_output_table 
        where prob > 0 and topicid = 6 order by prob desc limit 1000),
    temp as (select wordid, word2  from graph, topicwords where word1 = word)
    select temp.wordid as word1, topicwords.wordid as word2, 7 as topic from temp, topicwords where temp.word2=word);
insert into topicgraph(word1, word2,  topic) (
    with topicwords as (select word, wordid from helper_output_table 
        where prob > 0 and topicid = 7 order by prob desc limit 1000),
    temp as (select wordid, word2  from graph, topicwords where word1 = word)
    select temp.wordid as word1, topicwords.wordid as word2, 8 as topic from temp, topicwords where temp.word2=word);
insert into topicgraph(word1, word2,  topic) (
    with topicwords as (select word, wordid from helper_output_table 
        where prob > 0 and topicid = 8 order by prob desc limit 1000),
    temp as (select wordid, word2  from graph, topicwords where word1 = word)
    select temp.wordid as word1, topicwords.wordid as word2, 9 as topic from temp, topicwords where temp.word2=word);
insert into topicgraph(word1, word2,  topic) (
    with topicwords as (select word, wordid from helper_output_table 
        where prob > 0 and topicid = 9 order by prob desc limit 1000),
    temp as (select wordid, word2  from graph, topicwords where word1 = word)
    select temp.wordid as word1, topicwords.wordid as word2, 10 as topic from temp, topicwords where temp.word2=word);
--- pagerank for each topic
SELECT madlib.pagerank(
        'news_tf_vocabulary',               -- Vertex table
        'wordid',                           -- Vertix id column
        'topicgraph',                       -- Edge table
        'src=word1, dest=word2',            -- Comma delimted string of 
        'pagerank_out',                     -- Output table of PageRank
         NULL,                              -- Default damping factor (0.85)
         NULL,                              -- Default max iters (100)
         0.00000001,                        -- Threshold
        'topic');
INSERT INTO timenow(type, starttime, stoptime) 
SELECT 'Newsanalysis_5k_1k', now(), clock_timestamp();
\end{sqlls}

\subsection{\textit{PoliSci}}
\begin{sqlpyls}{Python UDF}
create function callner (tname character varying, 
colname character varying, 
filename character varying) 
returns character varying
    language plpython3u
as
$$
import os
import subprocess
with open(filename, 'w') as f:
    for row in plpy.cursor("SELECT " + colname + " FROM "+tname):
        f.write(row[colname]+'\n')
temp_file = filename.split(".")[0]
subprocess.call(['java', '-jar', 
    '/var/lib/postgresql/data/ner/target/NER-1.0-SNAPSHOT.jar', 
    '-i', filename, '-o', temp_file])
return temp_file
$$;
\end{sqlpyls}

\begin{sqlls}{SQL code}
drop table if exists keynews, keyusers, namedentity, timenow;
--- record start time 
INSERT INTO timenow( type, starttime, stoptime) 
    SELECT 'ner_start', now(), clock_timestamp();
create table keynews as (
    select news from newspaper 
    where news @@ to_tsquery('corona|covid|pandemic|vaccine') 
    limit 5000);
select callner('xw_keynews', 'news', 'news.txt');
CREATE TABLE namedentity (
  type text,
  entity text
);
COPY namedentity(type, entity)
FROM 'news'
DELIMITER ','
CSV HEADER;
create table keyusers as (
select distinct t.name as name, t.twittername as twittername
              from twitterhandle t,
                   namedentity e
              where LOWER(e.entity) = LOWER(t.name));
--- record ner time
INSERT INTO timenow( type, starttime, stoptime) 
    SELECT 'ner_end', now(), clock_timestamp();
select text from neo4j_tweet50000, keyusers 
    where text ilike '%' || keyusers.name || '%';
select * from neo4j_user_user50000 
    where userid2 in (
        select userid from neo4j_user50000 
        where username in (select twittername from keyusers));
--- record end time 
INSERT INTO timenow( type, starttime, stoptime) 
    SELECT 'all', now(), clock_timestamp();    
\end{sqlls}

\subsection{\textit{PatentAnalysis}}
\begin{sqlpyls}{Python UDFs}
create or replace function buildgraphfromtext(text character varying, keywords character varying[]) returns character varying[] language plpython3u as
$$
count = {}
key_phrase = set(keywords)
doc = [word.lower() for word in text.split(" ") if word in key_phrase]
for i in range(len(doc)):
    for j in range(i + 1, len(doc)):
        word_tuple = (min(doc[i], doc[j]), max(doc[i], doc[j]))
        if word_tuple in count:
            count[word_tuple] += 1
        else:
            count[word_tuple] = 1
result = []
for key in count:
    result.append([key[0], key[1], count[key]])
return result
$$;

create function keyphrase(tname character varying, colname character varying, num integer) returns character varying[]
    language plpython3u
as
$$
from subprocess import run
import os 
autophrase_dir = "/data/AutoPhrase/"
# store data to autophrase data directory
data_dir = "data/temp.txt"
count = 0
with open(autophrase_dir+data_dir, 'w') as fp:
    for row in plpy.cursor("SELECT " + colname + " FROM "+ tname):
        count += 1
        fp.write(row[colname] + os.linesep)
# call script
cmd = "sh auto_phrase.sh {} {}".format(data_dir, str(count/500))
run(cmd, shell=True, cwd=autophrase_dir)
# read data
result_dir = "{}models/{}/AutoPhrase_single-word.txt".format(autophrase_dir, "AWESOME")
key_phrase = []
count = 0
with open(result_dir) as fp:
    for line in fp:
        key_phrase.append(line.rstrip().split('\t')[1])
        count += 1
        if count > num:
            break
run("rm {}".format(data_dir), shell=True, cwd=autophrase_dir)
# print(key_phrase)
return key_phrase
$$;

create or replace function callbetweenness(tname character varying) returns character varying[]
    language plpython3u
as 
$$
import networkx as nx
graph = []
g = nx.Graph()
word2id = {}
id2word = {}
crt_id = 0
for row in plpy.cursor("SELECT * FROM "+ tname):
    g.add_edge(row['word1'], row['word2'], weight=row['cnt'])
betweenness = nx.betweenness_centrality(g).items()
degree = nx.pagerank(g).items()
return [i[0] for i in sorted(betweenness, key=lambda f: f[1], reverse=True)[:20]]
$$;
\end{sqlpyls}

\begin{sqlls}{SQL UDF}
create function unnest_2d_1d(anyarray) returns SETOF anyarray immutable strict parallel safe language sql as
$$
SELECT array_agg($1[d1][d2]) FROM generate_subscripts($1,1) d1, generate_subscripts($1,2) d2
GROUP BY d1 ORDER BY d1
$$;
\end{sqlls}
\begin{sqlls}{SQL code}
INSERT INTO timenow(type, starttime, stoptime)
SELECT 'PatentAnalysis_5k_500_start', now(), clock_timestamp();
--- tokenize using madlib 
CREATE table abstract5k as (select abstract as abs from sbir_award_data where abstract is not null limit 5000);
--- create graphelements 
create MATERIALIZED VIEW graph5k500 as (
with temp as (
    with keywords as (select keyphrase('abstract5k', 'abs', 500) as words)
    select unnest_2d_1d(buildgraphfromtext(abs, keywords.words)) as n
    from abstract5k, keywords),
temp2 as (select n[1] as word1, n[2] as word2, n[3]::INTEGER as cnt from temp)
select word1, word2, sum(cnt) as cnt from temp2 group by word1, word2
);
--- get keywords and graph from keywords 
select callbetweenness('graph5k500');
INSERT INTO timenow(type, starttime, stoptime)
SELECT 'PatentAnalysis_5k_500_end', now(), clock_timestamp();
\end{sqlls}
\newpage
\section{ADIL Script for Workloads}\label{appendix:script}
\subsection{\textit{NewsAnalysis}}
\begin{asmls}{ADIL Script for \textit{NewsAnalysis}}
/*specify configuration file*/
USE newsDB;
/* main code block */
create analysis NewsAnalysis as (
src :=  "http://www.chicagotribune.com/";
rawNews := executeSQL("News", "select id as newsid, news as  
                    newsText from newspaper where src = $src limit 1000");
processedNews := preprocess(rawNews.newsText, docid=rawNews.newsid, stopwords="stopwords.txt");
numTop := 10;
DTM, WTM := lda(processedNews, docid=true, topic=numTop, numKeywords=1000);
topicID := [range(0, numberTopic, 1)];
/* get the keywords for each topic */
wtmPerTopic := topicID.map(i => WTM where getValue(_:Row, i) > 0.00);
wordsPerTopic := wtmPerTopic.map(i => rowNames(i));
wordsOfInterest := union(wordsPerTopic);
G := buildWordNeighborGraph(processedNews, maxDistance=5, splitter=".", words=wordsOfInterest);
relationPerTopic := wordsPerTopic.map(words => (<n:String, m:String, count:Integer>) executeCypher(G,  "match (n)-[r]->(m) where n in $words and m in $words return n, m, r.count as count"));
graphPerTopic := relationPerTopic.map(r => ConstructGraphFromRelation(r, (:Word {id: r.n})-[:Cooccur{count: r.count}]-> (:Word{id: r.m})));
scores := graphPerTopic.map(g =>pageRank(g, topk=true, num=20));
aggregatePT := scores.map(i => sum(i.pagerank));
/* store a list to rDBMS as a relation*/
store(aggregatePT t, dbName="News", tableName="aggregatePageRankofTopk", columnName=[("id",t.index), ("pagerank",t.value)]);
\end{asmls}

\subsection{\textit{PoliSci}}
\begin{asmls}{ADIL for \textit{PoliSci}}
use newsDB as polystore;
create analysis politician as (
keywords := ["corona", "covid", "pandemic", "vaccine"];
temp := keywords.map(i => stringReplace("text-field: $", i));
t := stringJoin(" OR ", temp);
doc<text-field:String> := executeSOLR("allnews", """q= $t & rows=50""");
namedentity := NER(doc.text-field);
user := executeSQL("News", "select distinct t.name as name, t.twittername as twittername from twitterhandle t, $namedentity e where LOWER(e.name)=LOWER(t.name)");
userNameList := toList(user.name);
userNameP := userNameList.map(i => stringReplace("t.text contains '$' ", i));
predicate := stringJoin(" OR ", userNameP);
users<name:String> := executeCypher("tweetG", "match (u:User)-[:mention]-(n:User) where n.userName in $user.twittername return u.userName as name");
tweet<t:String> :=  executeCypher("tweetG", "match (t:Tweet) where $predicate return t.text as t");
);
\end{asmls}
\subsection{\textit{PatentAnalysis}}
\begin{asmls}{ADIL for \textit{PatentAnalysis}}
use newsDB as polystore;
create analysis patentanalysis as (
abstracts := executeSQL("Awesome", "select abstract from sbir_award_data where abstract is not null limit 1000");
docs := tokenize(abstracts.abstract, stopwords="C:\Users\xiuwen\IdeaProjects\awesome-new-version\stopwords.txt");
keywords := keyphraseMining(docs, 500);
wordsPair := buildWordNeighborGraph(docs, words=keywords); 
graph := ConstructGraphFromRelation(wordsPair, (:Word {value: $wordsPair.word1}) -[:Cooccur{count: $wordsPair.count}]->(:Word{value: $wordsPair.word2}));
betweenness := betweenness(graph);
pagerank := pageRank(graph);
);
\end{asmls}

\newpage \clearpage
\section{Python code for  workloads}
\subsection{\textit{NewsAnalysis}}
\begin{sqlpyls}{Python code for \textit{NewsAnalysis}}
import getopt
import io
import numpy as np
import sys
import time
from multiprocessing import Pool
import math
import networkx as nx
import sqlalchemy as sal
from gensim import corpora
from gensim.models.ldamulticore import LdaMulticore
from gensim.models.wrappers import LdaMallet
from sqlalchemy import text

# tokenize
def tokenize(doc):
    return doc.split(" ")

def build_graph_from_text(docs, dis, words):
    count = {}
    for doc in docs:
        for i in range(len(doc) - dis):
            if doc[i] in words:
                for j in range(1, dis):
                    tempPair = (min(doc[i], doc[i + j]), 
                                max(doc[i], doc[i + j]))
                    if doc[i + j] in words:
                        if tempPair in count:
                            count[tempPair] += 1
                        else:
                            count[tempPair] = 1
    return count

def split(list_a, chunk_size):
    for idx in range(0, len(list_a), chunk_size):
        yield list_a[idx:idx + chunk_size]

# LDA
def LDA(docs):
    id2word = corpora.Dictionary(docs)
    corpus = [id2word.doc2bow(text) for text in docs]
    model = LdaMulticore(corpus=corpus, num_topics=10, 
                         iterations=1000, id2word=id2word,workers=15)
    return model.show_topics(num_words=len(id2word))

def LDA_mallet(docs, threshold):
    path_to_mallet_binary = "/users/Xiuwen/Mallet/bin/mallet"
    id2word = corpora.Dictionary(docs)
    corpus = [id2word.doc2bow(text) for text in docs]
    model = LdaMallet(path_to_mallet_binary, corpus=corpus, num_topics=10, 
                        iterations=1000, id2word=id2word, 
                        random_seed=2, alpha=0.1, workers=96)
    matrix = model.get_topics()
    words = []
    for row in matrix:
        words_ids = np.argsort(row)[-threshold :]
        words.append(set([id2word[w] for w in words_ids]))
    return words

def page_rank(graph_data, num_of_point):
    G = nx.Graph()
    for i in graph_data:
        G.add_edge(i[0], i[1], weight=i[2])
    pr = nx.pagerank(G)
    return sorted(pr.items(), key=lambda val: val[1], 
                  reverse=True)[:num_of_point]

if __name__ == '__main__':
    num_of_docs = ""
    threshold = ""
    argv = sys.argv[1:]
    core = 1
    try:
        opts, args = getopt.getopt(argv, "i:t:c:")
    except getopt.GetoptError:
        print('query1.py -i <size> -t <threshold> -c <cores>')
        sys.exit(2)
    for opt, arg in opts:
        if opt == "-i":
            num_of_docs = arg
        elif opt == "-t":
            threshold = arg
        elif opt == "-c":
            core = int(arg)
    if num_of_docs == "" or threshold == "":
        print('query1.py -i <size> -t <threshold> -c <cores>')
        sys.exit(2)
    start = time.time()
    # read data
    engine = sal.create_engine('postgresql+psycopg2://')
    conn = engine.connect()
    sql = text("select newstext from xw_news_"+num_of_docs)
    result = conn.execute(sql)
    sql_exe = time.time()
    print("sql execution time: " + str(sql_exe - start))
    tokenized_docs = [tokenize(i[0]) for i in result]
    tk_exe = time.time()
    print("tokenize execution time: " + str(tk_exe - sql_exe))
    # read LDA results
    path = "/proj/awesome-PG0/data/"
    if num_of_docs=='5000':
        path = path + "5k/"
    else:
        path = path + "50k/"
    # path = "C://Users//xiuwen//Documents//"
    # get only partial words
    words_index_per_topic = []
    words_file = open(path+'sortedwords.txt', 'r')
    words_lines = words_file.readlines()
    for words in words_lines:
        words_index = [int(i) for i in words.strip().split(", ")][:int(threshold)]
        words_index_per_topic.append(words_index)
    alphabet_file = io.open(path +'alphabet.txt', 'r', encoding='utf-8')
    alphabet = alphabet_file.readline().strip().split(", ")
    words_per_topic = [set([alphabet[i] for i in index]) 
                       for index in words_index_per_topic]
    # get all words
    all_words = list(set.union(*words_per_topic))
    print("size of keywords after union: " + str(len(all_words)))
    lda_exe = time.time()
    print("lda execution time: " + str(lda_exe - tk_exe))
    pool = Pool(processes=core)
    jobs = []
    # split data to the number of cores partitions
    size = int(math.ceil(float(len(tokenized_docs)) / core))
    count_threads = []
    sublists = list(split(tokenized_docs, size))
    print(len(sublists))
    for alist in sublists:
        count_threads.append(pool.apply_async(
            build_graph_from_text, (alist, 5, all_words)))
    pool.close()
    pool.join()
    graph_elements = []
    total_counts = {}
    for c_count in count_threads:
        c_count_map = c_count.get()
        for key in c_count_map:
            if key in total_counts:
                total_counts[key] += c_count_map[key]
            else:
                total_counts[key] = c_count_map[key]
    for key in total_counts:
        graph_elements.append([key[0], key[1], total_counts[key]])
    print (graph_elements[:10])
    bg_exe = time.time()
    print("bg execution time: " + str(bg_exe - lda_exe))
    graph_data_per_topic = []
    # get graph data for each topic
    for i in range(10):
        words_in_this_topic = words_per_topic[i]
        temp_graph = [g for g in graph_elements if g[0] 
                      in words_in_this_topic and g[1] in words_in_this_topic]
        graph_data_per_topic.append(temp_graph)
    bsg_exe = time.time()
    print("bg execution time: " + str(bsg_exe - bg_exe))

    # get pagerank for each topic
    pagerank_all_topics = [page_rank(i, 20) for i in graph_data_per_topic]
    print(pagerank_all_topics)
    end = time.time()
    print("pr execution time: " + str(end - bsg_exe))
    print(end - start)
\end{sqlpyls}

\subsection{\textit{PoliSci}}
\begin{sqlpyls}{Python code for \textit{PoliSci}}
import time
import sqlalchemy as sal
from sner import Ner
import getopt
import sys

if __name__ == '__main__':
    num_of_docs = ""
    tweet = ""
    argv = sys.argv[1:]
    core = 1
    try:
        opts, args = getopt.\
            getopt(argv, "i:t:c:")
    except getopt.GetoptError:
        print('query2.py -i <size> -t <tweet>')
        sys.exit(2)
    for opt, arg in opts:
        if opt == "-i":
            num_of_docs = arg
        elif opt == "-t":
            tweet = arg
    if num_of_docs == "" or tweet == "":
        print('query1.py -i <size> -t <threshold> -c <cores>')
        sys.exit(2)

    start = time.time()
    # sql query without full text search index
    sql = "select news from usnewspaper where news " \
          "@@ to_tsquery('corona|covid|pandemic|vaccine') limit " \
          + num_of_docs
    engine = sal.create_engine('postgresql+psycopg2://')
    conn = engine.connect()
    result = conn.execute(sql)
    docs = [i[0] for i in result]
    # print([ner(i[0]) for i in result])
    print("full text search cost: " + str(time.time() - start))
    # NER
    nes = []
    tagger = Ner(host='localhost', port=9299)
    for d in docs:
        try:
            en = tagger.get_entities(d)
            nes.extend(en)
        except:
            continue
    key_nes = set([i[0].lower() for i in nes if i[1] != 'O'])
    # get senators
    sql = "select name, twittername from twitterhandle"
    conn = engine.connect()
    result = conn.execute(sql)
    senators_name_tn = [[i[0].lower(), i[1]] for i in result]
    # get userid-username
    sql = "select userid, username from xw_neo4j_user"+tweet
    conn = engine.connect()
    result = conn.execute(sql)
    user_id_name = [[i[0], i[1]] for i in result]
    # get user-tweet network
    sql = "select text from xw_neo4j_tweet"+tweet
    conn = engine.connect()
    result = conn.execute(sql)
    texts = [i[0].lower() for i in result]
    # get user-user network
    sql = "select userid1, userid2 from xw_neo4j_user_user"+tweet
    conn = engine.connect()
    result = conn.execute(sql)
    users_users = [[i[0], i[1]] for i in result]
    # get key users name and id
    key_names = set()
    key_users = set()
    for i in key_nes:
        for s in senators_name_tn:
            name = [i for i in s[0].lower().split(" ") if len(i) > 2]
            if i in name:
                key_names.add(i)
                key_users.add(s[1])
    key_users_ids = []
    for i in key_users:
        for j in user_id_name:
            if i == j[1]:
                key_users_ids.append(j[0])
        # get tweets that contain key users names
    key_tweets = []
    for t in texts:
        for i in key_names:
            if i in t.split(" "):
                key_tweets.append(t)
                break
    # get users that mention key user id
    second_key_users = []
    for i in key_users_ids:
        for j in users_users:
            if j[1] == i:
                second_key_users.append(j[0])
    print("total cost: " + str(time.time() - start))
    print(len(second_key_users))
    print(len(key_tweets))
\end{sqlpyls}

\subsection{\textit{PatentAnalysis}}
\begin{sqlpyls}{Python code for \textit{PatentAnalysis}}
import getopt
import math
import os
import sys
import time
from multiprocessing import Pool
from rake_nltk import Rake
from subprocess import run

import networkx as nx
import sqlalchemy as sal
from nltk.corpus import stopwords
from sqlalchemy import text

def tokenize(doc_list, stop_words):
    print(len(doc_list))
    return [[word.lower() for word in i.split(" ") if word not in stop_words] for i in doc_list]

def keyphrase_extract(docs, num):
    rake = Rake(max_length=1, include_repeated_phrases=False)
    rake.extract_keywords_from_text(docs)
    phrase = rake.get_ranked_phrases()
    key_phrase = phrase[:num]
    return key_phrase

def split(list_a, chunk_size):
    for idx in range(0, len(list_a), chunk_size):
        yield list_a[idx:idx + chunk_size]

if __name__ == '__main__':
    num_of_docs = ""
    num_of_keywords = ""
    argv = sys.argv[1:]
    core = 1
    try:
        opts, args = getopt.getopt(argv, "i:t:c:")
    except getopt.GetoptError:
        print('query3.py -i <size> -t <threshold> -c <cores>')
        sys.exit(2)
    for opt, arg in opts:
        if opt == "-i":
            num_of_docs = arg
        elif opt == "-t":
            num_of_keywords = arg
        elif opt == "-c":
            core = int(arg)
    if num_of_docs == "" or num_of_keywords == "":
        print('query3.py -i <size> -t <threshold> -c <cores>')
        sys.exit(2)
    start = time.time()
    engine = sal.create_engine('postgresql+psycopg2://postgres:Sdsc2018#@awesome-hw.sdsc.edu/postgres')
    conn = engine.connect()
    sql = text('''select abstract from sbir_award_data where abstract is not null  limit ''' + num_of_docs)
    result = conn.execute(sql)
    doc = [i[0] for i in result]
    sws = stopwords.words()
    pool = Pool(processes=core)
    jobs = []
    # split data to the number of cores partitions
    size = int(math.ceil(len(doc) / float(core))) + 1
    doc_threads = []
    sublists = list(split(doc, size))
    for alist in sublists:
        doc_threads.append(pool.apply_async(tokenize, (alist, sws)))
    pool.close()
    pool.join()
    tokenized_docs = [i for item in doc_threads for i in item.get()]
    doc_list = [" ".join(doc) for doc in tokenized_docs]
    # doc_string = " ".join(doc_list)
    key_phrase = keyphrase_autophrase(doc_list, int(num_of_keywords))
    print(len(key_phrase))
    g = nx.Graph()
    word2id = {}
    id2word = {}
    crt_id = 0
    graph_element = {}
    for doc in tokenized_docs:
        for i in range(len(doc)):
            if doc[i] not in key_phrase:
                continue
            for j in range(i + 1, len(doc)):
                if doc[j] in key_phrase:
                    word_tuple = (min(doc[i], doc[j]), max(doc[i], doc[j]))
                    if word_tuple in graph_element:
                        graph_element[word_tuple] += 1
                    else:
                        graph_element[word_tuple] = 1
    for edge in graph_element:
        g.add_edge(edge[0], edge[1], weight=graph_element[edge])
    print(g.number_of_edges())
    start_betweenness = time.time()
    betweenness = nx.betweenness_centrality(g).items()
    print([i[0] for i in sorted(betweenness, key=lambda f: f[1], reverse=True)[:20]])
    degree = nx.pagerank(g).items()
    print([i[0] for i in sorted(degree, key=lambda f: f[1], reverse=True)[:20]])
    end = time.time()
    print("betweeness+pagerank time: " + str(end-start_betweenness))
    print("total time: " + str(end-start))

\end{sqlpyls}

\newpage\clearpage
\section{Calibration results.}\label{appendix:benchmark}

\begin{figure}[ht]
    \centering
    \begin{subfigure}[b]{0.22\textwidth}
        \centering
        \includegraphics[width=\textwidth]{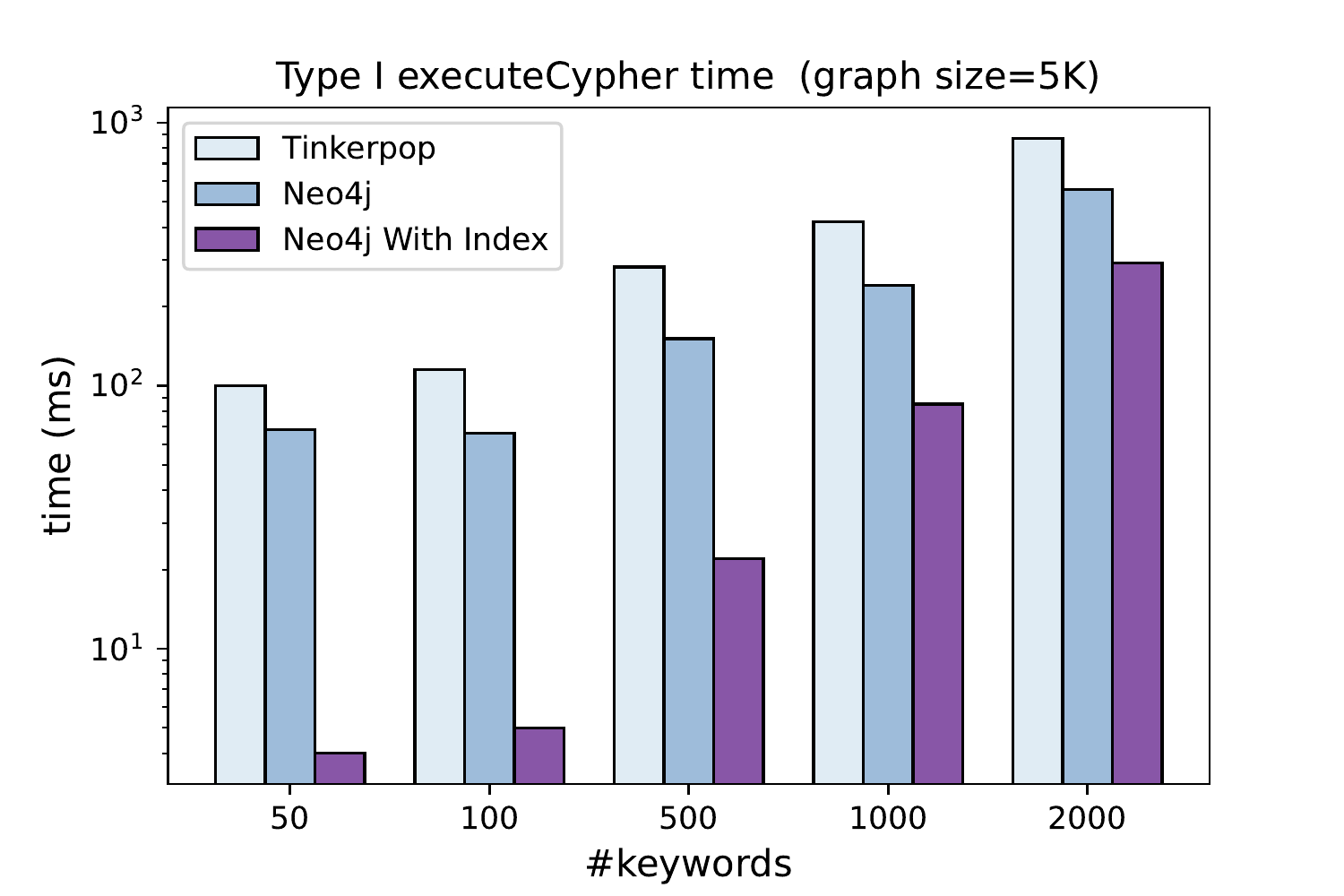}
        \caption{Graph size $=5K$.}
    \end{subfigure}
    \hfill
     \begin{subfigure}[b]{0.22\textwidth}
        \centering
        \includegraphics[width=\textwidth]{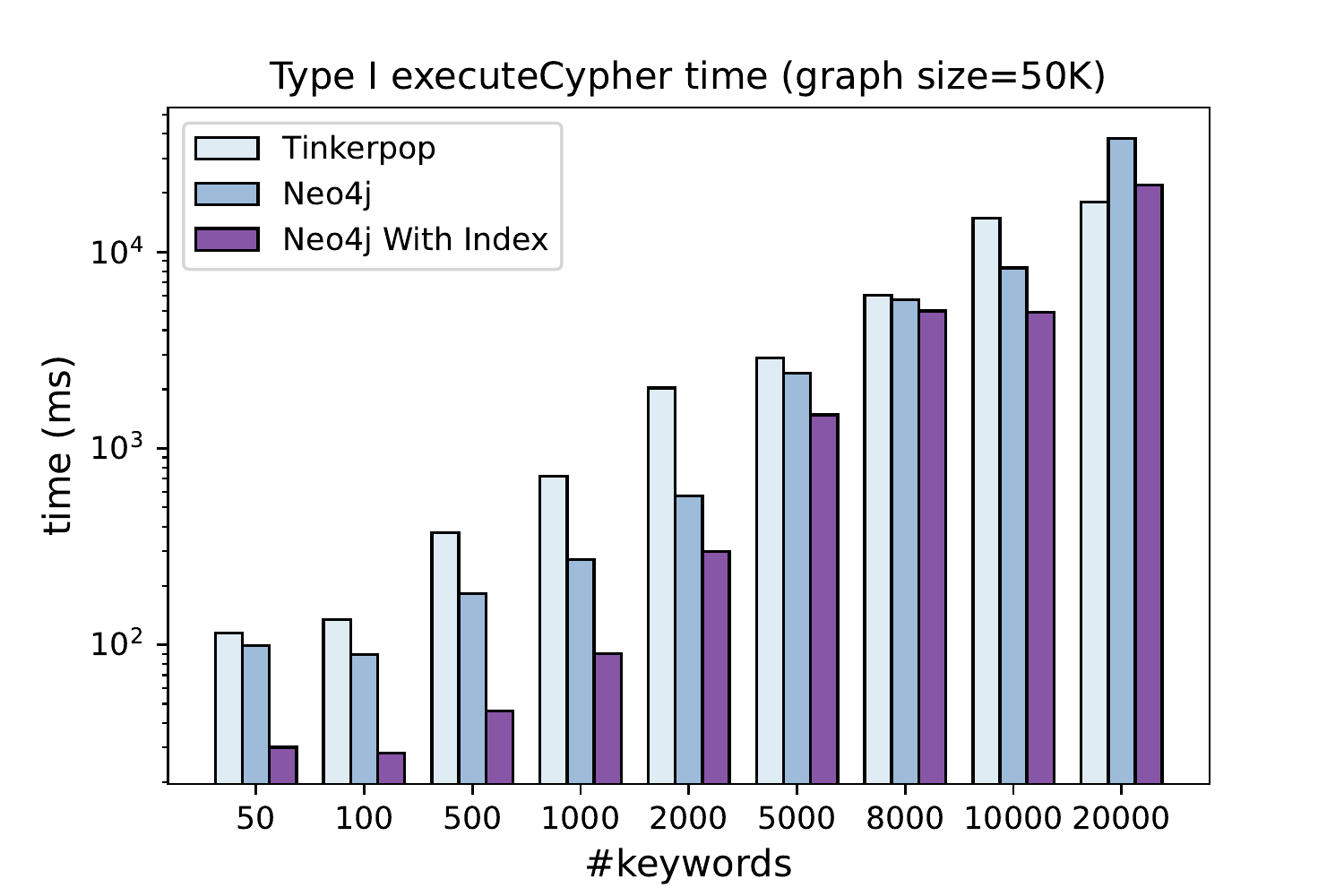}
        \caption{Graph size $=50K$.}
    \end{subfigure}
    \hfill
     \begin{subfigure}[b]{0.22\textwidth}
        \centering
        \includegraphics[width=\textwidth]{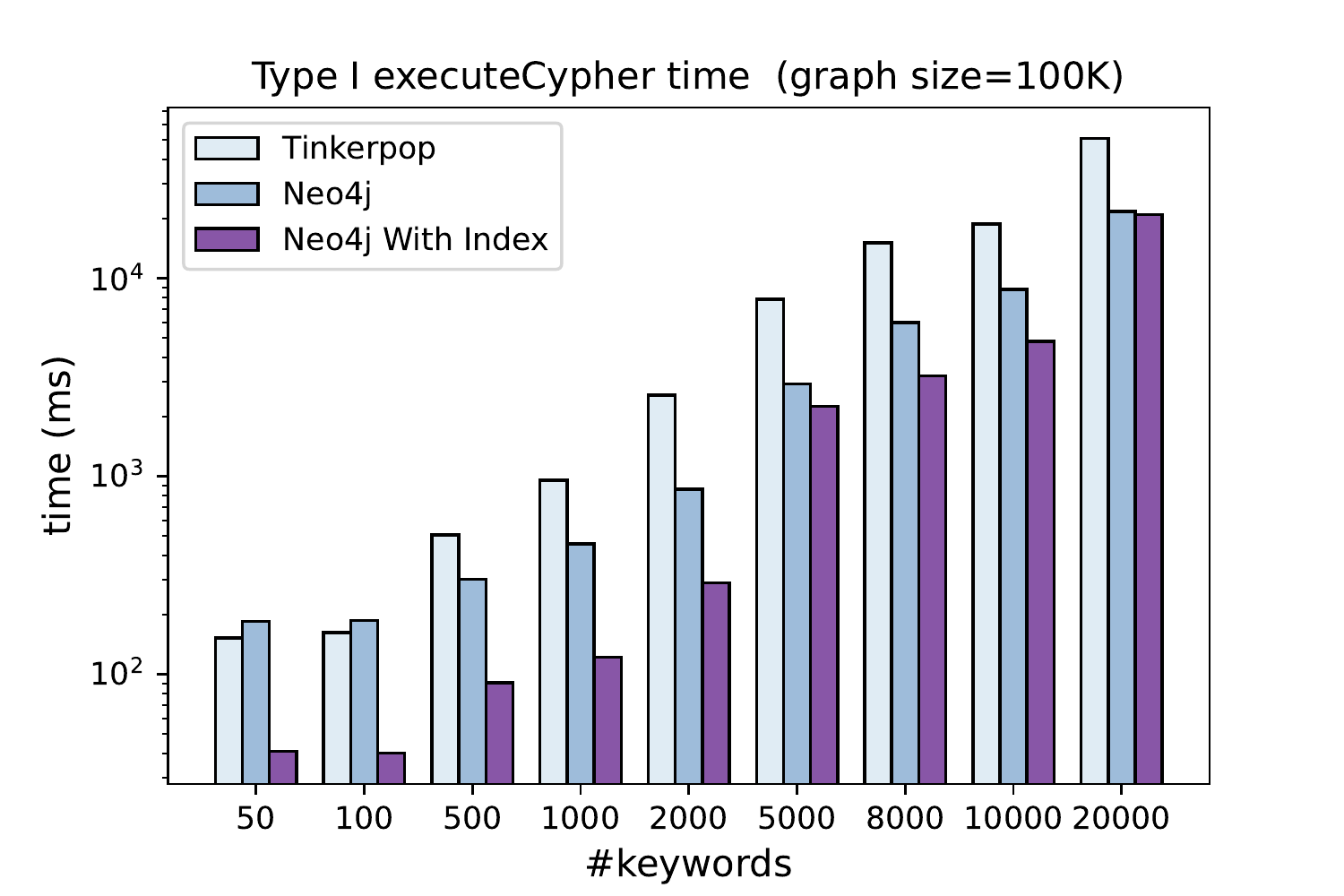}
        \caption{Graph size $=100K$.}
    \end{subfigure}
    \hfill
    \begin{subfigure}[b]{0.22\textwidth}
        \centering
        \includegraphics[width=\textwidth]{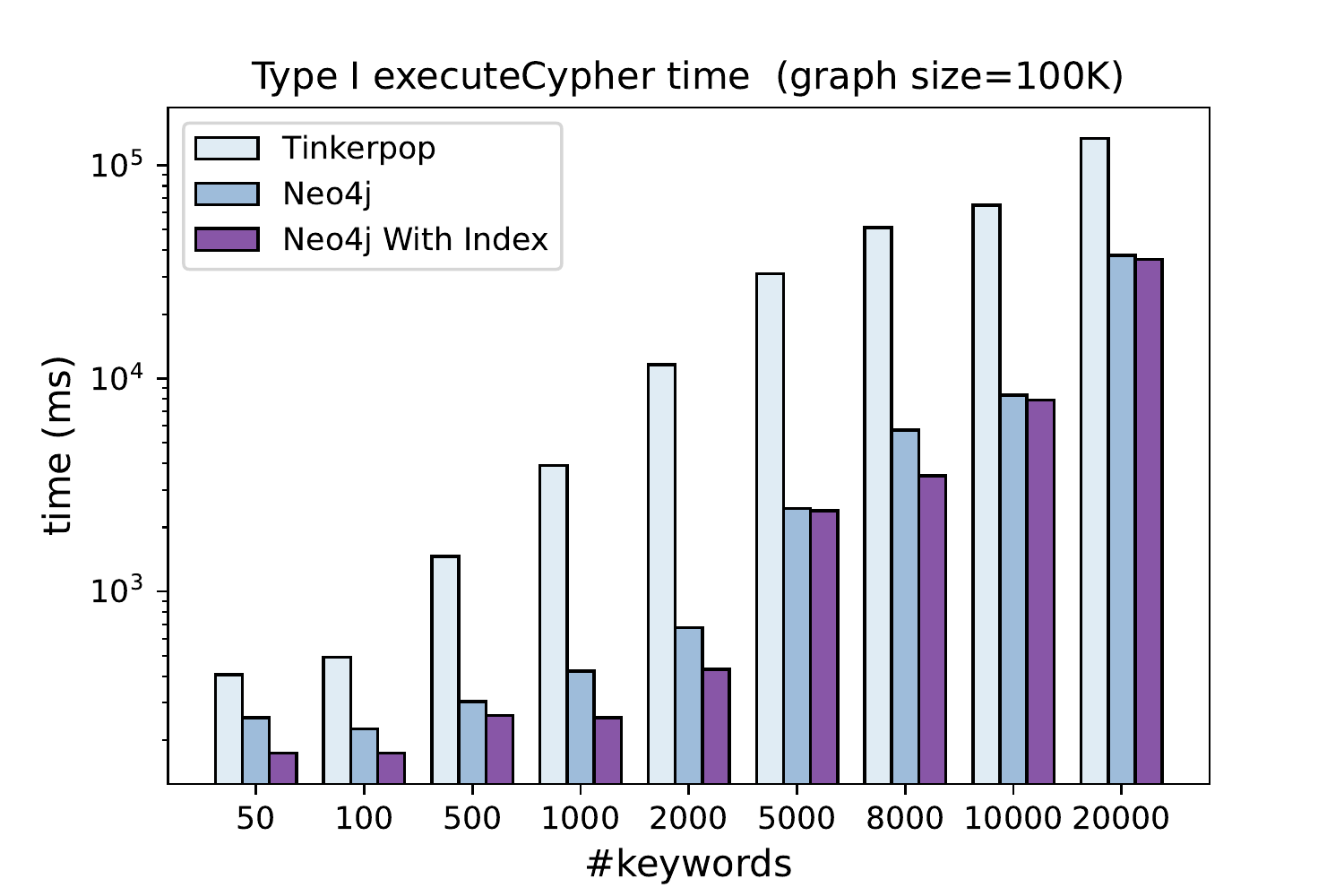}
        \caption{Graph size $=500K$.}
    \end{subfigure}
    \caption{Calibration results for Type I Cypher query w.r.t.~different graph sizes and \#keywords.}
    \label{fig:benchmarktype1}
\end{figure}

\begin{figure}[ht]
    \centering
    \begin{subfigure}[b]{0.22\textwidth}
        \centering
        \includegraphics[width=\textwidth]{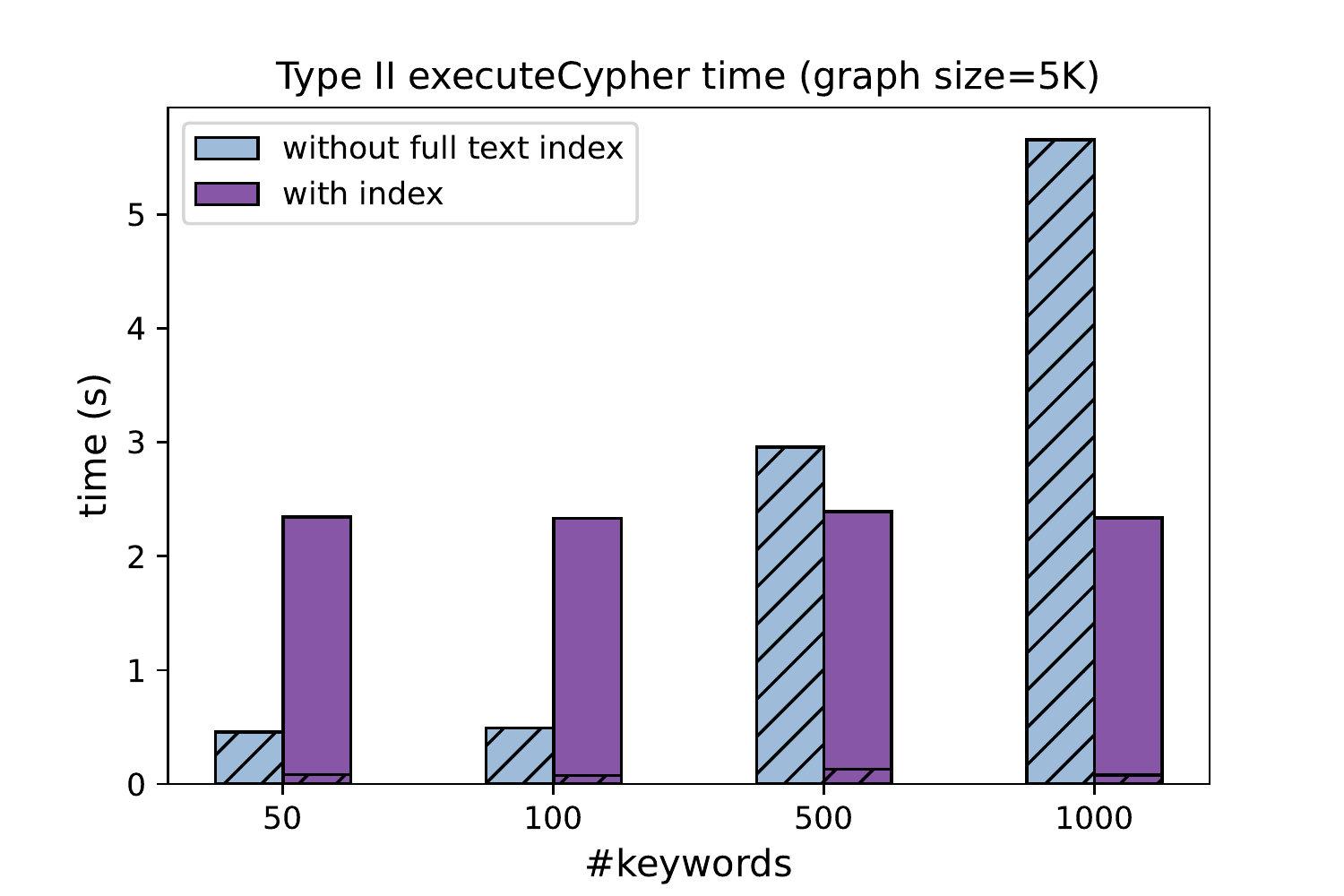}
        \caption{Graph size $=5K$.}
    \end{subfigure}
      \hfill
     \begin{subfigure}[b]{0.22\textwidth}
        \centering
        \includegraphics[width=\textwidth]{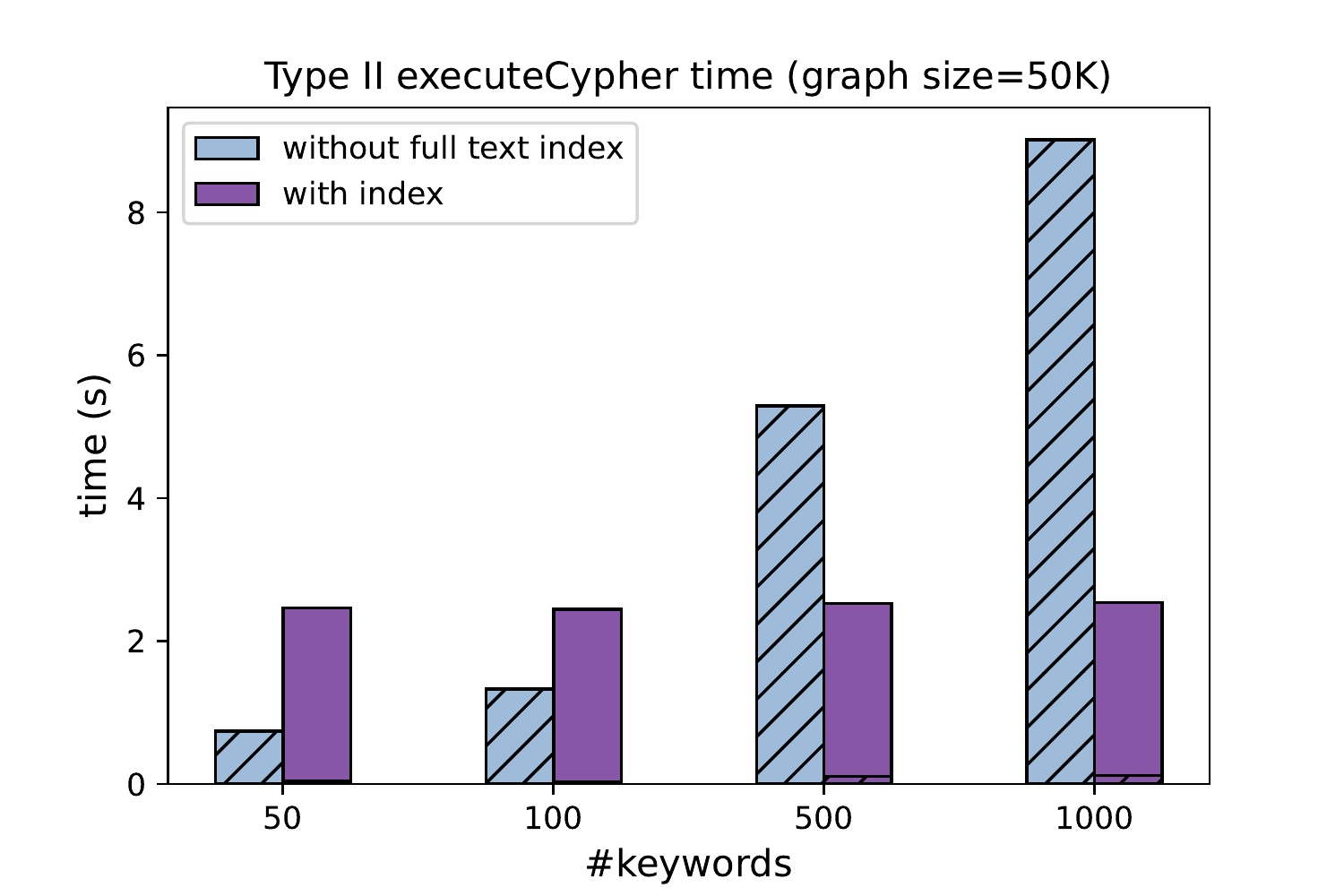}
        \caption{Graph size $=50K$.}
    \end{subfigure}
    \hfill
     \begin{subfigure}[b]{0.22\textwidth}
        \centering
        \includegraphics[width=\textwidth]{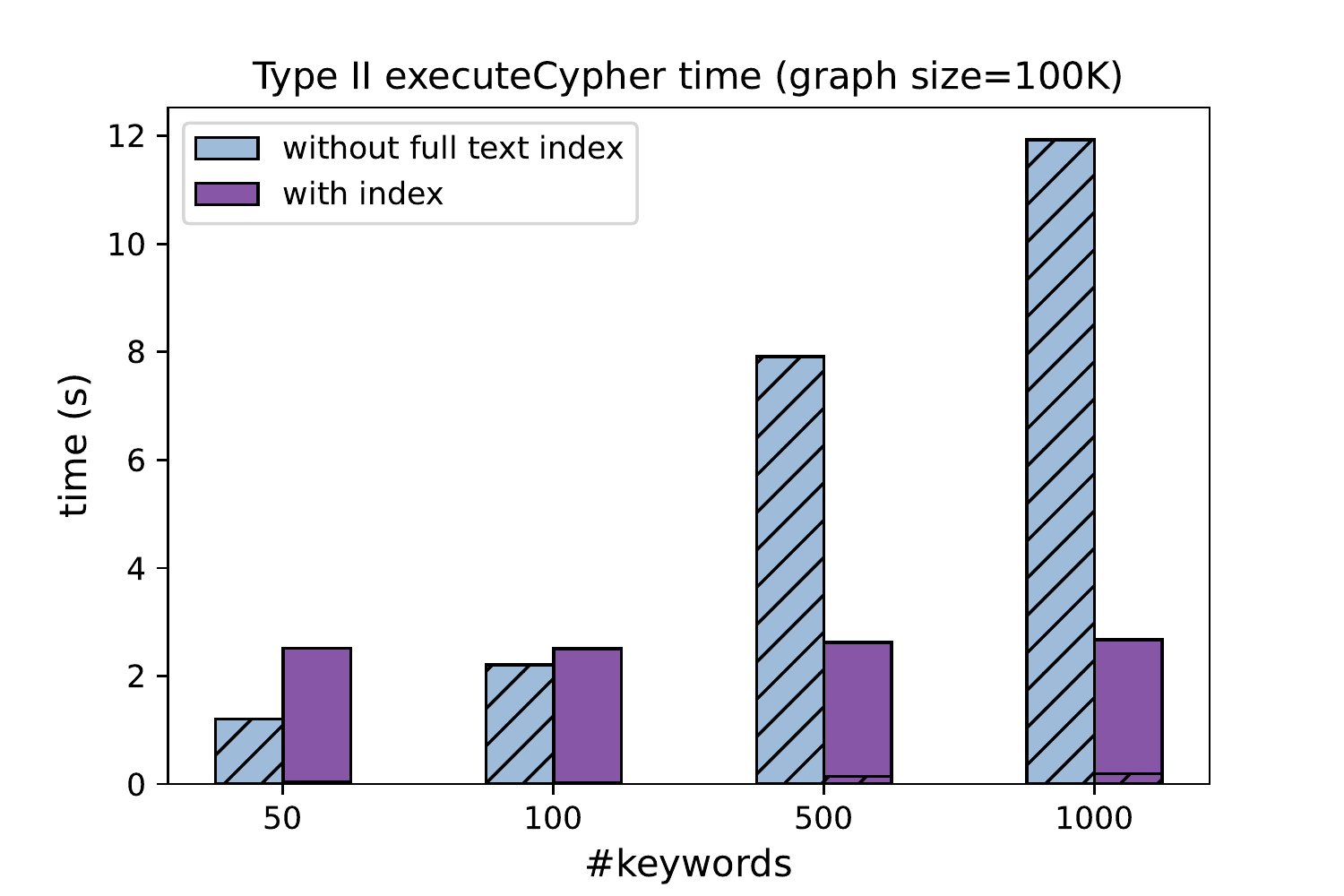}
        \caption{Graph size $=100K$.}
    \end{subfigure}
      \hfill
     \begin{subfigure}[b]{0.22\textwidth}
        \centering
        \includegraphics[width=\textwidth]{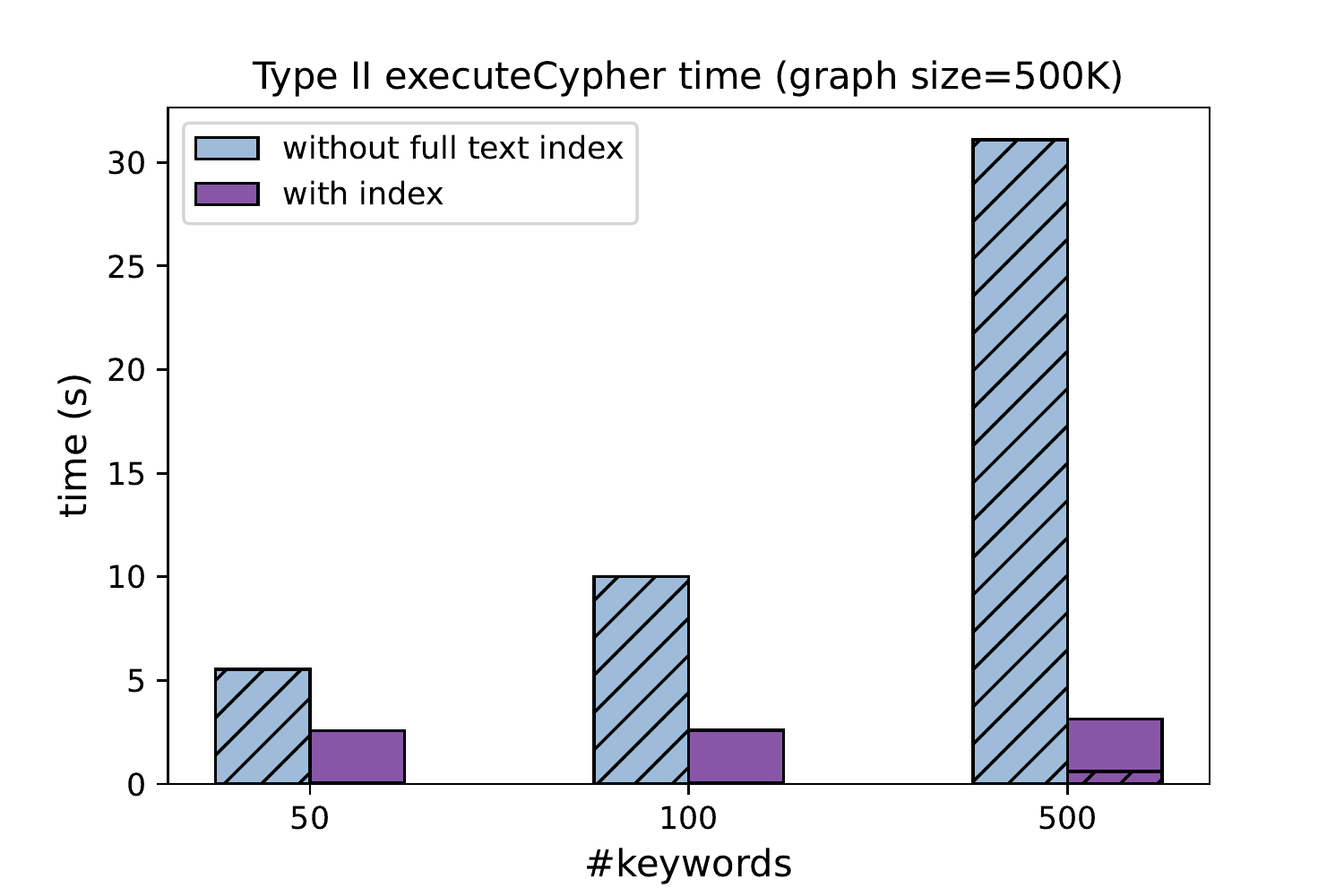}
        \caption{Graph size $=500K$.}
    \end{subfigure}
    \caption{Calibration results for Type II Cypher query w.r.t.~different graph sizes and \#keywords.}
    \label{fig:benchmarktype2}
\end{figure}

\begin{figure}[ht]
    \centering
    \begin{subfigure}[b]{0.21\textwidth}
        \centering
        \includegraphics[width=\textwidth]{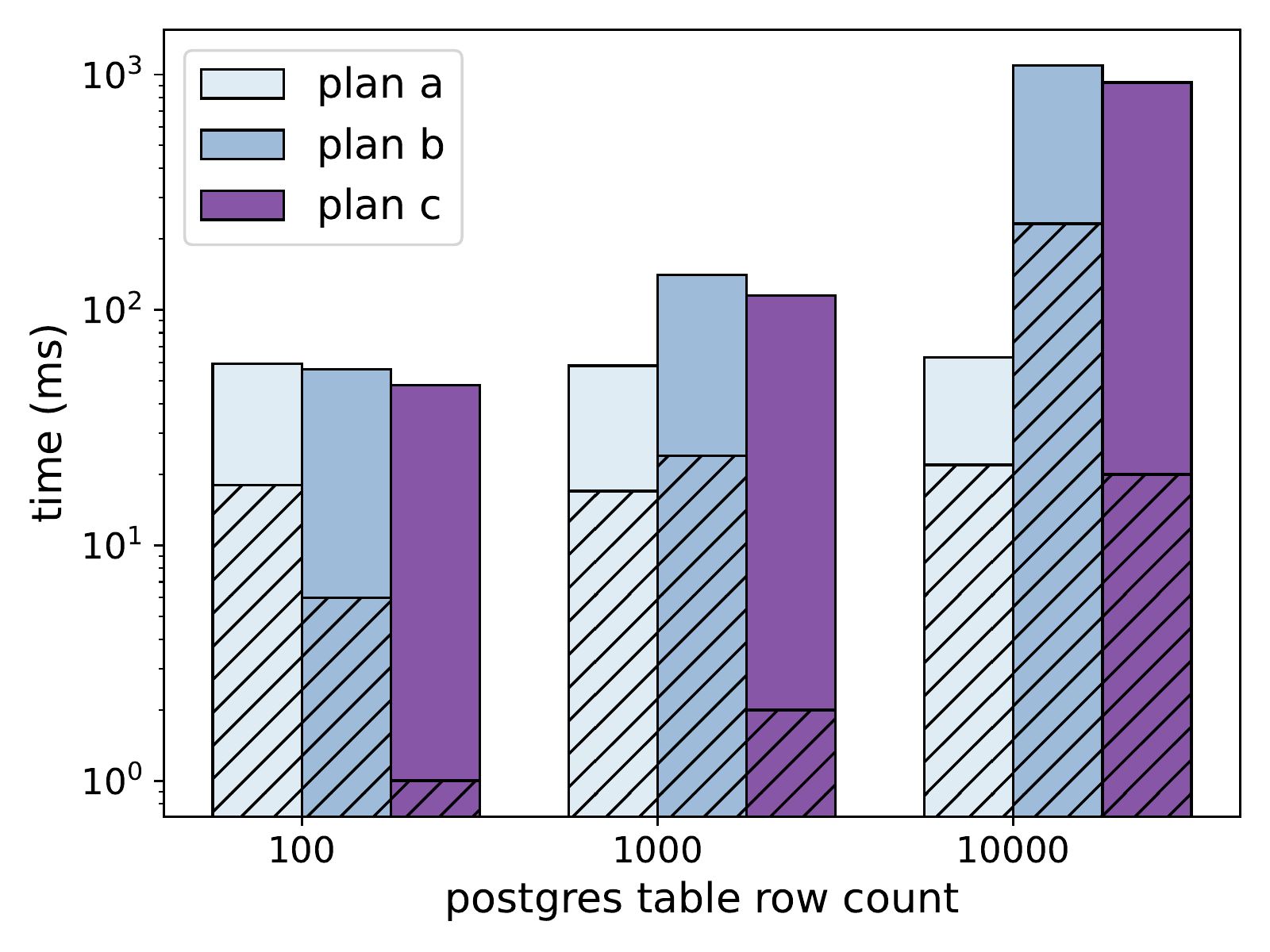}
        \caption{Row count of R: 100.}
    \end{subfigure}
    \hfill
     \begin{subfigure}[b]{0.21\textwidth}
        \centering
        \includegraphics[width=\textwidth]{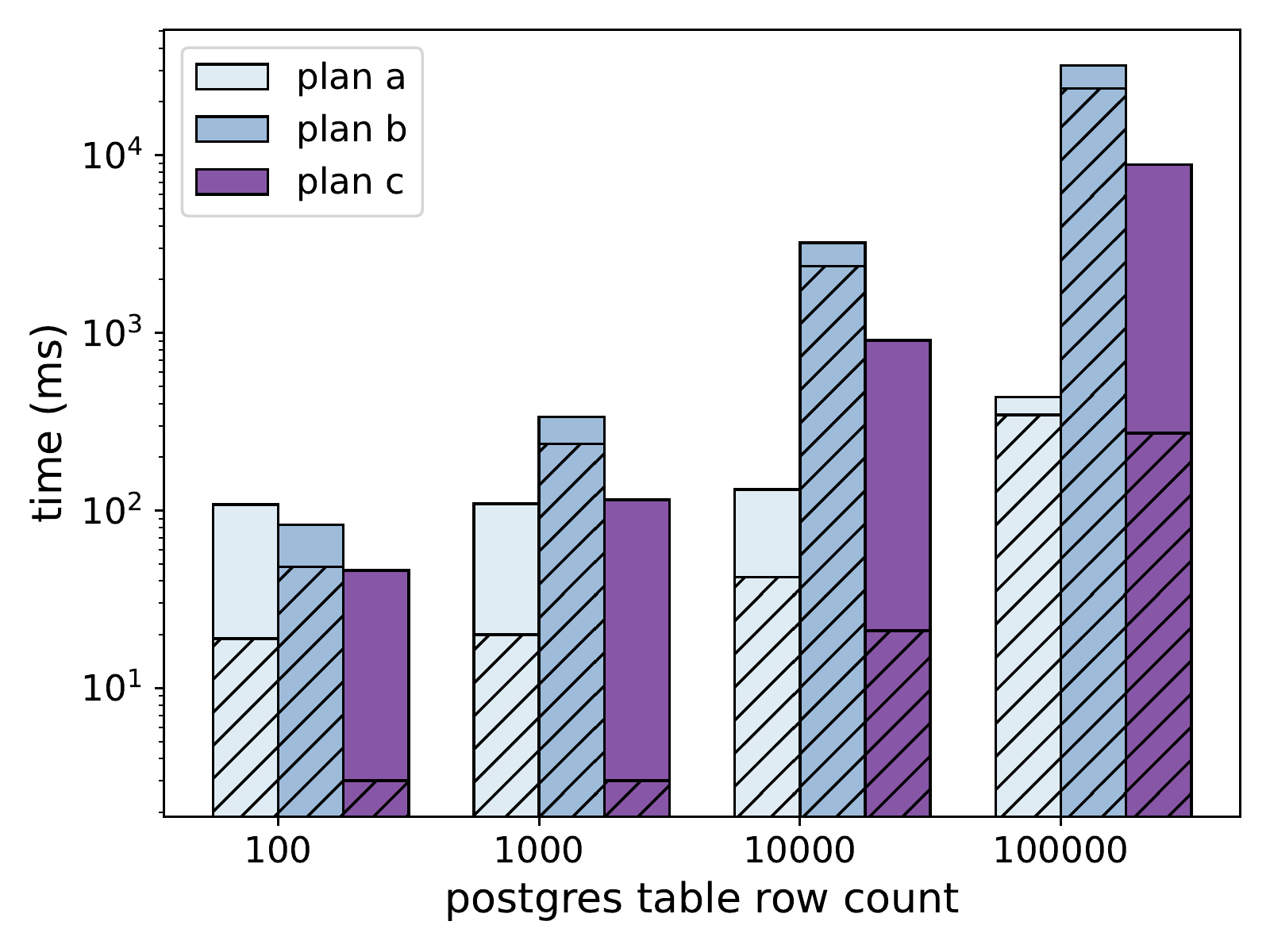}
        \caption{Row count of R: 1K.}
    \end{subfigure}
    \hfill
     \begin{subfigure}[b]{0.21\textwidth}
        \centering
        \includegraphics[width=\textwidth]{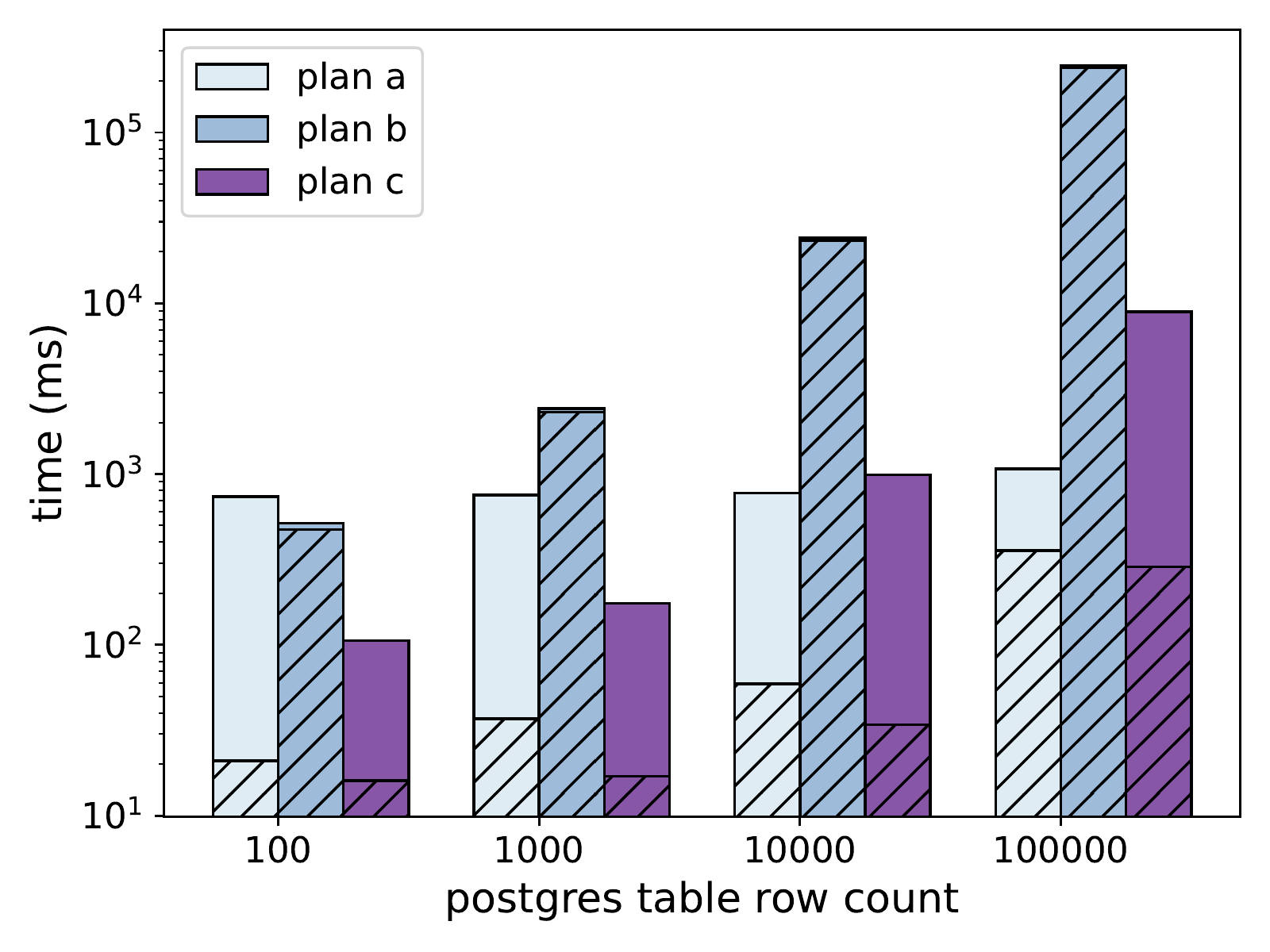}
        \caption{Row count of R: 10K.}
    \end{subfigure}
    \hfill
     \begin{subfigure}[b]{0.21\textwidth}
        \centering
        \includegraphics[width=\textwidth]{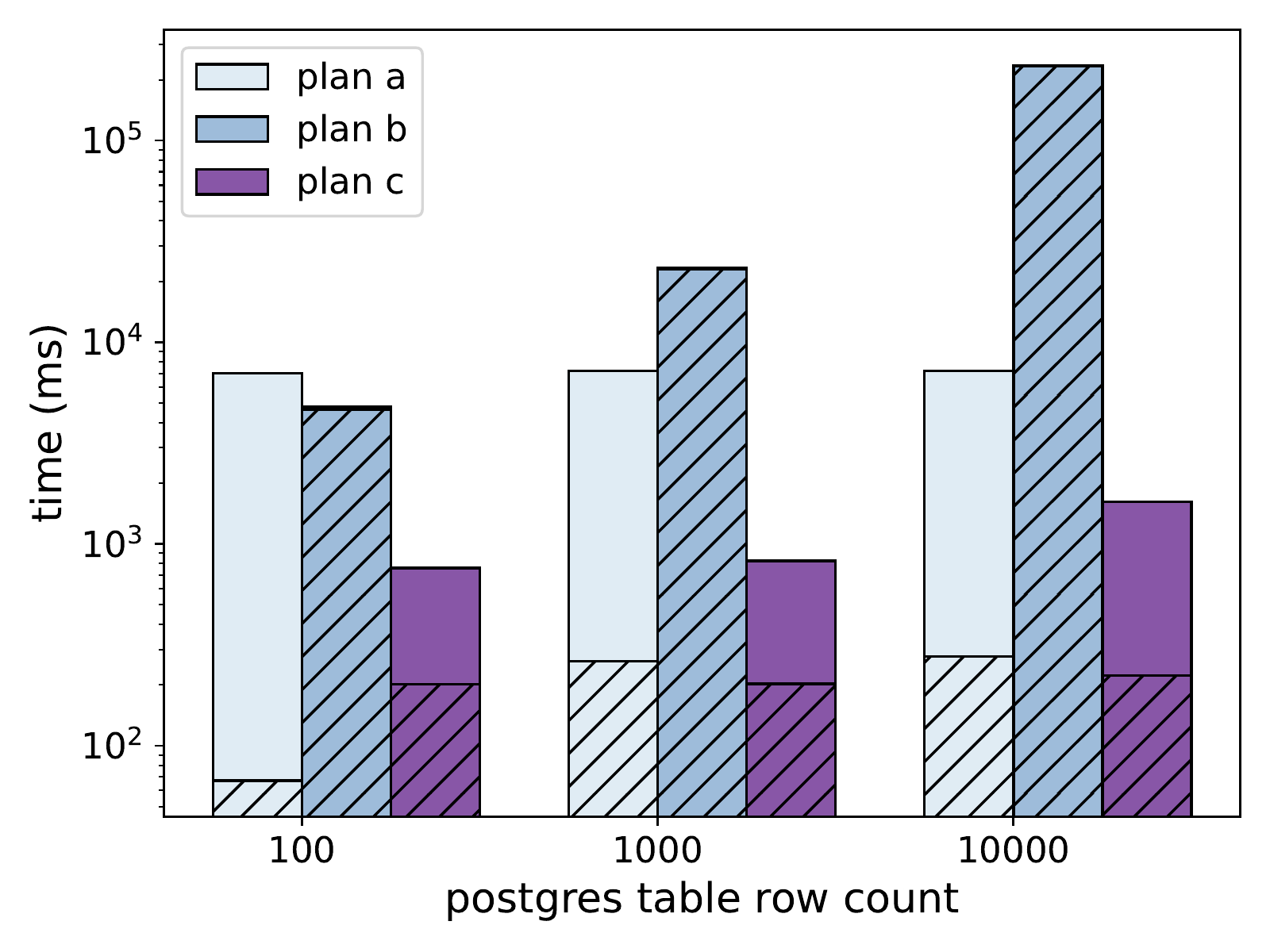}
        \caption{Row count of R: 100K.}
    \end{subfigure}
    \caption{Calibration results for cross engine \textit{executeSQL}}
    \label{fig:benchmark}
\end{figure}

\section{AWESOME logical and physical operators.}\label{appendix:ope}
\begin{table*}[h] 
\centering
\caption{AWESOME logical and physical operators.}\label{tab:operators}
\small
\begin{tabular}{p{0.05\textwidth}p{0.2\textwidth}llll} 
\toprule
Types& ADIL Keywords & Logical Operators & Physical Operators  & DP  & Buffering \\ 
\midrule
DBMS Query
& \begin{tabular}[c]{@{}l@{}}ExecuteCypher \\ExecuteSQL \\ExecuteSolr\end{tabular} & \begin{tabular}[c]{@{}l@{}}ExecuteCypher \\ExecuteSQL \\ExecuteSolr \\CreateRelation\end{tabular} 
& \begin{tabular}[c]{@{}l@{}}
ExecuteCypher@Neo4j\\
ExecuteCypher@Tinkerpop \\
ExecuteSQL@Postgres \\
ExecuteSQL@SQLite\\
ExecuteSolr\\
AwesomeTable2Postgres\\
AwesomeTable2SQLite\\
PostgresTable2SQLite \\
PostgresTable2Awesome\\
SQLiteTable2Postgres \\
SQLiteTable2Awesome
\end{tabular}  
& \begin{tabular}[c]{@{}l@{}}EX \\EX \\EX \\EX \\EX \\ST\\ST\\ST\\ST\\ST\\ST \end{tabular} 
& \begin{tabular}[c]{@{}l@{}} B \\ B \\B \\B \\ B \\SI\\SI\\B\\SO\\B\\SO \end{tabular}
\\ \midrule
Graph Operations 
& \begin{tabular}[c]{@{}l@{}}BuildWordNeighborGraph \\BuildGraphFromRelation  \\PageRank\\Betweenness \end{tabular} & \begin{tabular}[c]{@{}l@{}} CollectGraphElementsFromDocs\\CollectGraphElementsFromRelation \\CreateGraph \\PageRank\\ Betweenness \end{tabular} 
& \begin{tabular}[c]{@{}l@{}}CollectGraphElementsFromDocs\\CollectGraphElementsFromRelation \\CreateAwesomeGraph\\ CreateNeo4jGraph\\CreateTinkerpopGraph\\CreateJGraphTGraph\\PageRank@Neo4j \\PageRank@Tinkerpop\\PageRank@JGraphT
\\Betweenness@Neo4j \\Betweenness@Tinkerpop
\\Betweenness@JGraphT~\\ Neo4jGraph2JGraphT\\ 
Neo4jGraph2Tinkerpop\\ 
Neo4jGraph2Awesome\\ 
TinkerpopGraph2Awesome\\
TinkerpopGraph2JGraphT
\end{tabular} 
&  \begin{tabular}[c]{@{}l@{}} PR \\PR \\PR \\ EX \\ EX \\EX\\ EX\\EX \\EX \\EX\\EX\\EX\\ST\\ST\\ST\\ST\\ ST  \end{tabular}
& \begin{tabular}[c]{@{}l@{}}
SS\\SS\\SI\\SI\\SI\\SI\\B\\B\\B\\B\\B\\B\\B\\B\\B\\B\\B
\end{tabular}
\\ \midrule
Relation Operations &  GetColumns & GetColumns 
& \begin{tabular}[c]{@{}l@{}}GetColumnsFromPostgresT\\GetColumnsFromSQLiteT\\GetColumnsFromAwesomeT\\ Records2List \end{tabular}  
&\begin{tabular}[c]{@{}l@{}}EX\\EX\\ST\\DP \end{tabular}   
&\begin{tabular}[c]{@{}l@{}}B\\B\\SS\\SS\end{tabular}
\\\midrule
Text Operations  
&\begin{tabular}[c]{@{}l@{}}LDA\\SVD\\TopicModel\\KeyphraseMining\\Tokenize\\NER \end{tabular}  
& \begin{tabular}[c]{@{}l@{}}LDA\\SVD\\TopicModel\\KeyphraseMining\\NLPAnnotator(tokenize) \\ NLPAnnotator(ssplit)\\NLPAnnotator(pos)\\NLPAnnotator(lemma)\\NLPAnnotator(ner) \end{tabular}
& \begin{tabular}[c]{@{}l@{}}CreatDocumentsFromRecords\\CreatDocumentsFromList\\LDA\\SVD\\KeyphraseMining\\NLPAnnotator(tokenize) \\ NLPAnnotator(ssplit)\\NLPAnnotator(pos)\\NLPAnnotator(lemma)\\NLPAnnotator(ner)\end{tabular} 
& \begin{tabular}[c]{@{}l@{}}
PR\\PR\\EX\\EX\\EX\\PR\\PR\\PR\\PR\\PR\end{tabular} 
& \begin{tabular}[c]{@{}l@{}}
SS\\SS\\B\\B\\B\\SS\\SS\\SS\\SS\\SS\end{tabular} 
\\ \midrule
Matrix Operations 
& \begin{tabular}[c]{@{}l@{}}GetValue\\ GetRows\\ GetColumns\\ColumnKeys\\RowKeys\end{tabular}         
& \begin{tabular}[c]{@{}l@{}} GetValueByIndices\\GetValueByKeys\\GetRowsByIndices \\GetRowsByKeys\\GetColsByIndices\\GetColsByKeys\\ ColumnKeys\\RowKeys\end{tabular}                     
& \begin{tabular}[c]{@{}l@{}} GetValueByIndices\\GetValueByKeys\\GetRowsByIndices \\GetRowsByKeys\\GetColsByIndices\\GetColsByKeys\\ ColumnKeys\\RowKeys\end{tabular} 
&\begin{tabular}[c]{@{}l@{}}ST\\ST\\ST\\ST\\ST\\ST\\ST\\ST\end{tabular} 
&  \begin{tabular}[c]{@{}l@{}}B\\B\\B\\B\\B\\B\\B\\B\end{tabular} 
\\ 
\midrule
Other Functions       
& \begin{tabular}[c]{@{}l@{}}Sum\\Range\end{tabular}        
& \begin{tabular}[c]{@{}l@{}}SumList\\SumColumn\\SumMatrix\\SumVector\\Range\end{tabular}
& \begin{tabular}[c]{@{}l@{}}SumList\\SumColumn\\SumMatrix\\SumVector\\Range\end{tabular}
& \begin{tabular}[c]{@{}l@{}} PR\\PR\\PR\\PR\\ST \end{tabular}
& \begin{tabular}[c]{@{}l@{}} SI\\SI\\SI\\SI\\SO \end{tabular}
\\ \midrule
Data Movement & Store  & \begin{tabular}[c]{@{}l@{}}Store2Postgres \\ Store2Neo4j \\ Store2CSV\end{tabular} & 
\begin{tabular}[c]{@{}l@{}} List2Postgres\\AwesomeTable2Postgres\\AwesomeTable2Neo4j\\
AwesomeGraph2Neo4j\\AwesomeTable2List\\AwesomeGraph2List
\\StoreList2CSV
\end{tabular} 
& \begin{tabular}[c]{@{}l@{}}
ST\\ST\\ST\\ST\\DP\\DP\\ST  \end{tabular}
& \begin{tabular}[c]{@{}l@{}} 
SI\\SI\\SI\\B\\SS\\SO\\SI\end{tabular}\\
\bottomrule
\end{tabular}
\end{table*}